\documentclass[a4paper,fleqn,usenatbib]{mnras}
\usepackage[T1]{fontenc}
\usepackage{ae,aecompl}
\usepackage{amstext}
\usepackage{url}
\usepackage[]{natbib,amsmath,amssymb,times,refname,bm}
\bibpunct{(}{)}{;}{a}{}{,}
\usepackage{tabularx,amsmath,amssymb,hyperref}
\usepackage{graphicx,epsfig,color,latexsym}

\newcommand{\ii}{\mathrm{i}}
\newcommand{\bea}{\begin{eqnarray}}
\newcommand{\eea}{\end{eqnarray}}
\newcommand{\be}{\begin{equation}}
\newcommand{\ee}{\end{equation}}
\newcommand{\rund}[1]{\left(#1\right)}

\newcommand{\eck}[1]{\left[ #1 \right]}

\def\elabel#1{\label{eq:#1}}
\renewcommand{\exp}{\mathrm{exp}}

\title[elliptical plasma lens]{Two families of elliptical plasma lenses}

\author[Er and Rogers]%
       {Xinzhong Er$^1$ \thanks{xer@ynu.edu.cn}
         Adam Rogers$^2$ \thanks{RogersA@BrandonU.CA}\\
$^1$ South-Western Institute for Astronomy Research, Yunnan University, Kunming, P.R.China\\
$^2$Department of Physics and Astronomy, Brandon University, Brandon, MB, R7A 6A9, Canada\\
\\
}
\begin{document}
\maketitle

\begin{abstract}
  Plasma lensing is the refraction of low-frequency electromagnetic
  rays due to free electrons in the interstellar medium. Although the
  phenomenon has a distinct similarity to gravitational lensing,
  particularly in its mathematical description, plasma lensing
  introduces other additional features, such as wavelength dependence,
  radial rather than tangential image distortions, and strong
  demagnification of background sources. Axisymmetrical models of
  plasma lenses have been well-studied in the literature, but density
  distributions with more complicated shapes can provide new and
  exotic image configurations and increase the richness of the
  magnification properties. As a first step towards non-axisymmetrical
  distributions, we study two families of elliptical plasma lens,
  softened power-law and exponential plasma distributions. We perform
  numerical studies on each lens model, and present them over a parameter
  space. In addition to deriving elliptical plasma lens formulae, we
  also investigate the number of critical curves that the lens can
  produce by studying the lens parameter space, in particular the
  dependence on the lensing ellipticity. We find that the introduction
  of ellipticity into the plasma distribution can enhance the lensing
  effects as well as the complexity of the magnification map.

\end{abstract}
\begin{keywords} gravitational lensing: strong - gravitational lensing: micro, Interstellar medium
\end{keywords}

\vspace{1.0\baselineskip}

\section{Introduction}

Compact radio sources exhibit intervals of rapid change in their
flux-density, which have been attributed to extreme scattering events
(ESEs). The first ESEs were discovered decades ago \citep{ESE0}, but
the construction of detailed physical models to describe all aspects
of the phenomenon remains open. The millisecond duration pulses known
as fast radio bursts
\citep[FRBs;][]{2007Sci...318..777L,2018NatAs...2..842P,frbreview}, may also be
related to plasma lensing in a dense medium, though it remains unknown if this environment is located in the host galaxy, or is an intervening structure along the line of sight. The chromaticity of FRBs
suggests that strong refraction by plasma structures in the intervening
interstellar medium (ISM) may be related to the origin of these
events. Additionally it has been observed that the 2-dimensional dynamic power
spectra of some pulsars contain remarkably
organized parabolic structures
\citep[e.g.][]{stinebring,2007ASPC..365..254S}, which can be explained
by highly-anisotropic diffractive scattering of radio waves from the
pulsar \citep[e.g.][]{2004MNRAS.354...43W,2006ApJ...637..346C}.
Moreover, the radio pulsar time delay has been attributed to plasma
structures in the ISM \citep[e.g.][]{2017MNRAS.464.2075S}. Plasma
distributions enclosing compact objects have also been shown to alter
their appearance \citep{rogers17a}.

Refractive plasma lens models have been developed that describe plasma
density distributions analytically. As the prototypical example,
\citet{cleggFey} studied the properties of a one dimensional Gaussian
plasma lens. This model has been widely applied through the literature
and forms the basis for analytical models of isolated plasma
distributions \citep[e.g.][]{2018MNRAS.481.2685D}. It has been suggested that the plasma lensing in the
host galaxies of FRBs can modulate the amplitudes of the bursts
\citep{FRBplasma1}. This analysis was undertaken using a
one-dimensional Gaussian lens. In addition, \cite{newESE2017,2017ApJ...845...90V, kerr2018} present the case of an ESE with typical U-shaped and
W-shaped light curves, which was modelled using a superposition of two
one-dimensional Gaussians to build up a dual-lobed morphology for a
slice across the plasma density. The small scale magnetic field in the
black widow pulsar system PSR B1957+20 has been revealed due to
lensing in which plasma stripped from the pulsar's companion acts to
magnify the pulsar during eclipses
\citep{2018arXiv180910812L,2018arXiv180809471S}. \cite{coles2015}
models the Astronomical Unit (AU) scale plasma lensing for pulsar observations, and also
suggests an electron density on the order of $\sim10$ cm$^{-3}$. In
addition, model independent inversion methods have also been developed
and applied to data directly. The dynamic spectrum of the ESE in the
radio loud quasar PKS 1939-315 was used to determine the column
density profile of the plasma lens \citep[see, for example,]
[]{ESE3,ESE4}.

Despite these successes, there are lingering mysteries that remain
regarding the nature of plasma lensing. It is difficult to interpret
the high density and pressures within isolated plasma lenses. The
dispersion measure (DM), i.e. the integrated column density of free
electrons between the observer and the source, necessary to account
for the lensing effect is too large ($\sim10^3$ cm$^{-3}$) for a
structure in the ionized interstellar medium \citep{cleggFey}, and
cannot exist in pressure balance with the ambient interstellar medium
in the Milky Way \citep[e.g.][]{2002astro.ph..7156C}. Some suggestions
have been put forth, e.g., it requires modest electron density if
highly elongated plasma sheets are seen from an edge-on perspective
\citep{romani87,plasmaSheets,ESE2,ESE5,2010ApJ...708..232B}. The
plasma sheet has also been found in numerical simulations of the
supernova-driven turbulence \citep[e.g.][]{2012ApJ...750..104H}. On
the other hand, previous studies on plasma lens models mainly focus on
the axi-symmetrical Gaussian distribution of the electrons, which is a
smooth density profile. However, there is the more realistic
possibility of asymmetry and clumpiness in the electron density. The
ellipticity and small scale variations of the lens may increase the
lensing efficiency, e.g. the lens with low electron density may also
generate high magnifications. Moreover, the distribution of the
ionized interstellar medium may be affected by the stellar wind or the
explosion of the supernovae, and forms a highly elongated shape. Thus,
the investigation of different density profiles, especially the
elliptical models is required. In particular, the shape of shock fronts
can be approximated by the highly elliptical models and explicit
analytical models can help us.

\citet{ESE3} have demonstrated the usefulness of real-time, multi-wavelength monitoring of ESEs in progress. Their work provides a new ESE detection method, which formed the basis for the Australia Telescope ESE project \citep[ATESE;][]{ESE3, tuntsov17} that is positioned to detect many more such events in progress and map the column density through the lenses with dedicated monitoring. Such observations have been used to eliminate the possibility of an isotropic spherical lens structure, favoring anisotropic projected densities instead, such as shells and filaments. However, this does not mean that all such ESEs are formed by lenses with the same morphologies. In fact, the spherical Gaussian lens \citep{cleggFey} remains the most widely applied lens model to date in the ESE literature and remains relevant today \citep{FRBplasma1}. It is well within the possibility that the distinction between an elliptical and spherical lens could be feasible by an analogous observational campaign. More monitoring of ESEs in progress is required.

The ``non-parametric'' ESE modeling performed by \citet{ESE4} shows that more observations are required to distinguish between axisymmetric and highly anisotropic lenses. The elliptical lens is a versatile model that interpolates between spherical and highly anisotropic charge distributions.
Thus, elliptical lenses naturally bridge two extremes that represent the most successful and interesting options for modeling ESEs.

The deflection by the plasma on a given frequency is similar in
practice to gravitational lensing. Thus, well-established theory in
gravitational lensing \citep[e.g.][]{SEF,perlickBook} can be adopted
for studying plasma lensing, although there is some difference in the
plasma lens case. First of all, gravitational lensing mainly occurs on
cosmological scales \citep[e.g.][]{2010A&A...516A..63S}, caused by
massive objects such as galaxies \citep[e.g.][]{2009ApJ...703L..51K}
or galaxy clusters \citep[e.g.][]{2006ApJ...648L.109C}. The exception
is microlensing by stars within the Galaxy
\citep[e.g.][]{2012RAA....12..947M}.  In reality, the microlensing
case is similar to the plasma lensing, since in both scenarios the
deflection occurs on small scales and depends on the relative motion
between the lens and the source. Mass models for gravitational lenses
are well studied from galaxy dynamics
\citep[e.g.][]{1987gady.book.....B, 2012MNRAS.419..656C,
  2012MNRAS.423.1073B, 2019arXiv190309282L} and cosmological
simulations
\citep[e.g.][]{nfw97,2015MNRAS.451.1247S,2016MNRAS.456..739X}. In
  contrast, for plasma lenses, only simple symmetrical models for the
electron density have been largely considered. Model degeneracy has
been found between symmetric dual-component plasma lens distributions
and an asymmetric model used to fit the ESE in PKS 1939-315
\citep{ESE4,rogers&er2019}. The parametric model shows that this
distribution is one of a larger family in which two lenses act on a
source.

In general, plasma lenses are diverging, however they can
cause large magnifications if they are offset from the background
source along the observers line of sight. It thus will enrich the
variability of the light curves, e.g. a volcano shape in the pulsar
light curves \citep[e.g.][]{coles2015}.  Moreover, the deflection due
to the plasma lens is chromatic. Only in the radio bands can
significant signatures be observed. In extremely low frequency
observations, ray optics may no longer apply and wave effects may
become significant. Wave asymptotes in plasma lensing have also been
studied recently \citep{2018arXiv181009058G}.

In current observations, the frequency ranges mainly from a few MHz to
a few GHz \citep[e.g.][]{2010ApJ...708..232B,coles2015}.  Therefore,
wave effects may not strong enough to be detected
\citep[e.g.][]{nakamuraDeguchi,nambu1}. Even in geometric optics,
  the optical properties of plasma lens models need to be
  systematically studied for magnification and critical curves
  \citep{2002astro.ph..7156C}. For example, the Gaussian or
  axisymmetric profiles have been discussed by
  \citet[e.g.][]{2004ARA&A..42..275S}. In this work, we will follow
similar methodology of geometric optics as in gravitational lensing to study the magnification and critical curves in plasma lensing.
We study two families of
analytical models for the electron distribution. In particular, we
focus on the ellipticity of the plasma density, and how it will
increase the magnification. Elliptical models of gravitational lenses
have been widely studied both analytically and numerically
\citep[e.g.][]{2001astro.ph..2341K}. In gravitational lensing, it has
been shown that the elliptical distribution and the small scale
variations, i.e. sub-structures can significantly changed the lensing
properties, such as increase the cross section of the multiple images.
The elliptical gravitational lenses show high efficiency of generating
multiple images and high magnification. Two kinds of elliptical models
have been studied in the gravitational lens literature, which
introduce ellipticity into the lens potential
\citep[e.g.][]{1987ApJ...321..658B, 1990A&A...231...19S} or by means
of an elliptical mass density \citep[e.g.][]{1993ApJ...417..450K,
  1994A&A...284..285K, 1998ApJ...495..157K,
  2015A&A...580A..79T}. Unlike in gravitational lensing, the plasma effective lensing ``potential" is proportional to
the projected density of electrons. The ellipticity in the potential
and electron density will give the same result. Thus, we will modify
the lens potential into an elliptical form in this work. We present
the elliptical exponential plasma lenses and softened power-law (SPL)
lenses and discuss the properties of the magnification in
Section\,\ref{sec:ellp}. We summarize our conclusions in Section
\ref{sec:conclusions}. A gallery of criticals and caustics for a
selection of elliptical lens examples is shown in the appendix.

In our study we will consider pulsars or AGN, which will be treated as point sources in our calculations. Another
approximation is the effect of cosmological distances (e.g. the
difference between angular diameter distance and luminosity distance)
will not taken into account as we assume that the lenses are within
our Galaxy.

\section{Basic formulae}

We will outline the basic formulae regarding the two families of
plasma lenses in this work. More details can be found in
\citet{ErRogers18}. The notation follows the general gravitational
lens formalism discussed in \citet{SEF, narayan}. For astrophysically
relevant situations, the lens is considered weak. Due to the large
distances between the source and lens ($D_\text{ds}$) and the
distances from lens and source to the observer ($D_\text{d}$ and
$D_\text{s}$ respectively) compared to the diameter of the lens
  structure, the thin lens approximation can be adopted \citep[see e.g.][]{narayan}. We introduce
angular coordinates $\theta=\sqrt{\theta_1^2+\theta_2^2}$ with respect
to the line-of-sight, and those on the source plane as $\beta$. They
are related through the lens equation
\be
\beta = \theta - \alpha = \theta - \nabla_\theta \psi(\theta),
\ee
where $\alpha$ is the deflection angle, $\psi$ is the effective lens
potential and $\nabla_\theta$ is the gradient on the image plane.

\subsection{Exponential Lenses}
\label{sec:Exp}
The exponential lenses are a widely used model to describe ESEs, since
the Gaussian lens model introduced by \citet{cleggFey} to model
observations of the extragalactic sources 0954+654 and 1741-038 is a
special case of the exponential model. We follow the description of
exponential models in \citet{ErRogers18} and the Gaussian lens in
\citet{cleggFey}, and specify the projected electron distribution on the lens plane $N_e(\theta)$. This choice is made to compare with the most common parameterization in the literature \citep{cleggFey}. We adopt a form for the electron column
density in the lens plane,
\be
N_e(\theta)=N_0\,{\rm exp}\rund{-{\theta^h\over h\sigma^h}}\quad\quad(\theta>0),
\ee
with $N_0$ the maximum electron column density within the lens and
$\sigma$ as the width of the lens for $h>0$
\citep{newESE2017, ErRogers18}. The projected electron density gives
the potential
\be
\psi(\theta)= \theta_0^2 \exp\left( -\frac{\theta^h}{h\sigma^h} \right)
\elabel{exp-pot}
\ee
and deflection angle
\be
\alpha_\text{exp}(\theta)=-\theta_\text{0}^2
\frac{\theta^{(h-1)}}{\sigma^h}\exp\left( -\frac{\theta^h}{h \sigma^h} \right)
\label{deflExp}
\ee
with the characteristic angular scale
\be
\theta_0 = \lambda \left(\frac{D_\text{ds}}{D_\text{s} D_\text{d}}
\frac{1}{2\pi} r_\text{e} N_\text{0} \right)^\frac{1}{2},
\ee
where $\lambda$ is the observing wavelength and $r_e$ is the classical
electron radius. The wavelength of a photon $\lambda$ can vary in the
gravitational field via the gravitational redshift effect, which
introduces an additional complication to the deflection angle. Since
we are only discussing lensing from plasma, the gravitational
deflection will not be taken into account in this work.

For each $h$ value, we can define a critical limit below which
the exponential lenses produce no critical curves, and therefore only
a single image: $\theta_0<f(h)\sigma$, and
\be
f(h) = \eck{F^{h-2\over h} (F+1-h) e^{-F\over h}}^{-1/2},
\ee
where the factor $F$ is
\be
F={1\over 2}\eck{3(h-1) + \sqrt{(h-1)(5h-1)}}.
\ee
One can find more detail about the critical limit in \citet{ErRogers18,rogers&er2019}.

\subsection{Power-Law Lenses}
\label{sec:PL}
The optical properties of the power-law (PL) lens are determined by the charge volume density $n_\text{e}(r)$ (Sec. \ref{PL1}). The general approach to setting up the lensing equations is to project the volume density to the lens plane to produce $N_\text{e}(\theta)$, the column density. Therefore, with a model for the charge volume density $n_\text{e}$ we can find the column density and derive the lens potential, deflection angle and magnification. However, rather than putting the ellipticity directly into the volume density $n_\text{e}$, we can instead parameterize the 2D projection, the column-density $N_\text{e}(\theta)$ (the dispersion measure (DM) of the lens; Sec. \ref{PL2}). This approach of parameterizing the projected electron density follows \citet{cleggFey}, who originally used it to establish the successful Gaussian lens. However, as we will show, the two approaches produce power-law lens models that are simple transformations of one another. This behaviour is radically different from the two related elliptical gravitational lens models, one of which includes ellipticity in the potential \citep{1993ApJ...417..450K} and the other includes ellipticity in the mass distribution \citep{1998ApJ...495..157K}. Let us investigate both cases below.

\subsubsection{Volume-Density Power-Law}
\label{PL1}
The family of power-law lenses is produced by a three-dimensional electron volume density given by
\be
n_\text{e}(r) = n_\text{0}\frac{R_0^h}{r^h}
\ee
with the power-law index $h$ and the characteristic radius $R_\text{0}$
at which $n_\text{e}(R_\text{0}) = n_\text{0}$.
This electron density profile produces an effective lens potential \citep{ErRogers18}:
\begin{equation}
\psi(\theta) = \left\{
\begin{array}{ll}
 -\theta_{0}^2 {\rm \ln} \theta, & h=1\\
\\
 \dfrac{\theta_{0}^{h+1}}{(h-1)} \dfrac{1}{\theta^{h-1}}, & h \neq 1
\end{array}\right.
\label{softpl}
\end{equation}
The lens potential gives the deflection angle
\be
\alpha_\text{PL}(\theta) = - \frac{\theta_\text{0}^{h+1}}{\theta^h}
\label{deflPL}
\ee
which is written in terms of the characteristic angular scale
\be
\theta_0 = \left( \lambda^2 \frac{D_\text{ds}}{D_\text{s} D_\text{d}^h}
\frac{r_\text{e} n_\text{0} R_\text{0}^h}{\sqrt{\pi}}
\frac{\Gamma\left( \frac{h}{2} + \frac{1}{2} \right)}{\Gamma \left( \frac{h}{2} \right)} \right)^\frac{1}{h+1}.
\ee
The deflection angle for this density distribution was first evaluated
by \citet{1966PhRvL..17..455M}, and \citet{BKT09} present a study from
gravitational lensing by a compact object embedded in a non-uniform
plasma.

In order to soften the singularity at the origin, it is trivial to
include a finite core $\theta_{\rm c}$ by simply making the
transformation $\theta \rightarrow \sqrt{\theta^2 +
  \theta_\text{c}^2}$ in the deflection angle. Similar to the exponential lens, we derive a critical value of
the core size required to produce a critical curve. This limit is given
in terms of the characteristic angular scale, such that
\be
\theta_{\rm crit} = \theta_0\eck{2\rund{{3\over h} +1 }^{-{h+3\over 2}} }^{1\over h+1}.
\ee
For $\theta_c>\theta_{\rm crit}$, the power-law lens cannot form critical curve.
The detail about the core size can be found in \citet{rogers&er2019}.

\begin{figure}
  \includegraphics[width=8.0cm]{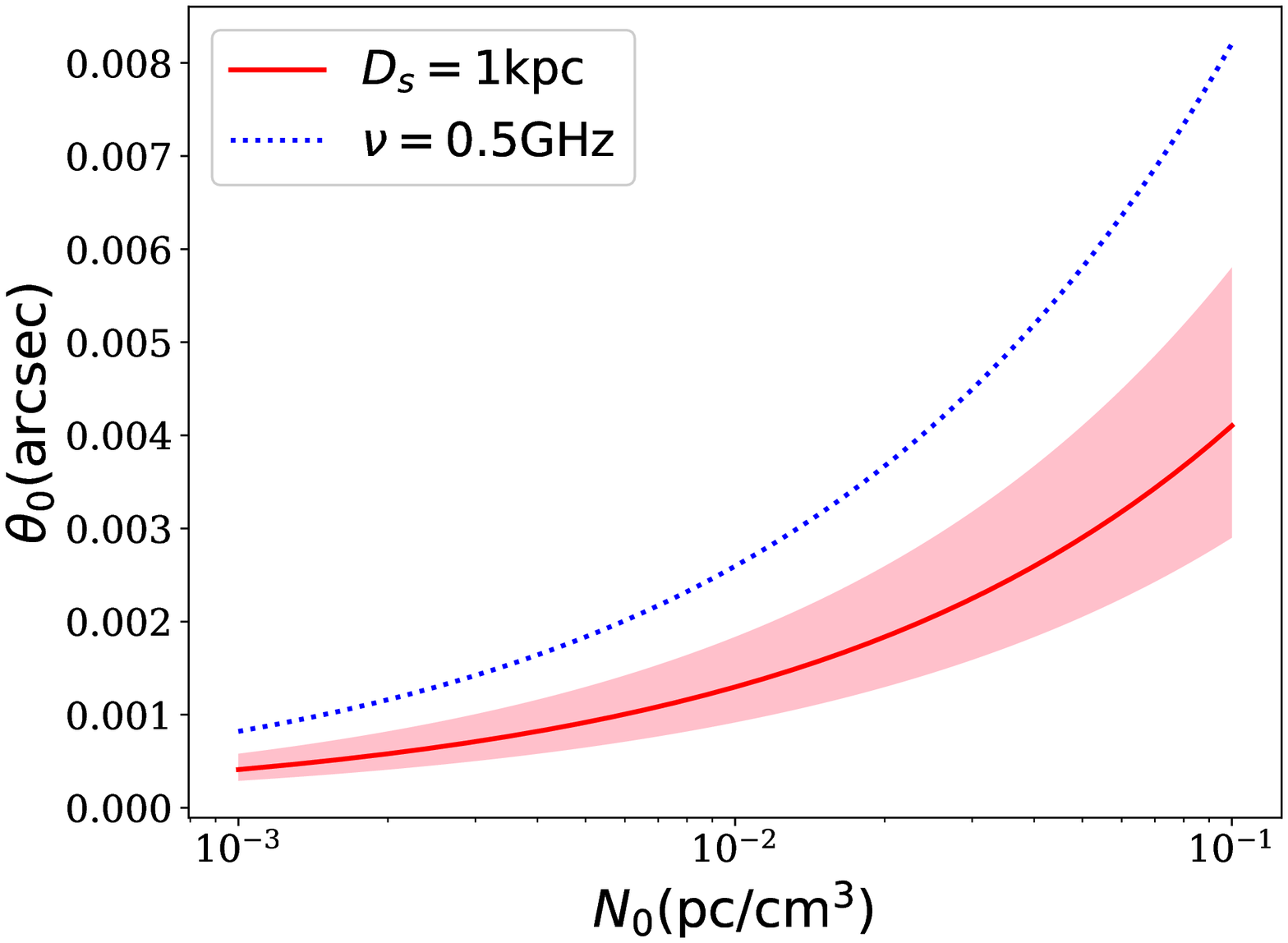}
  \includegraphics[width=8.0cm]{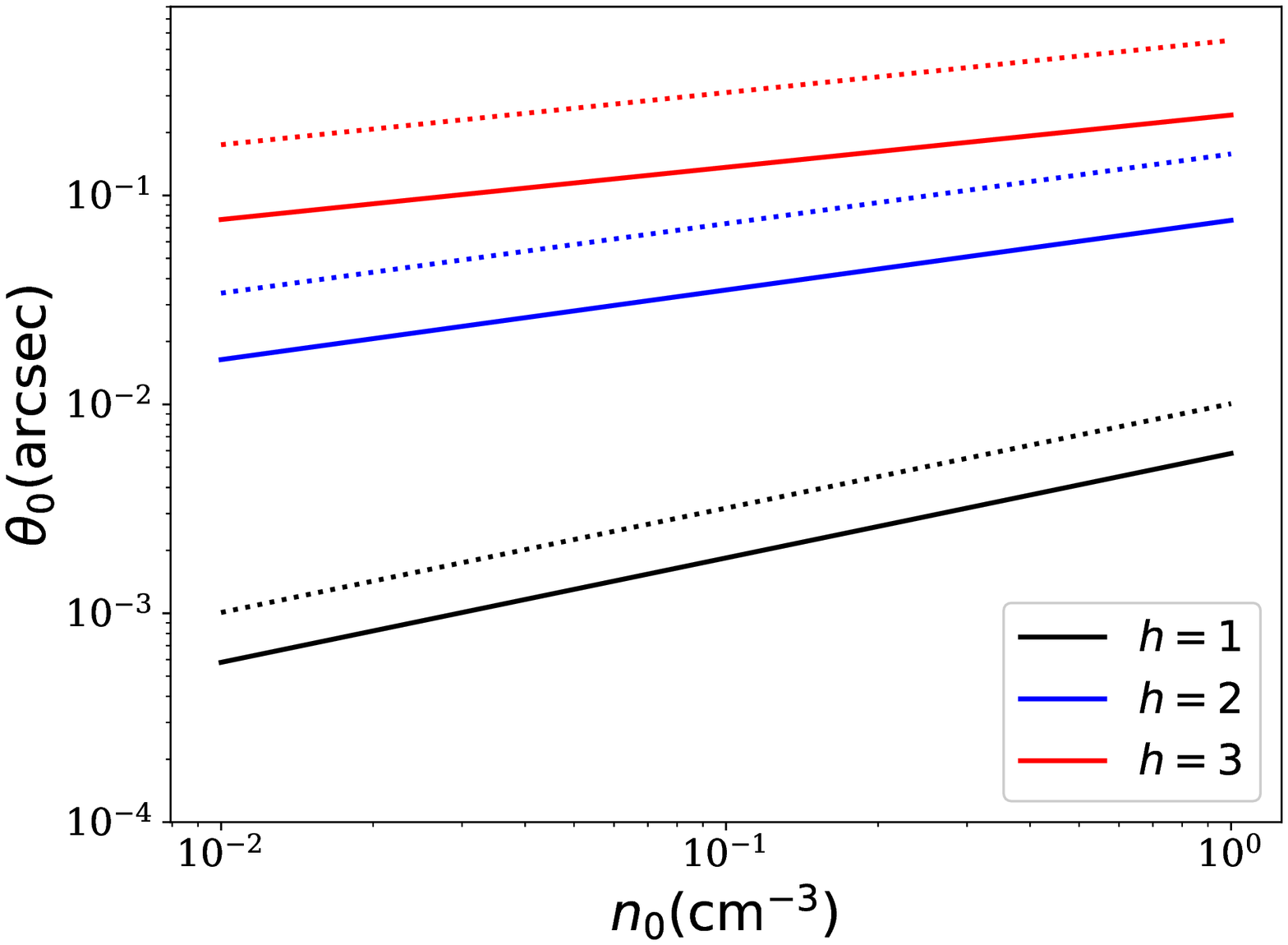}
  \caption{The characteristic angular scale $\theta_0$ as a function
    of electron density for circular plasma lens models. Top panel
    shows the exponential models: the red solid (blue dotted) line
    presents the relation for $\nu=1$ GHz ($0.5$GHz) with source
    distance $D_s=10$kpc. The lens is placed at the mid-point between
    the source and the observer. The shaded area denotes a variety of
    source distances between $5$ and $20$kpc. The bottom panel shows
    the power-law models with different power index $h$. The solid
    (dotted) lines present results of $R_0=0.1(0.3)$pc. }
  \label{fig:theta0}
\end{figure}

The characteristic angular radius $\theta_0$ indicates the lens
strength for the circularly symmetric lens. We thus show $\theta_0$ as
a function of electron density for the two families of models in
Fig.\,\ref{fig:theta0}. In the exponential model (top panel),
$\theta_0$ depends on the density and distance. We can see that the
projected electron density increases by two order of magnitudes, but $\theta_0$
only increases by a factor of about $8$. A similar dependence can be seen
in the power-law models (bottom panel): the electron density does not
have a large impact on the lens strength $\theta_0$. On the other
hand, the power index can dramatically change the lens strength for
the power-law lens, e.g. $\theta_0$ in model with $h=3$ (red lines) is
about $200$ times larger than that of $h=1$ (black lines). Therefore,
we need to consider the density profile as well in the plasma,
e.g. radial gradient and ellipticity rather than only the maximum
density itself. As we will present in the following, the lens
ellipticity can improve the lens strength as well as the richness of
magnification properties. For simplicity, we will use arcsec for
the unit in the following discussion if not mentioned.

\subsubsection{Column-Density Power-Law}
\label{PL2}

Now let us consider the family of power-law lenses given by a two-dimensional electron column density that has a power-law form
\be
N_\text{e}(\theta) = N_\text{0}\frac{\theta_R^H}{\theta^H},
\ee
with the power-law index $H$ and the angular radius $\theta_\text{R}$
at which $N_\text{e}(\theta_\text{R}) = N_\text{0}$. This electron column density produces an effective lens potential
\begin{equation}
\psi(\theta) = \frac{\lambda^2}{2 \pi}\frac{D_{ds}}{D_d D_s} r_\text{e} N_0 \theta_\text{R}^{H}\frac{1}{\theta^H}
\end{equation}
Notice here that $\psi(\theta) \propto 1/\theta^H$ whereas the volume-density case has a deflection angle with this dependence instead. Following the usual approach, the gradient of the potential gives
\begin{equation}
\alpha=-\frac{\theta_0^{H+2}}{\theta^{H+1}},
\end{equation}
where we have defined 
\begin{equation}
\theta_0=\left(\frac{\lambda^2}{2 \pi}\frac{D_{ds}}{D_d D_s} r_\text{e} H N_0 \theta_\text{R}^{H} \right)^\frac{1}{H+2}.
\end{equation}
The deflection angle differs from the volume density PL lens result by an extra factor of $\theta_0/\theta$. $\theta_0$ can have the same definition if we use the substitutions below. This gives distinct behaviour for the magnification,
\begin{equation}
\mu^{-1}=1-H\frac{\theta_0^{H+2}}{\theta^{H+2}}-(H+1)\frac{\theta_0^{2(H+2)}}{\theta^{2(H+2)}}.
\end{equation}
Note that this is quadratic in $\theta_0^{H+2}/\theta^{H+2}$, so we can easily find the critical curves of this lens using the quadratic formula,
\begin{equation}
\frac{\theta_0}{\theta_\text{crit}}= \left(  -\frac{H\pm{(H+2)}}{2(H+1)}  \right)^\frac{1}{H+2}
\end{equation}
We can discard the additive solution in the above expression, which will not produce any real solutions for any value of $H$. This means the negative case produces an overall positive quantity which gives real solutions in the root. The remaining solution produces
\begin{equation}
\theta_\text{crit}=\theta_0 \left[  H+1 \right]^\frac{1}{H+2}.
\end{equation}
These equations are identical to the deflection angle and magnification for the volume-density lens provided that we make the substitution $H+1 \rightarrow h$. The lensing potential and characteristic angular scale can be reproduced provided we make the substitutions $R_0/D_d  \rightarrow \theta_R$ and
\begin{equation}
N_0 \rightarrow 2 \frac{n_0 R_0}{(h-1)}\sqrt{\pi}\frac{\Gamma\left( \frac{h}{2}+\frac{1}{2}\right)}{\Gamma\left(\frac{h}{2}\right)}.
\end{equation}

Since the lensing magnification and image formation properties of this charge column-density lens are effectively identical to the charge volume-density lens, we will restrict the remainder of our analysis to the volume-density lens case.

\section{Elliptical plasma lensing models}
\label{sec:ellp}

We generalize the lens models of previous section by introducing an elliptical
coordinate $\Theta=\sqrt{\theta_1^2q+\theta_2^2/q}$, where $q$ is the axis ratio \citep{1993ApJ...417..450K,1994A&A...284..285K}.
The ellipticity of the density distribution can be given by $\epsilon=(1-q)/(1+q)$.  The value of the axis ratio
$q$ is only taken from the interval $0<q\le1$, as it is equivalent for
$\theta_1-$ and $\theta_2-$axis. Thus in the following calculations, $q<1$ is
assumed, and the circularly symmetric model is obtained in the
limiting case $q\rightarrow1$. We only consider the iso-elliptical
distribution, i.e. the ellipticity is the same for each radius.

\subsection{Exponential model}

For the exponential models, the elliptical potential can be obtained
by substituting $\Theta$ for $\theta$ in Eq.\,\ref{eq:exp-pot}. The
deflection angle is
\be
\alpha = - 2 A(h)\,\rund{\theta_1q + \ii \theta_2/q},
\ee
where the pre-factor is given by
\be
A(h)={\theta_0^2\Theta^{h-2} \over 2\sigma^h} {\rm exp} \rund{-{\Theta^h\over h\sigma^h} }.
\ee
The lensing convergence and shear are
\begin{flalign}
\kappa = &A(h)\left[-q-1/q -\rund{ {(h-2)\over \Theta^2  }-{\Theta^{h-2} \over \sigma^h}}(\theta_1^2q^2 + \theta_2^2/q^2) \right] ,\nonumber\\
\gamma =
&A(h)\left[-q+1/q -\rund{ {(h-2)\over \Theta^2  }-{\Theta^{h-2} \over \sigma^h}}(\theta_1^2q^2 - \theta_2^2/q^2) \right. \nonumber \\
&\left. +2\ii \rund{ {\Theta^{h-2}\theta_1\theta_2 \over \sigma^h} -(h-2){\theta_1\theta_2\over \Theta^2}} \right].
\end{flalign}
We use complex notation for the two shear components. The
imaginary part represents the second component of the shear
\citep{1994A&A...284..285K}.

The lens magnification is defined as the Jacobian determinant of the thin lens equation. This can be stated in terms of the convergence,
\be
\mu^{-1} = 1- 2\kappa + \frac{\partial \alpha_1}{\partial \theta_1} \frac{\partial \alpha_2}{\partial \theta_2} - \frac{\partial \alpha_1}{\partial \theta_2} \frac{\partial \alpha_2}{\partial \theta_1}.
\elabel{mu-derive-1}
\ee
Using this expression we derive the elliptical exponential lens magnification for general $h$,
\begin{flalign}
\mu^{-1} = 1-2A(h)\left[ -q-\frac{1}{q}+\left(\frac{\theta_1^2q^2+ \theta_2^2/q^2}{\Theta^2} \right)\left(2-h+\frac{\Theta^h}{\sigma^h} \right)\right] \nonumber\\
+4A^2(h)\left( h-1-\frac{\Theta^h}{\sigma^h} \right).
\elabel{mu-exponential_check}
\end{flalign}
For $h=2$, the expression becomes an
elliptical Gaussian lens, and the magnification can be written as
\be
\mu^{-1} = 1-2A(2)\rund{-q-{1\over q}+\dfrac{\theta_1^2q^2+\theta_2^2/q^2}{\sigma^2}}
+4A(2)^2\eck{1-\dfrac{\Theta^2}{\sigma^2}}
\elabel{mu-gaussian}
\ee
One can obtain the maximum demagnification of Gaussian lens at the origin
\be
\mu_{\rm origin}\leq {1\over (1+(\theta_0/\sigma)^2)^2}.
\ee
In the circular limit, the demagnification is the strongest at the
origin. The ellipticity further increases the demagnification, which
can vary based on the trajectory of the source behind the
lens. Moreover, the frequency dependence of $\mu_{\rm origin}$ can
provide further constraints on the model parameters. For lenses with
$h<2$ the magnification at the origin vanishes and the lens has an
exclusion region. For lenses with $h>2$, the magnification at the
origin approaches unity, producing a W-shaped light curve, as
  opposed to a U-type light curve consistent with the result derived
  in \citet{ErRogers18} for symmetric exponential lenses and
  equivalent to \citet{rogers&er2019} for a binary lens.

\begin{figure}
  \includegraphics[width=7.0cm]{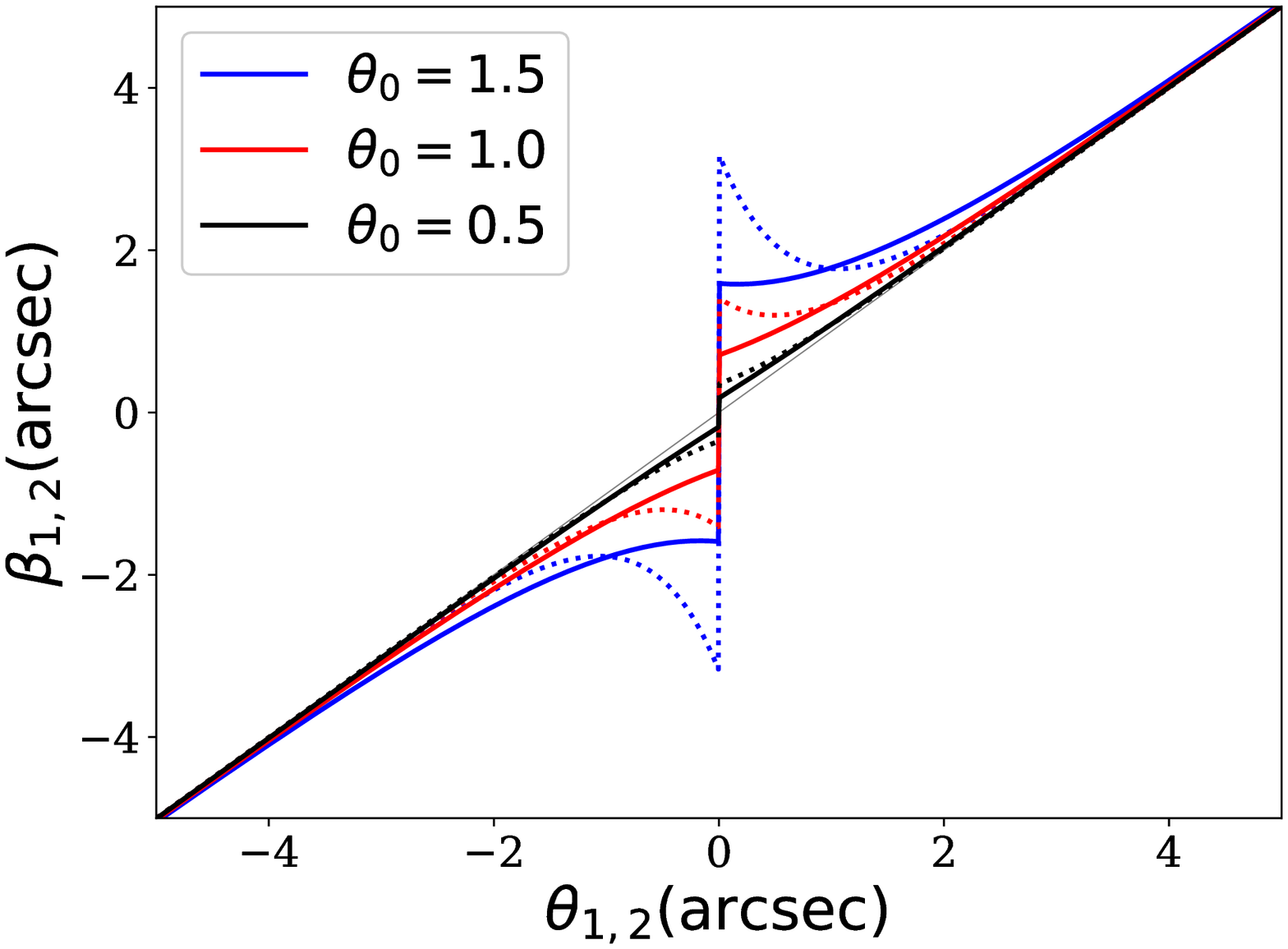}\hspace{-0.3cm}
  \includegraphics[width=7.0cm]{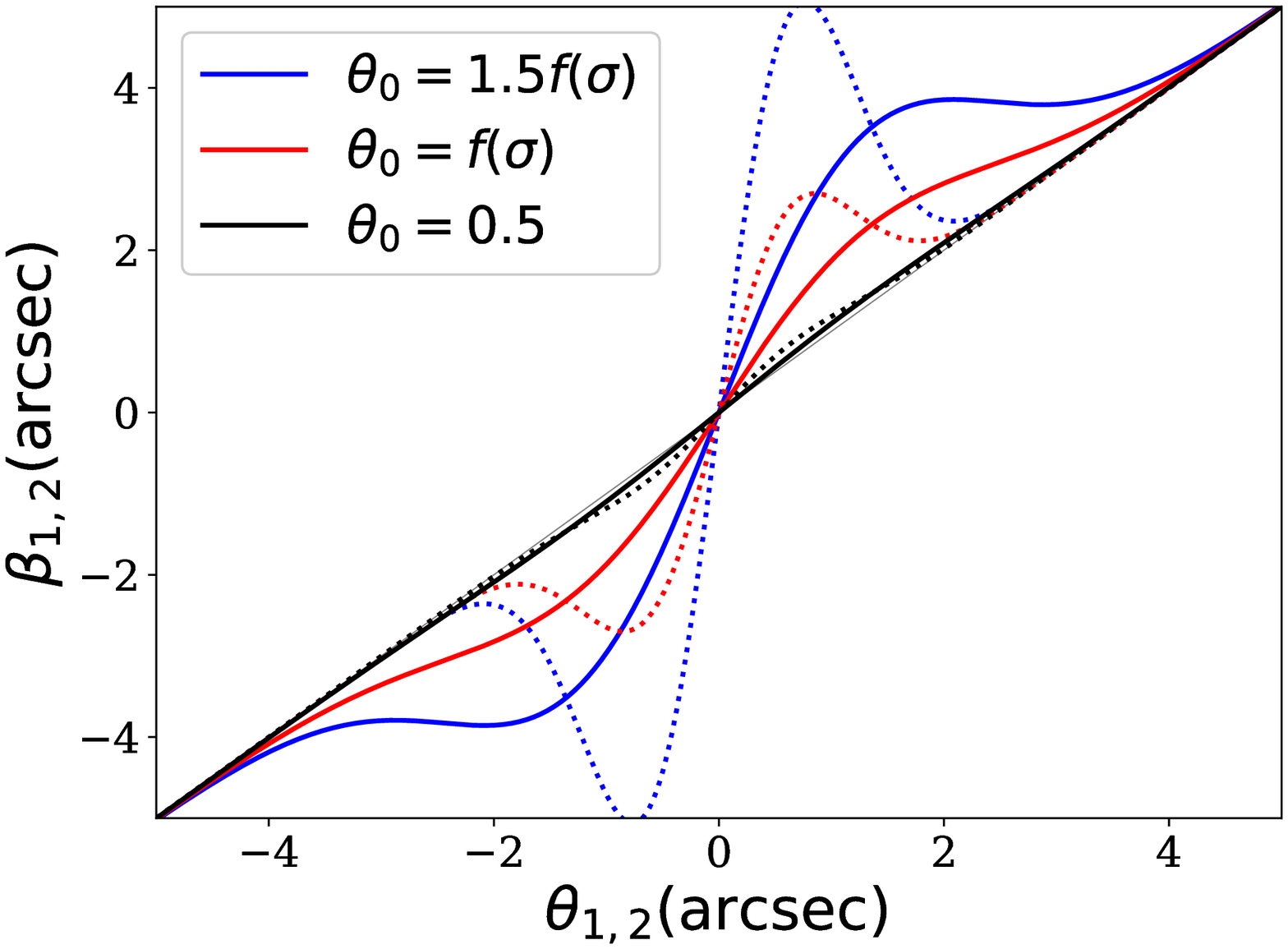}\hspace{-0.3cm}
  \includegraphics[width=7.0cm]{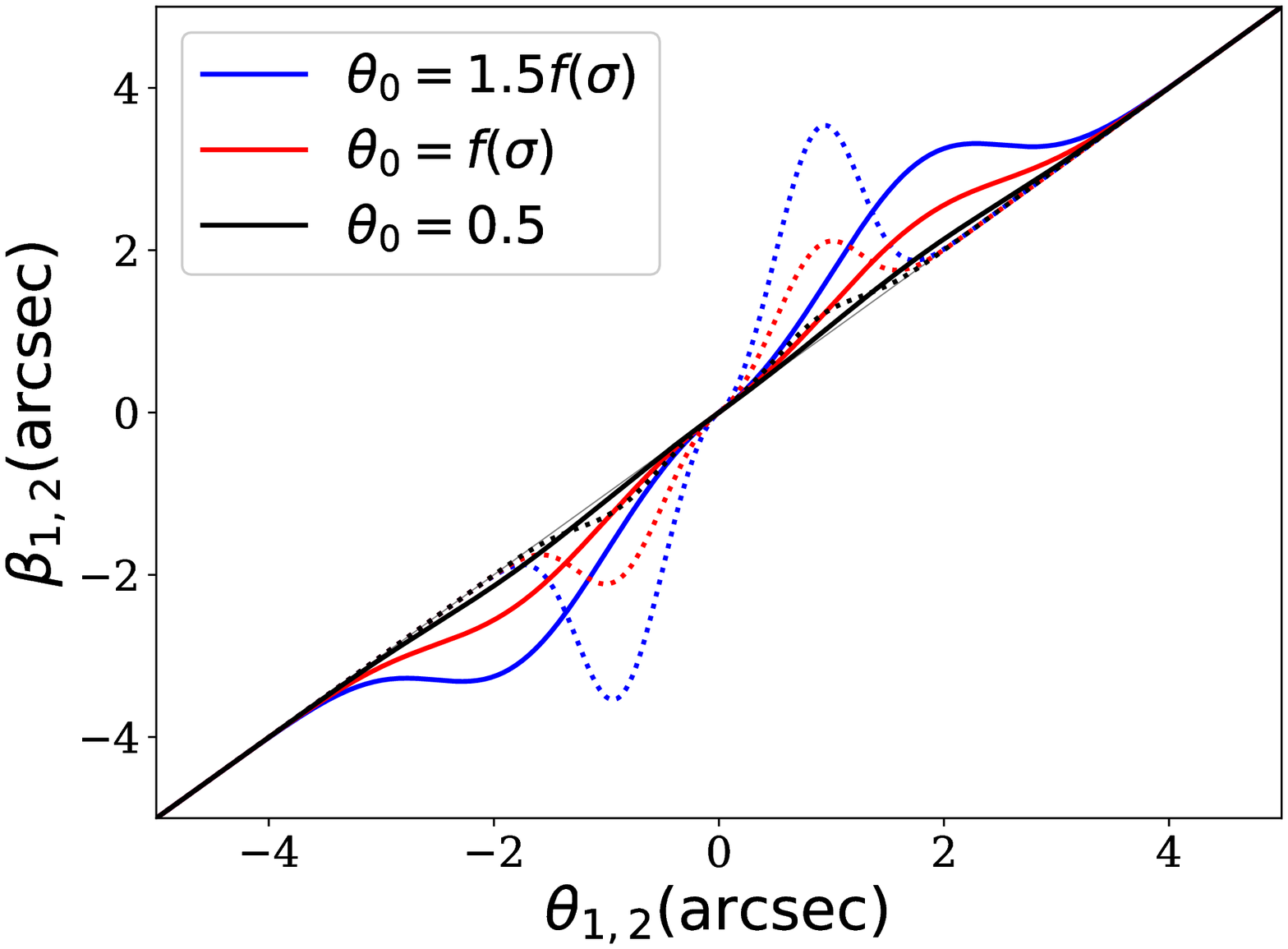}
  \caption{The Young diagram of the exponential models with $\sigma=1.0$,
    $q=0.5$. Grey line indicates the identity. From top to bottom, we
    present the exponential lenses with $h=1,2,3$ respectively. The
    solid (dotted) lines indicate the mapping along the major (minor)
    axis of the lens. The formula of critical condition $f(\sigma)$
    can be found in \citet{ErRogers18}.}
  \label{fig:exp-alpha}
\end{figure}

We first show the Young diagram of exponential models for $h=1,2,3$ in
Fig.\,\ref{fig:exp-alpha}. The blue, red and black line represents the
super-, critical and sub-critical case in the circular lens models
respectively \citep{ErRogers18}. The solid line shows the source
position as a function of the image plane position along the major
($\theta_1-$) axis, while the dotted line shows the same plot except along
the minor ($\theta_2-$) axis. As we can see that the deflection along the
major axis is mild and without rapid variation, and the deflection
along the minor axis shows rapid change even for the sub-critical
cases. We therefore expect strong changes in magnification along the
minor axis, even for the sub-critical lens.

\begin{figure*}
  \centerline{
  \includegraphics[width=7.0cm]{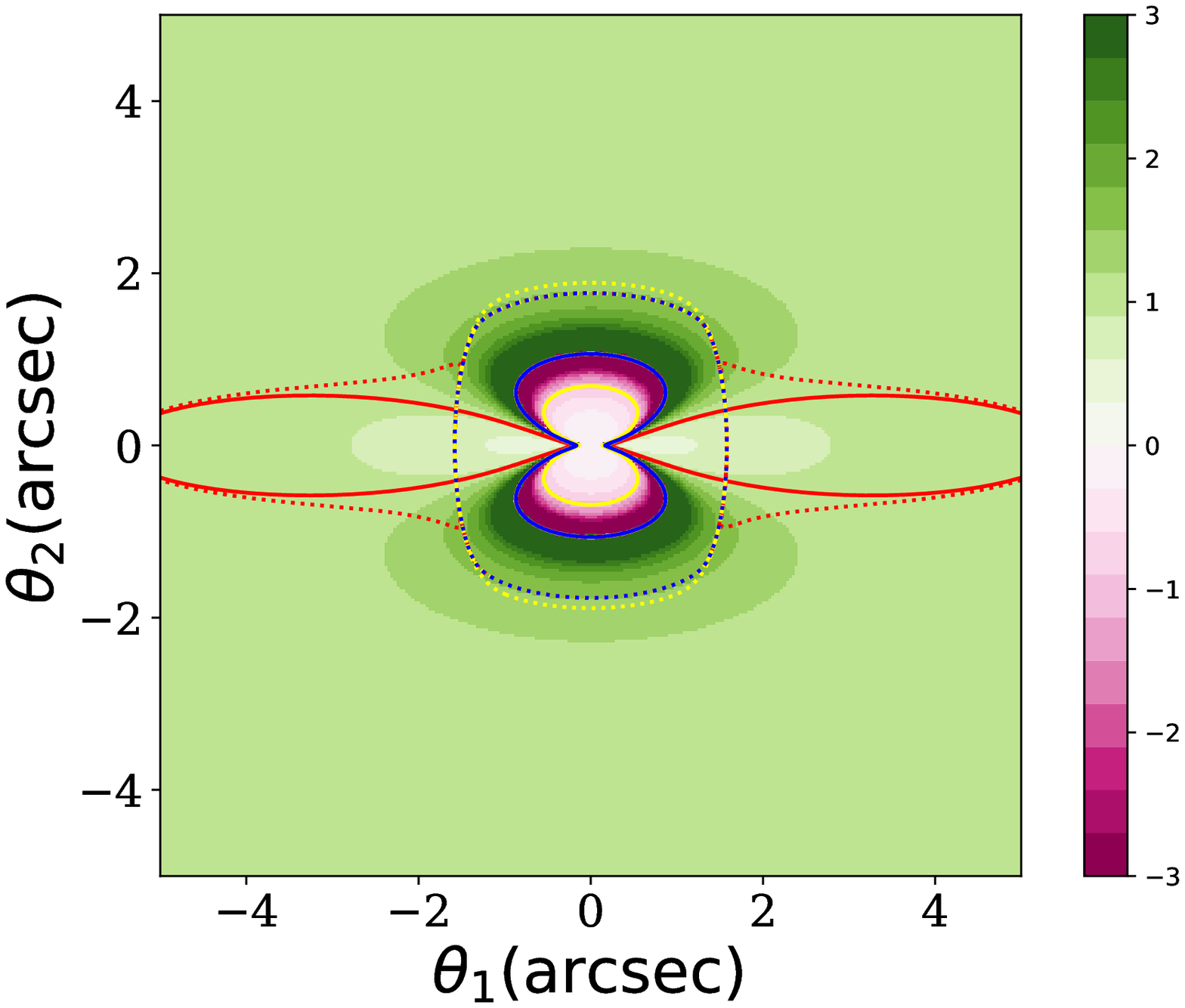}\hspace{-0.6cm}
  \includegraphics[width=7.0cm]{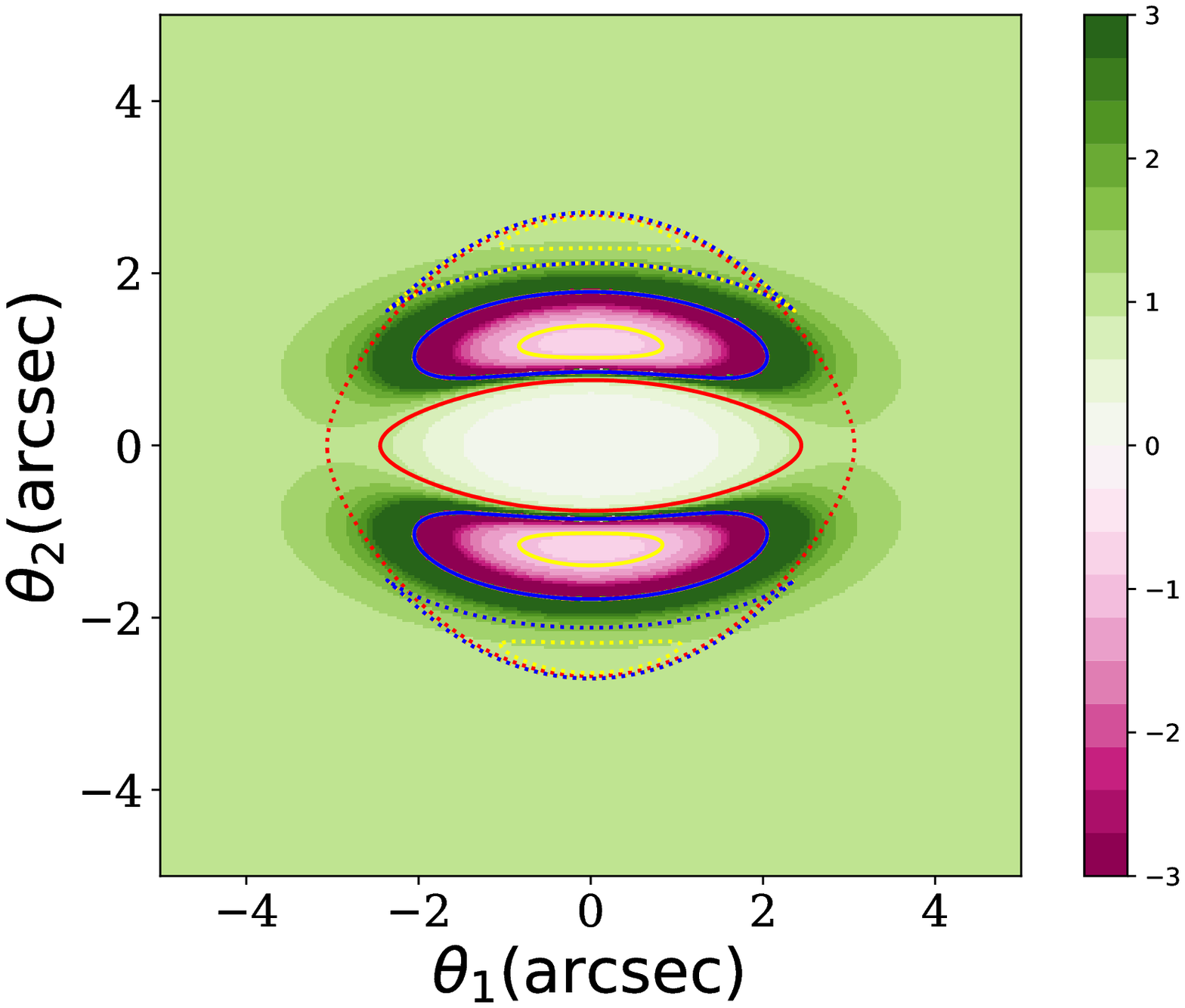}\hspace{-0.6cm}
  \includegraphics[width=7.0cm]{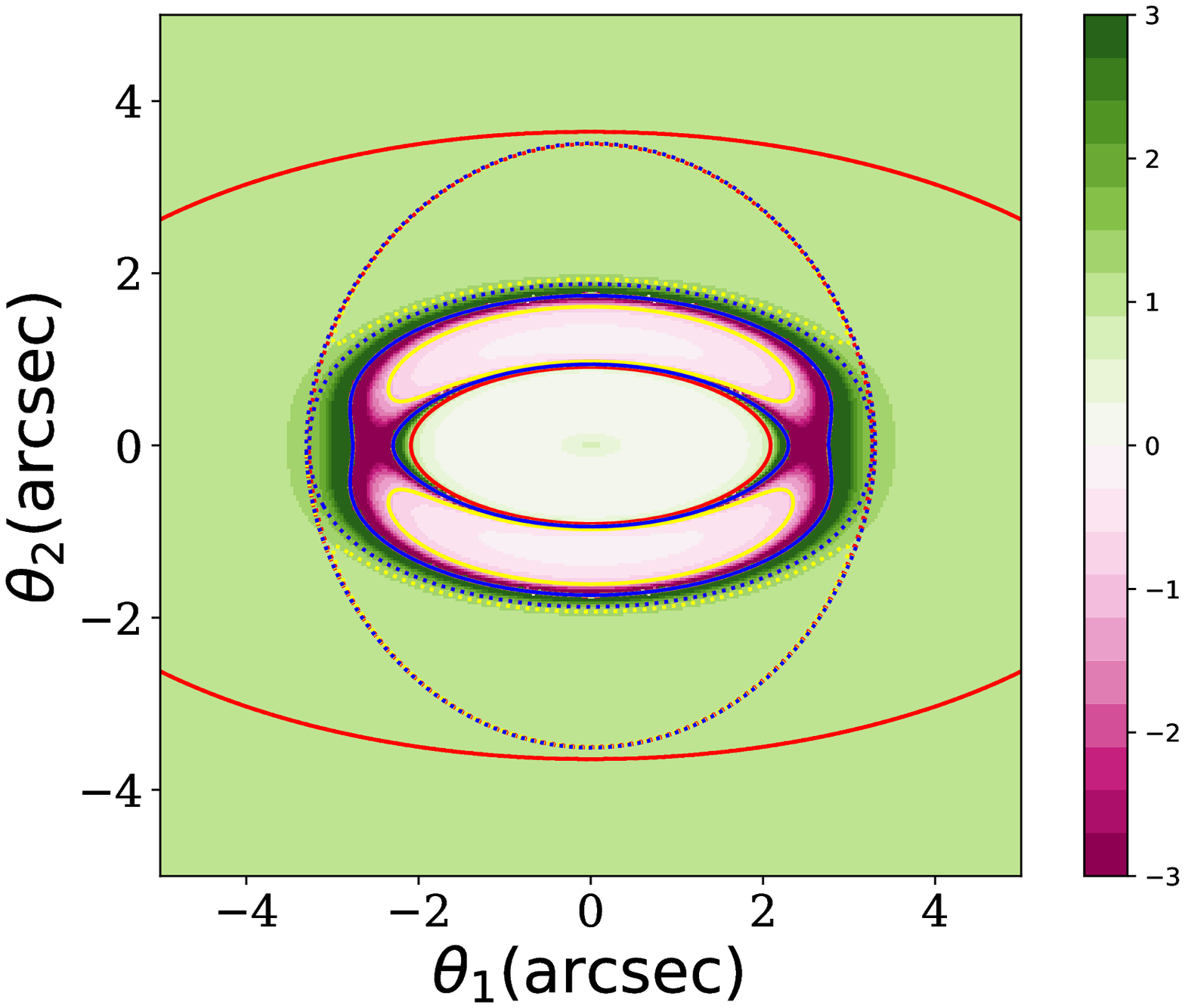}}
  \caption{The two dimensional magnification map of the exponential
    models with lens parameters $q=0.5,\theta_0=1.5,\,\sigma=1.0$, and
    from left to right: $h=1,2,3$. The green (pink) region indicates
    the positive (negative) magnification, and the white region
    indicate the demagnification region. The solid (dotted) blue line
    presents the critical curve (caustics) of the lens on the image
    (source) plane. We truncate the magnification to $\pm3$ for better
    visibility. }
  \label{fig:expmag}
\end{figure*}

\begin{figure*}
  \centerline{
    \includegraphics[width=7.0cm]{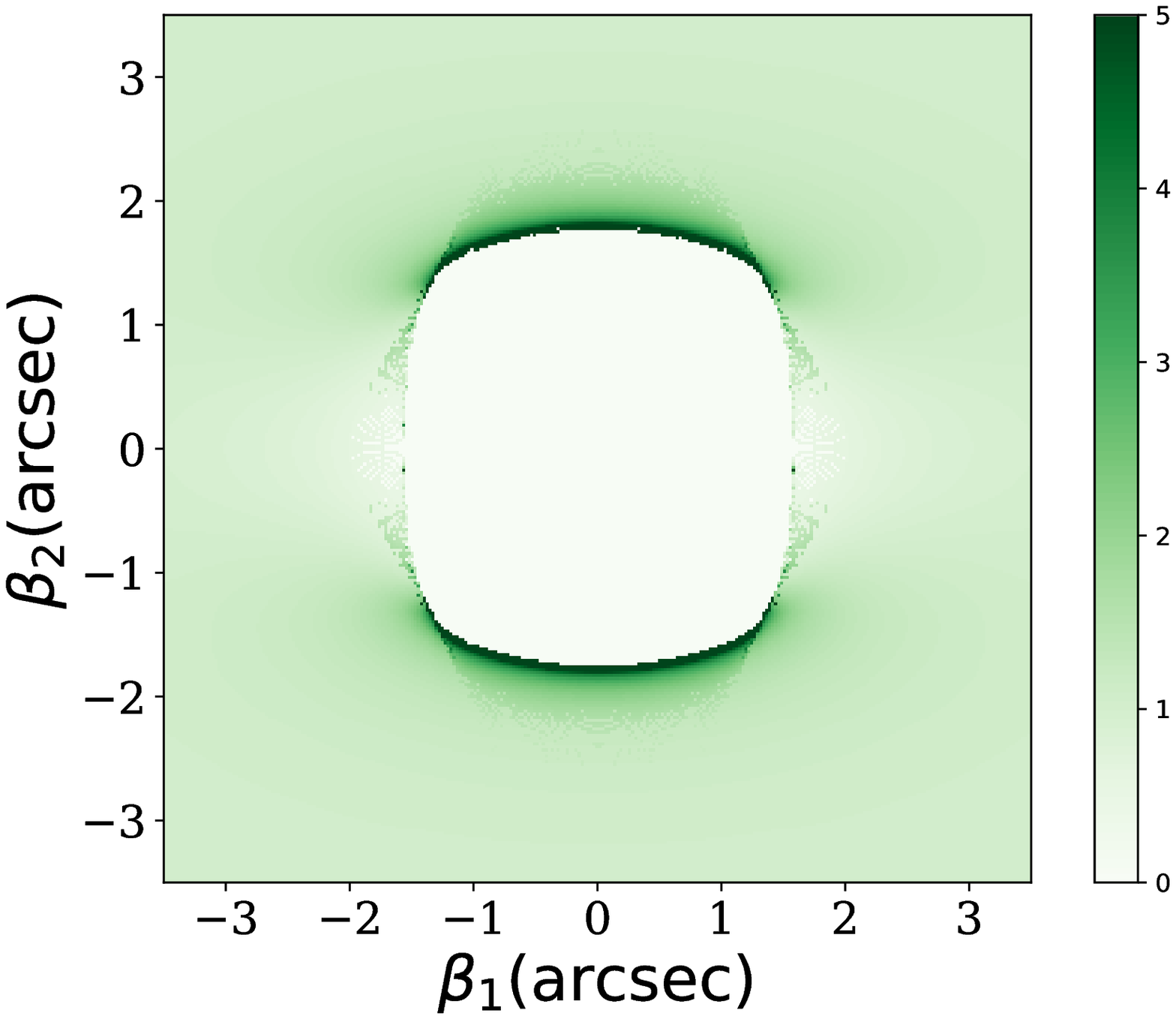}\hspace{-0.6cm}
    \includegraphics[width=7.0cm]{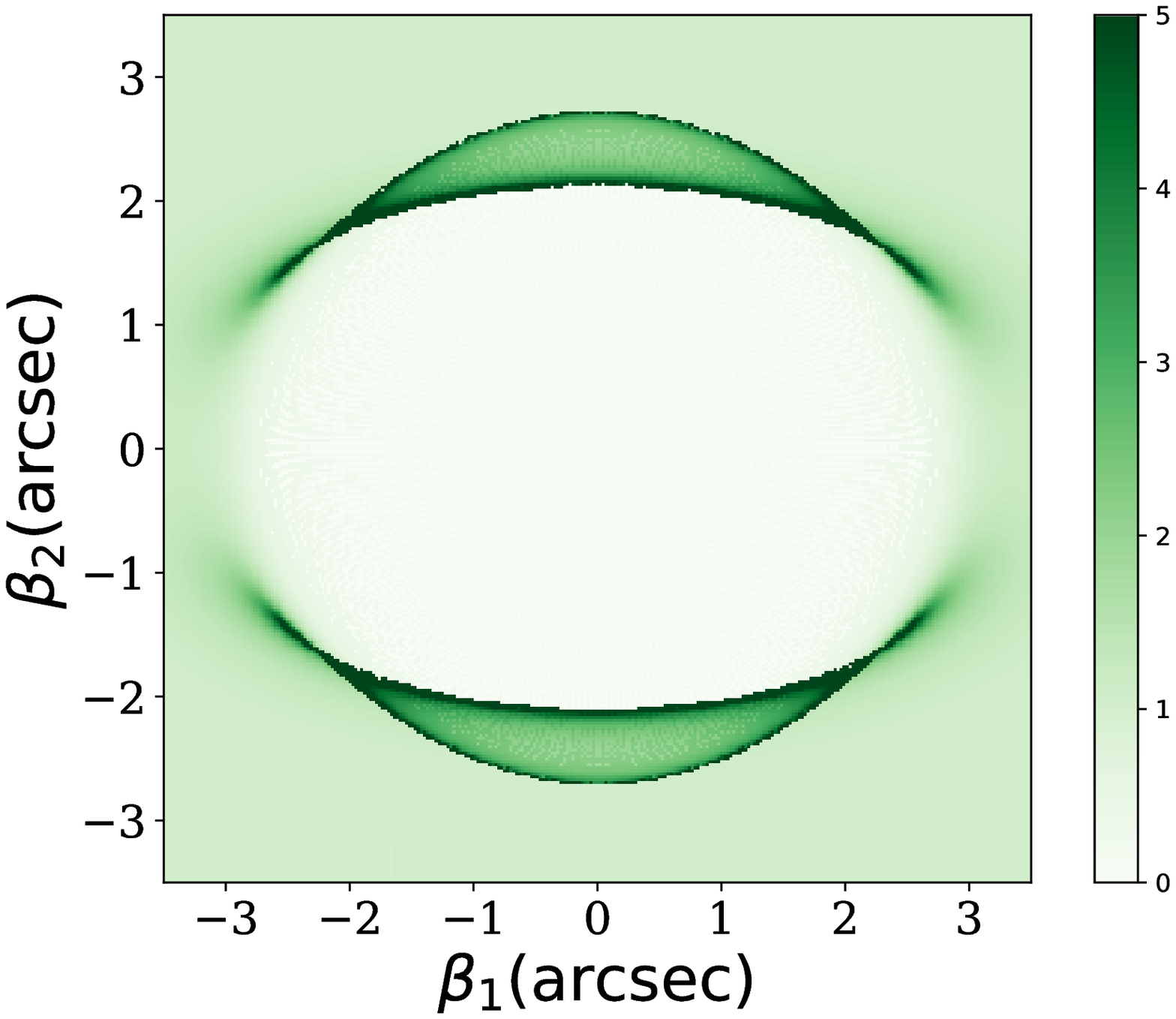}\hspace{-0.6cm}
    \includegraphics[width=7.0cm]{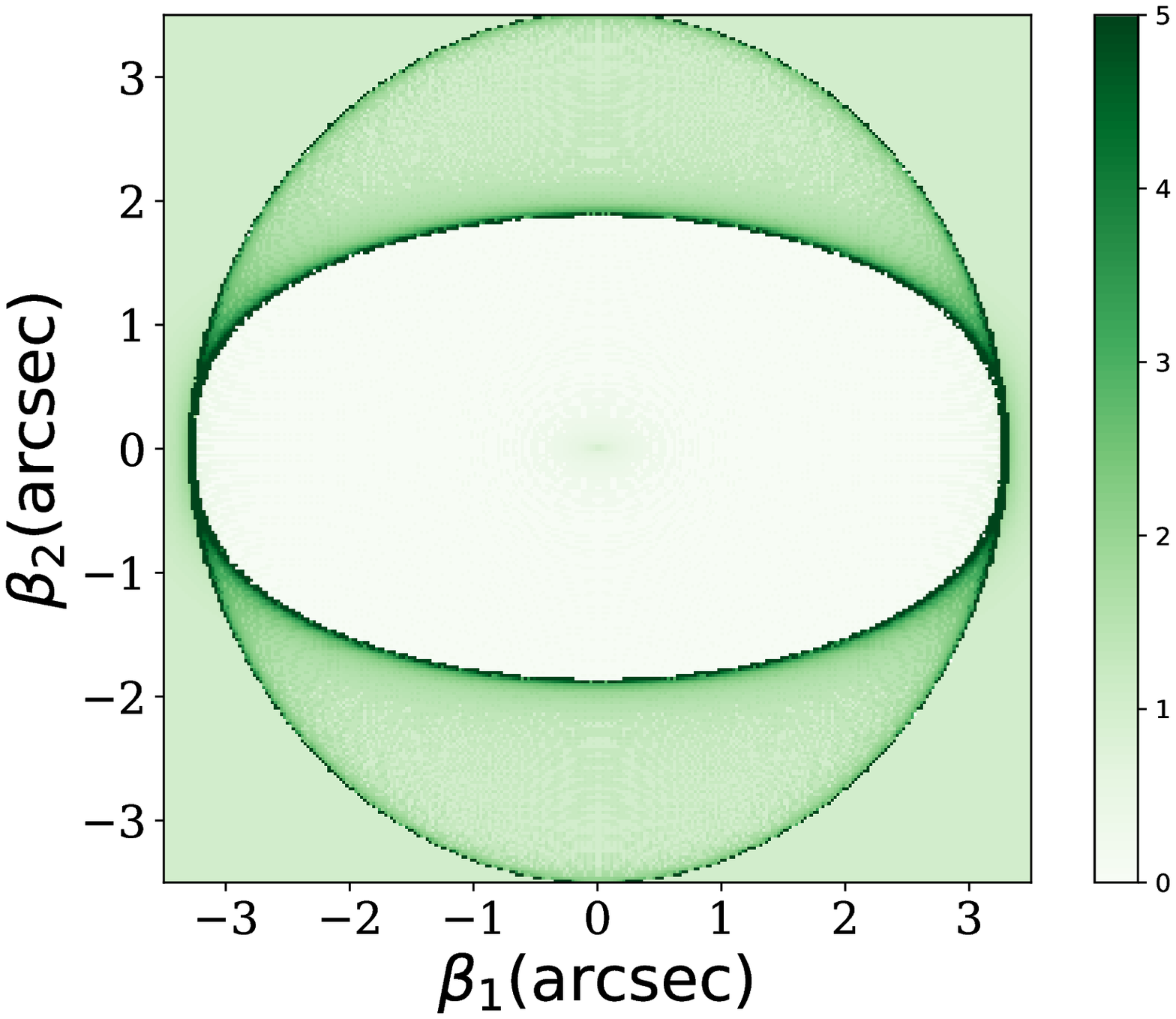}}
  \caption{The magnification map of the exponential model on the
    source plane with the same lensing parameters as in
    Fig.\,\ref{fig:expmag}. The magnification is truncated to $5$ for
    better visibility. Notice that the coordinates have different
    ranges with Fig.\,\ref{fig:expmag}.}
  \label{fig:exp-musource}
\end{figure*}

In Fig.\,\ref{fig:expmag}, two dimensional magnification maps on the
lens plane are shown for $h=1,2,3$ respectively. The green colour
stands for the positive magnification, with pink for the negative. The
solid (dotted) blue lines show the critical curves (caustics). The red
and yellow curves indicate the boundary of $\mu=+1$ and $\mu=-1$ on
the lens plane (solid lines) and on the source plane (dotted lines)
respectively. In the light colour regions, the images will be
demagnified. The lens with $h=1$ has a dumbbell shape critical, while
the lenses of $h=2,3$ have two separated kidney-like arcs. The
demagnification region of $h=1$ extended along the major axis, while
that of $h=2,3$ are mainly concentrated near the lens origin. With
different trajectories of the background source, the elliptical lenses can
produce a rich series of magnification curves, and those along the
minor axis are similar to that produced by the circular lens. In the
right panel of Fig.\,\ref{fig:expmag}, there is a light green region
at the origin, where the magnification is close to unity. It has been
observed that such a spike can cause model degeneracy \citep{rogers&er2019}.
A corresponding magnification map on the source plane is
shown in Fig.\,\ref{fig:exp-musource}. We focus on the center region
of the lens as this map is computationally expensive. There is a
large, extended exclusion region around the center of the lens. Along
the minor axis, there is a sharp boundary, while along the major axis,
the variations are smooth. One can extract the magnification curve
given a trajectory of the background source. Those near the minor axis
are similar to that generated by a spherical lens.

In the appendix, a set of criticals and caustics are displayed
selected from the lens parameter space. Some of them show similar
shapes to dual-component plasma lenses \citep{rogers&er2019}.

\begin{figure}
  \includegraphics[width=6.0cm]{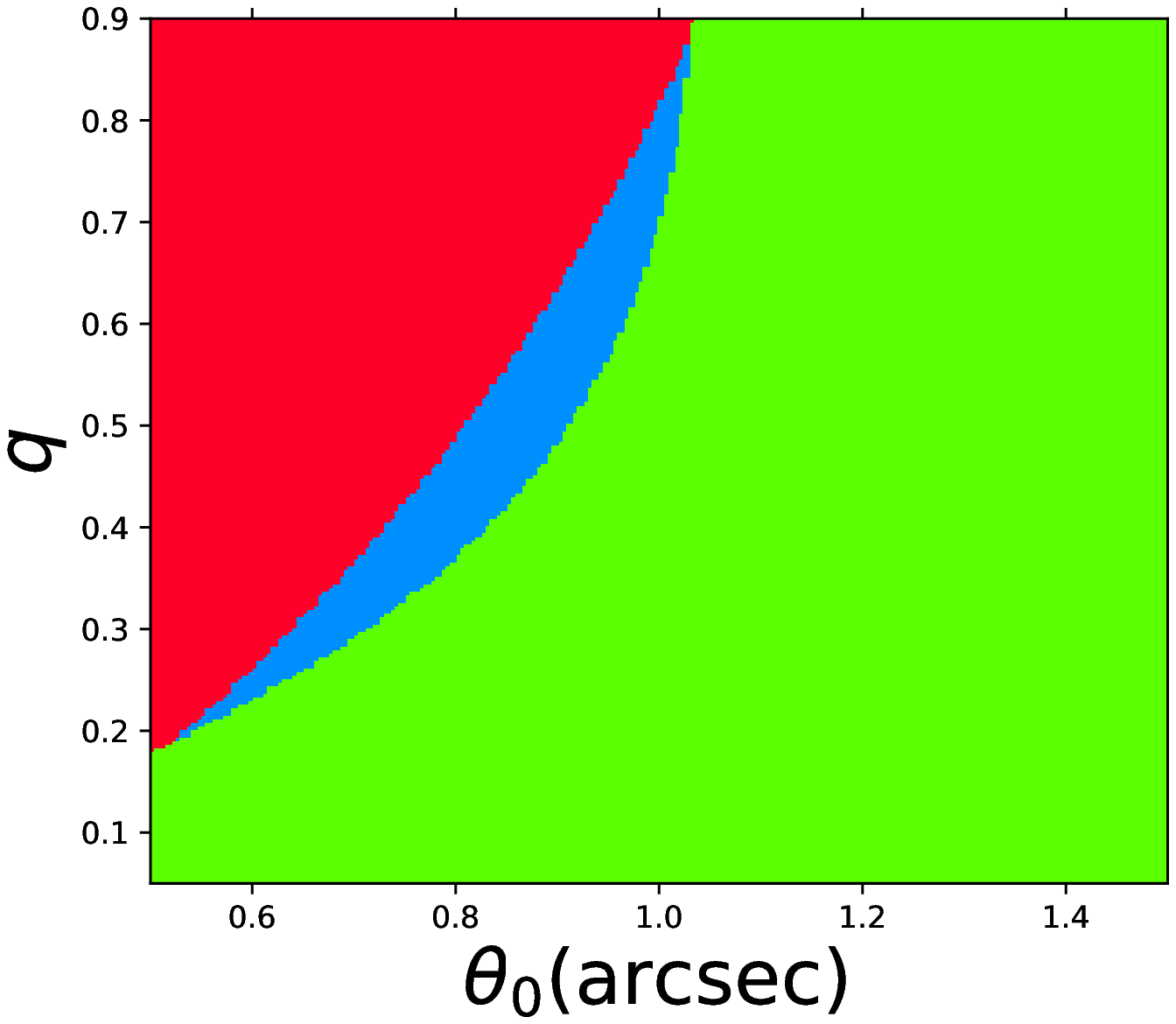}
  \includegraphics[width=6.0cm]{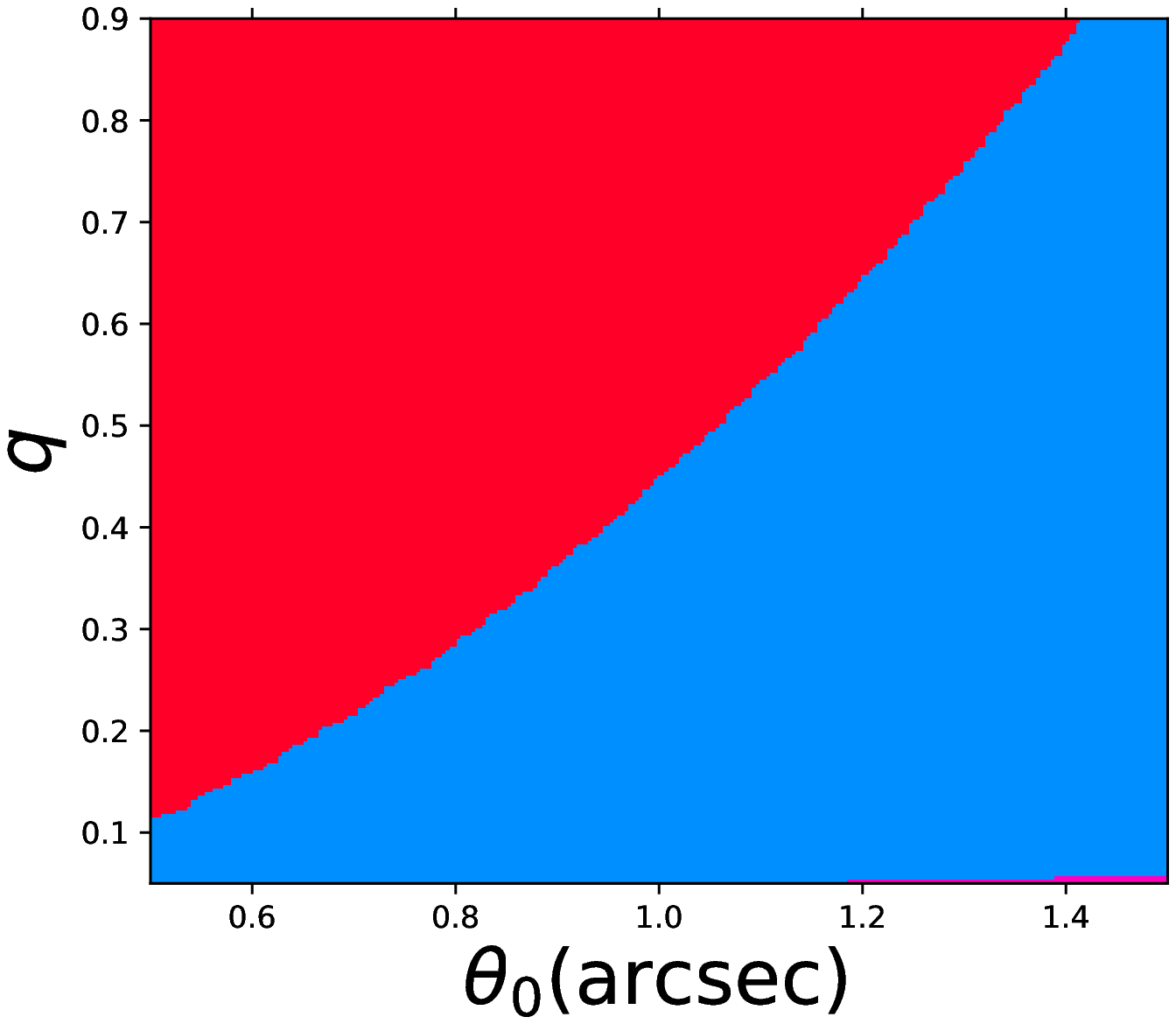}
  \includegraphics[width=6.0cm]{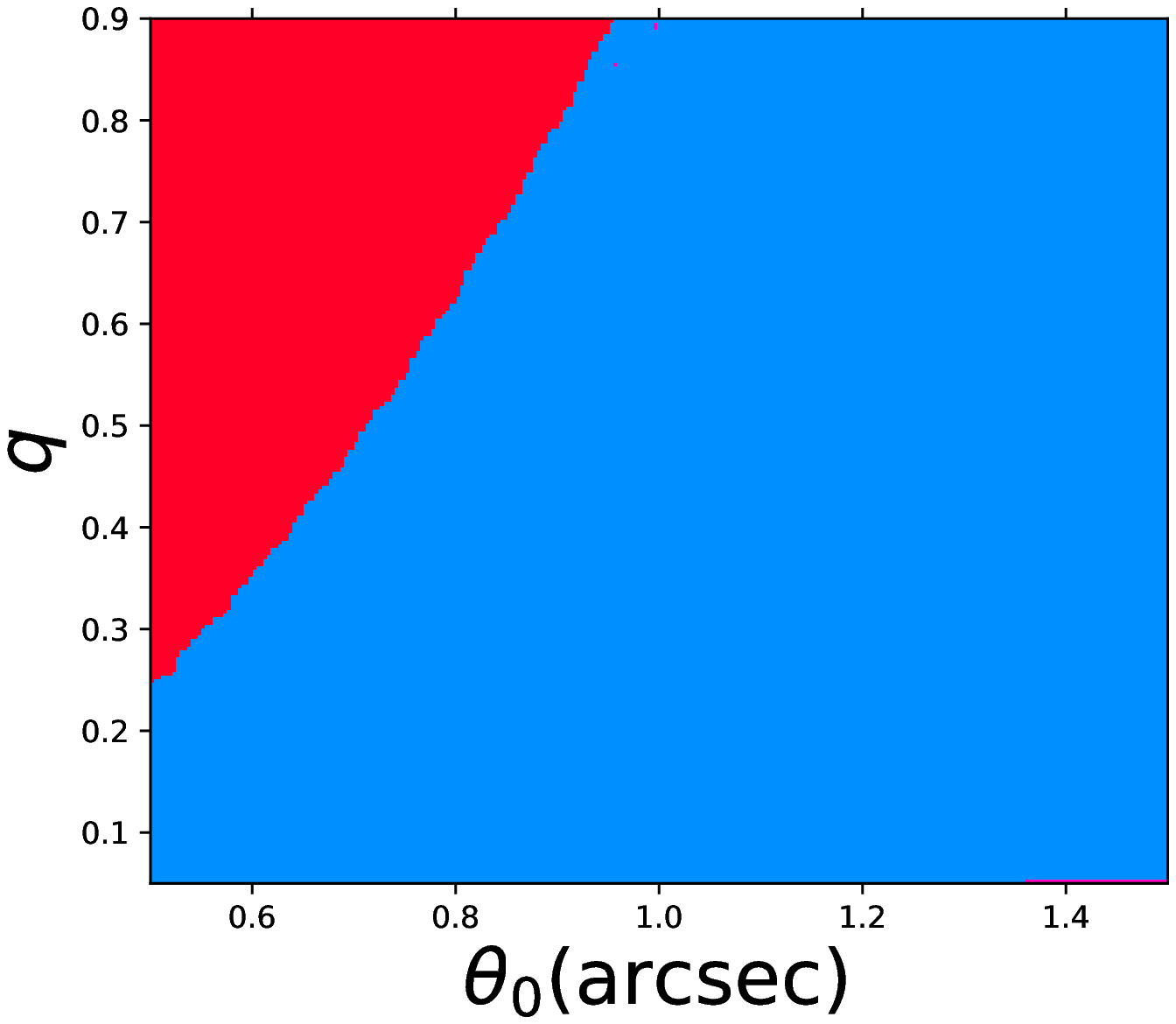}
  \caption{The figures show the number of critical curves for the
    elliptical exponential lens as a function of $\theta_0$ and axis
    ratio $q$ ($\sigma=1$). From top to bottom panel, we present the
    case of $h=1, h=2$ and $h=3$ families respectively. The maps are
    color-coded such that the number of critical curves per lens
    configuration is shown for 0-red, 1-green, 2-blue critical curves.}
    \label{fig:expccs}
\end{figure}

The critical curves provide useful properties of the lensing. They
indicate strong variations in the magnification map, corresponding to
spikes on the light curve of a background source. The critical curve
also sections the lens plane into regions that count the number of
multiple images. Moreover, the number of critical curves in plasma
lensing tightly correlates with the density profile of the electron in
axisymmetric lens \citep{ErRogers18}, such as the ratio between
$\theta_0$ and $\sigma$ in exponential models. In order to
characterize the behavior of the elliptical lens, we explore the
parameter space of the lens model as a function of $q$ and $\theta_0$.
For each ($q$, $\theta_0$) pair, we calculate the determinant of the
Jacobian metric on a finite coordinate grid $(\theta_1,\theta_2)$ and
contour the determinant to reveal the curves over which it
vanishes. This set of curves are the lens criticals. We then count the
number of contours that a given ($q$, $\theta_0$) pair provides. The
parameter space map is shown in Fig.\,\ref{fig:expccs}. From top to
bottom, we show lenses with $h=1,2,3$ respectively. In general, the
lens families of higher index $h$ are more efficient in generating
critical curves.  For $h=1$, the lenses generate one critical in most
cases, although the shape of criticals can vary dramatically (see
appendix). Two criticals can be generated when the two parts of a
dumbbell demagnification region break apart from one another. In the
parameter space, these situations only occur within a small area (blue
region). For $h=2,3$, the lenses generate two symmetric criticals on
both side of the major axis. The two arc-shape criticals shrink and
degrade to the two ellipses and eventually approach the circular lens
case with the increasing of $q$.

\begin{figure}
  \includegraphics[width=7.0cm]{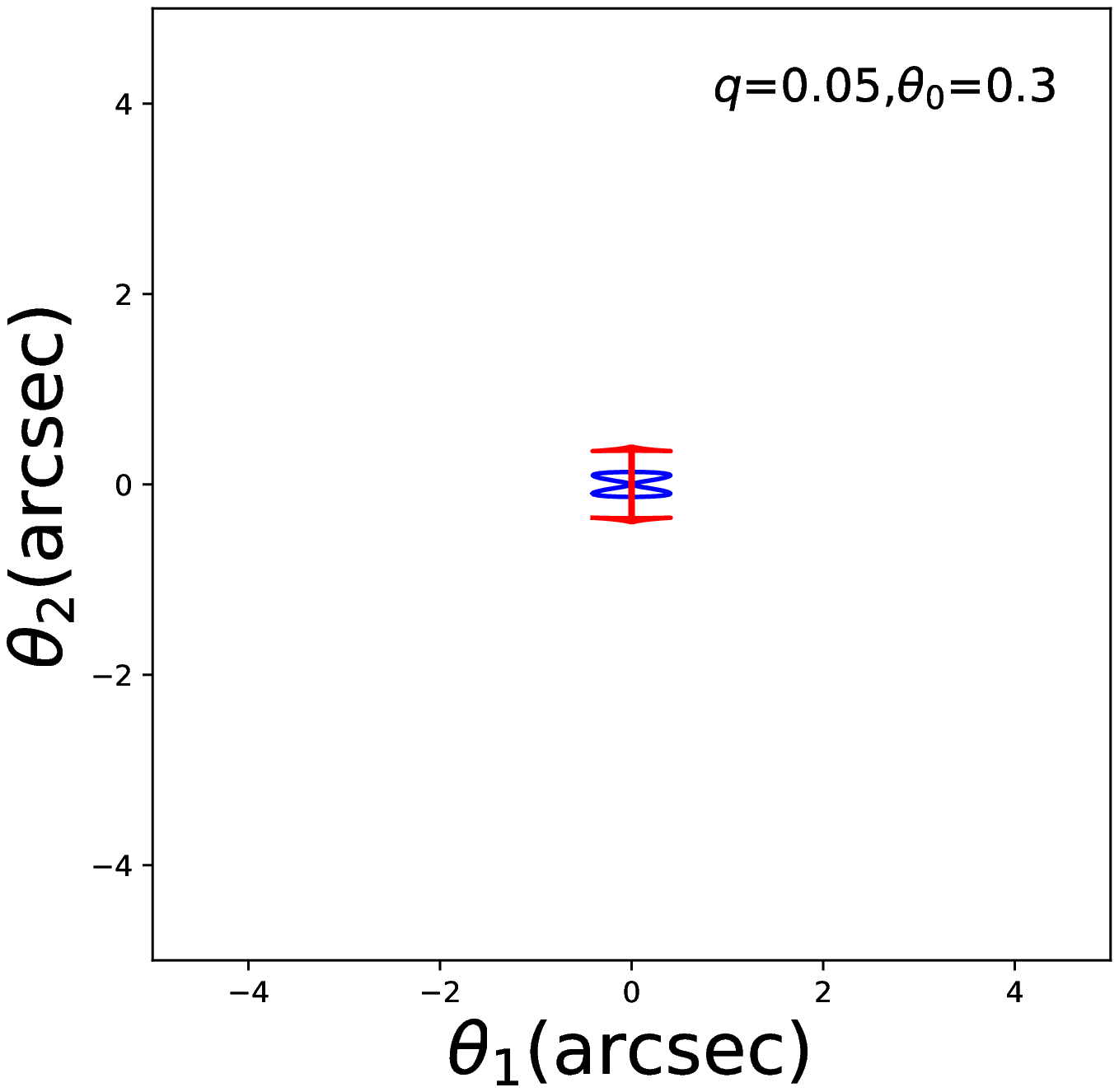}
  \includegraphics[width=7.0cm]{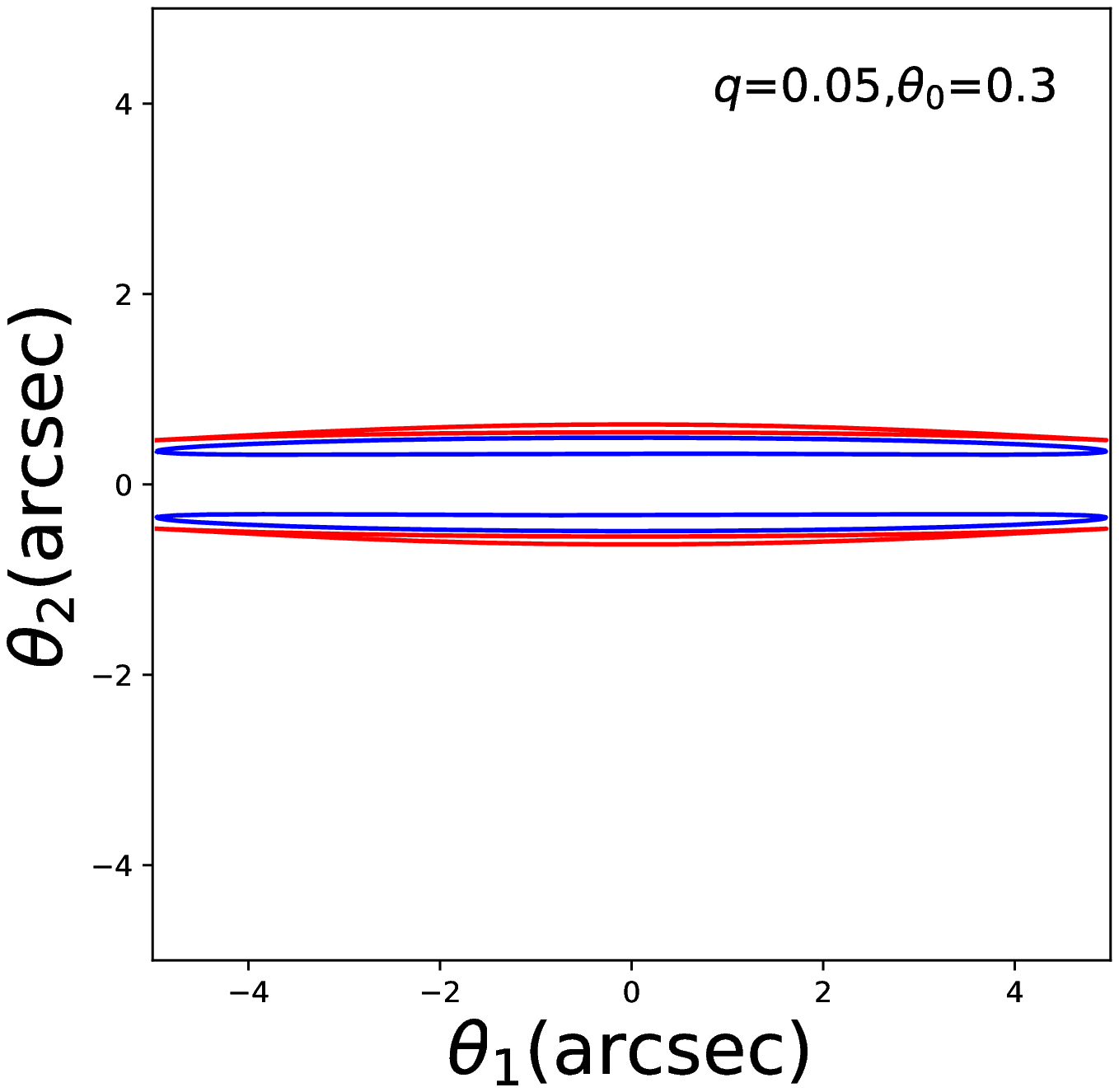}
  \caption{Two sub-critical elongated exponential lenses. The top
    (bottom) one show the lens of $h=1$ ($h=3$). The blue (red) curves
    are the critical curves (caustics). }
  \label{fig:expccq005}
\end{figure}

We also present the lens with an extremely high ellipticity
($q=0.05$), which is different from the ellipticity of galaxy or dark
matter halo
\citep[e.g.][]{2006MNRAS.367.1781A,2015MNRAS.454.1432S}. The
ellipticity significantly increases the lensing efficiency. The
elongated lens with sub-critical $\theta_0$ (much smaller than the
critical value) can also generate critical curves, as we show in
Fig.\,\ref{fig:expccq005}.  For such a plasma lens, the required
central density $N_0$ to have critical curves will be one order
smaller than that of a circular lens.


%



\subsection{Power-law model}

%
%
%
%
%
%

We further generalize the power-law model to the elliptical softened
power-law (ESPL) model. The lens potential can be
generalized from Eq.\,\ref{softpl} by making the substitution $\theta
\rightarrow \Theta=\sqrt{\theta_1^2q + \theta_2^2/q+\theta_c^2}$.
The deflection angle given by the ESPL lens is
\be
\alpha(\theta)= -\dfrac{\theta_0^{h+1}}{\Theta^{h+1}}\rund{\theta_1q+\ii \theta_2/q},
\label{eq:alphapl}
\ee
and the lensing convergence and shear
\begin{flalign}
\kappa = -B(h) & \eck{ q+1/q - (h+1) {\theta_1^2q^2+\theta_2^2/q^2\over \Theta^2} };\\
\gamma = -B(h) & \left[ q-1/q - (h+1) {\theta_1^2q^2-\theta_2^2/q^2 \over \Theta^2}
-\ii{2(h+1)\theta_1\theta_2 \over \Theta^2} \right],
\label{eq:kappashear-pl}
\end{flalign}
where pre-factor $B(h)=\theta_0^{h+1}/(2\Theta^{h+1})$.
The inverse magnification can be written as
\begin{flalign}
  \mu^{-1} = &1+ 2B(h)\eck{q+1/q-(h+1)\dfrac{\theta_1^2q^2+\theta_2^2/q^2}{\Theta^2}}
  \nonumber \\
  &- 4hB(h)^2 +4(h+1)B(h)^2\dfrac{\theta_c^2}{\Theta^2},
\label{eq:mag-spl}
\end{flalign}

The demagnification at the lens origin has an upper limit as well
\be
\mu_{\rm origin} \leq \dfrac{1}{\rund{1+{\theta_0^{h+1} \over \theta_c^{h+1}}}^2}.
\ee
It depends on the ratio between the core radius $\theta_c$ and $\theta_0$,
and a large core radius weakens the lensing effects and provides a potential
explanation for the observations with weak demagnification. The ellipticity
however will increase the lensing demagnification.
We further simplify the mathematics by considering the
singular case, i.e. $\theta_c=0$, and regress the magnification to one
dimension on the $\theta_1$- and $\theta_2$-axis. The critical curve crosses the two axis at
\be
\theta_{1c}=\theta_0 \rund{\dfrac{h}{q^{(h-1)/2}}}^{1\over h+1},\;\;
\theta_{2c}=\theta_0 \rund{h q^{(h-1)/2}}^{1\over h+1}.
\ee
For the softened model, the existence of the critical curve depends on
$\theta_c$ and can be solved numerically. For the lenses of $h=1$, the
critical curve cross both two axis at the same length without
dependence on the axis ratio $q$, i.e. $\theta_{1c}=\theta_{2c} =\theta_0$.

\begin{figure}
  \includegraphics[width=7.0cm]{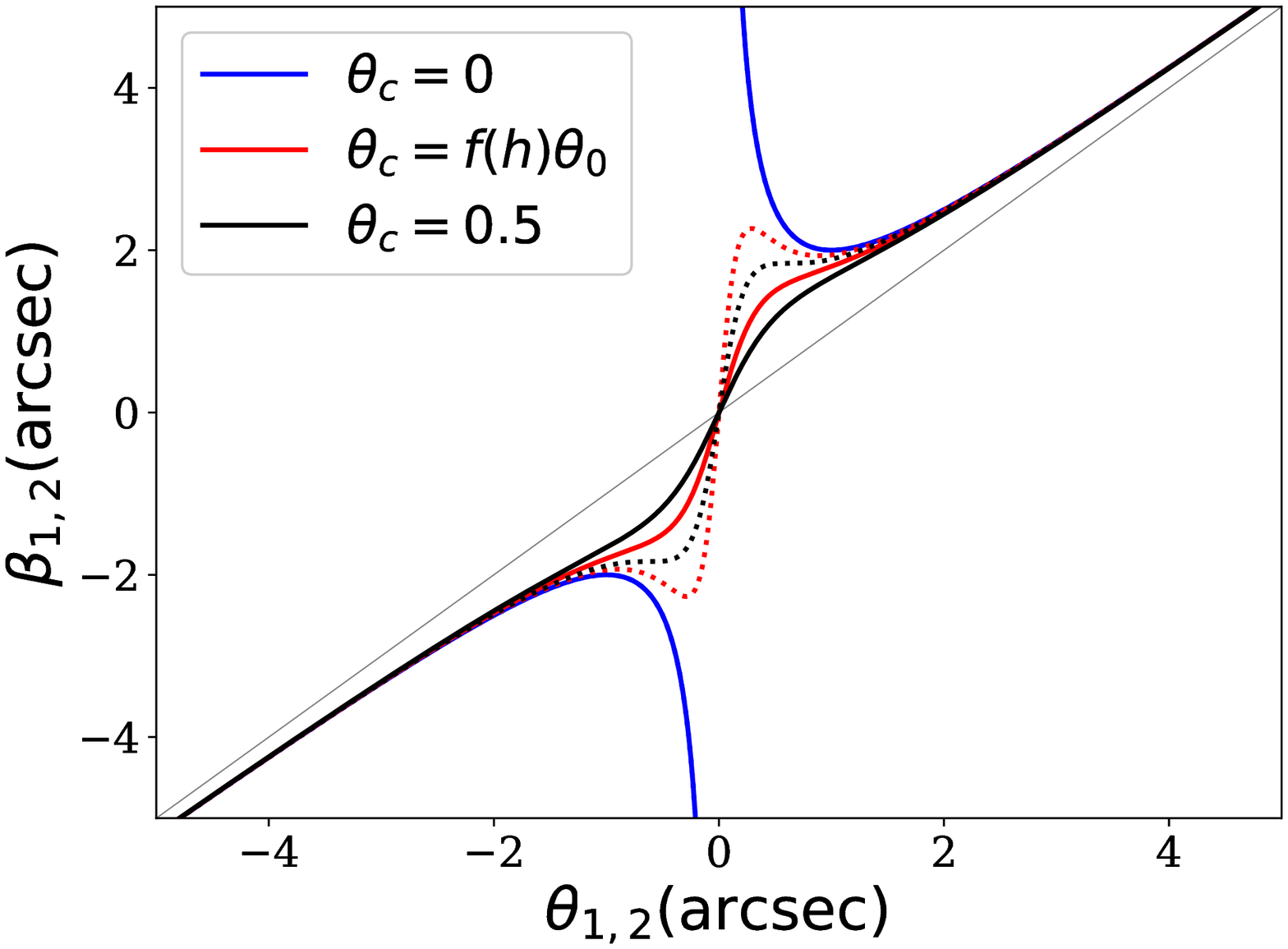}\\
  \includegraphics[width=7.0cm]{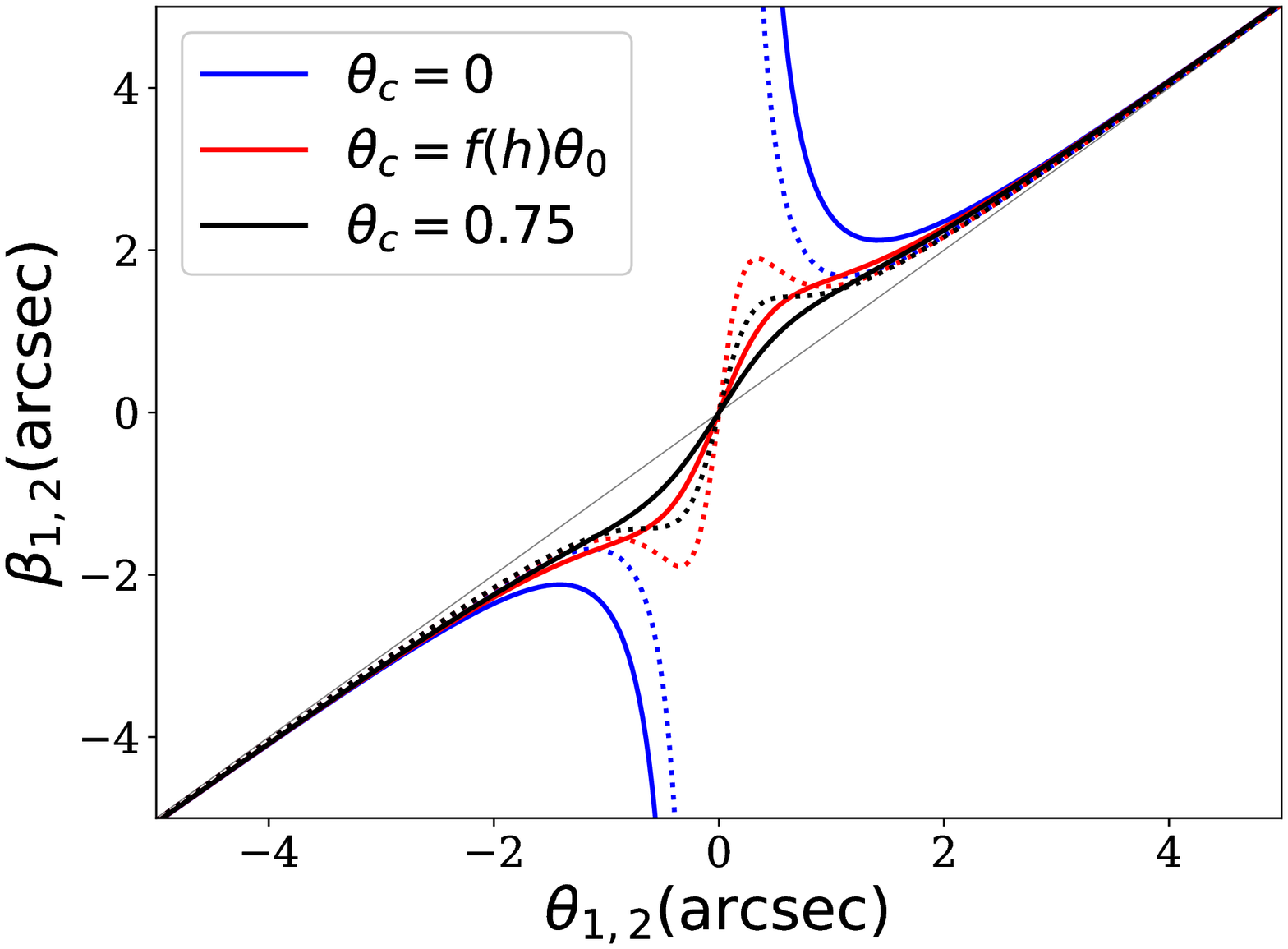}
  \caption{Young diagram of elliptical softened power law models with
    parameters: $\theta_0=1$, $q=0.5$, $h=1$ (top) and
    $h=2$ (bottom). The solid (dotted) lines present the source
    positions on the major (minor) axis.}
  \label{fig:spl-alpha}
\end{figure}

\begin{figure*}
  \centerline{
  \includegraphics[width=7.0cm]{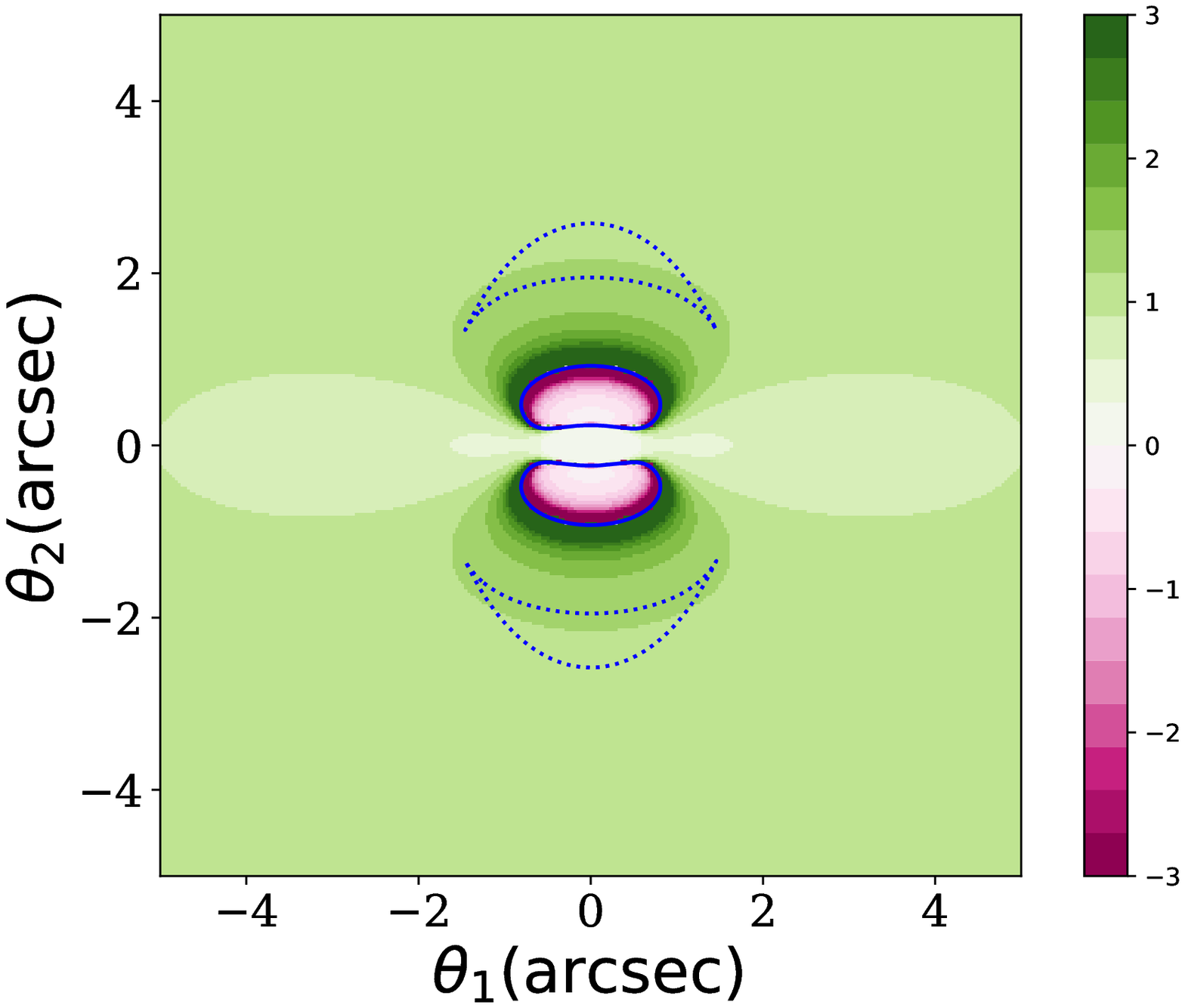}\hspace{-0.6cm}
  \includegraphics[width=7.0cm]{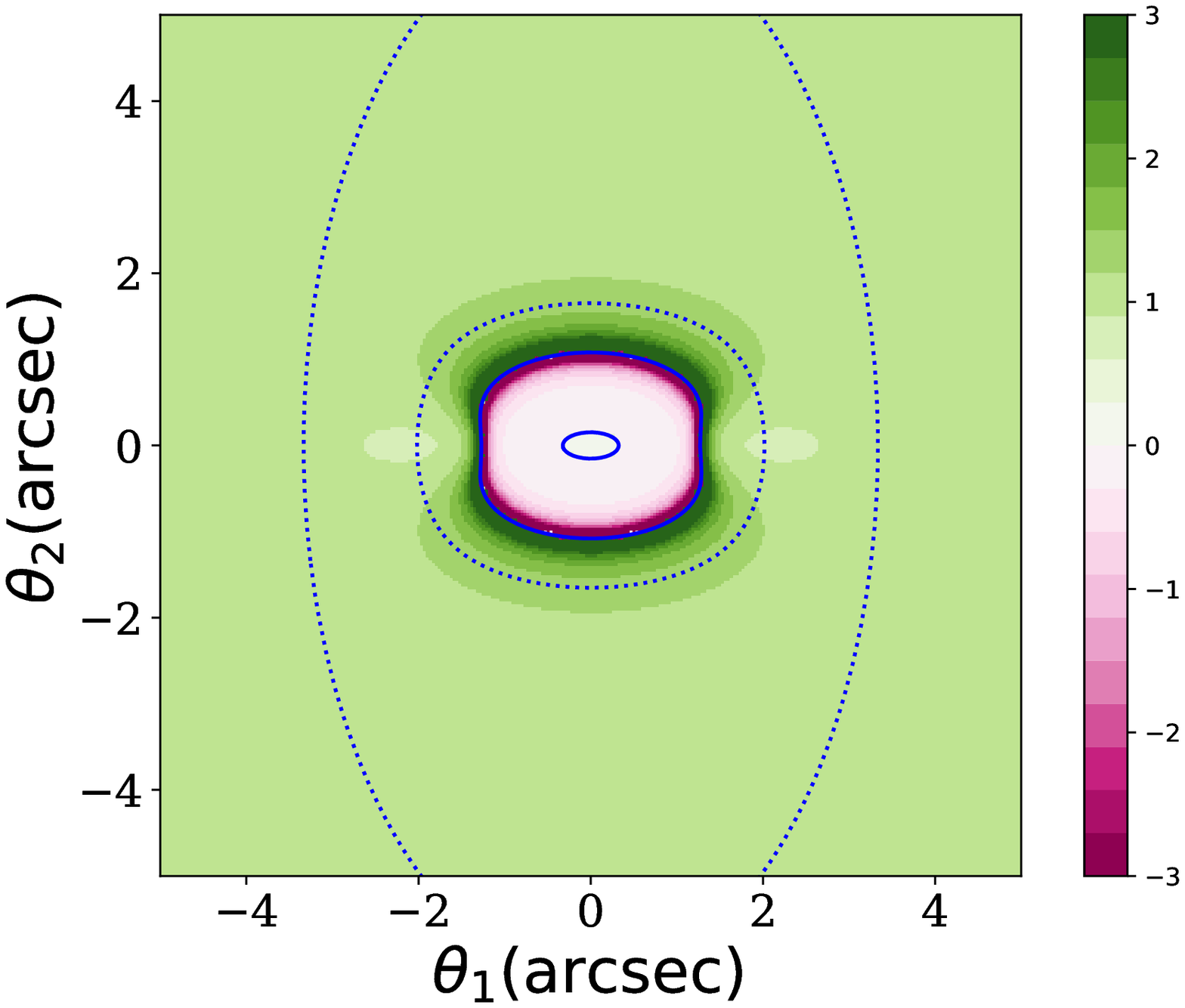}\hspace{-0.6cm}
  \includegraphics[width=7.0cm]{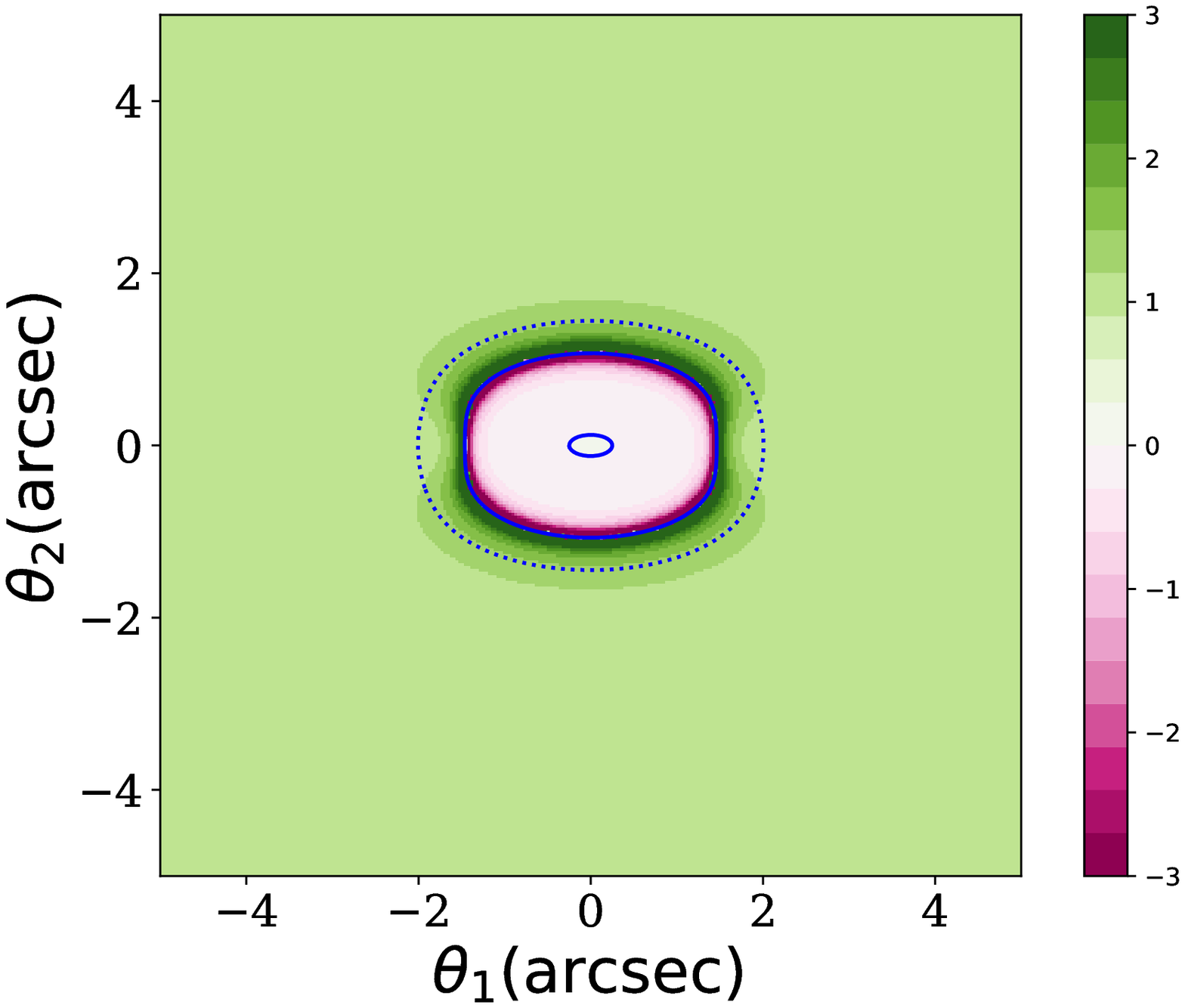}}
  \caption{The two dimensional magnification map of softened power-law lens
    with parameters $\theta_0=1$, $q=0.5$, $\theta_c=0.3$, $h=1$
    (left), $h=2$ (middle) and $h=3$ (right). The same color code is
    applied as in Fig.\,\ref{fig:expmag}. }
    \label{fig:splmag}
\end{figure*}

In Fig.\,\ref{fig:spl-alpha}, the Young diagram of ESPL lenses with
axis ratio $q=0.5$ is shown. The same colour code is applied here: the
blue, red and black lines represent the super-, critical and
sub-critical case in the circular lens models respectively. The solid
and dotted line shows the source position as a function of the image
plane position along the major and minor axes, respectively.  Similar as in
the one-dimensional case, increasing the core size weakens the lensing
effects but causes more complicated lensing behaviour, which can be
seen from the number of criticals. The deflection angles along the
major axis are slightly smoother than that on the minor axis as well.
The turnover where the slope of the line vanishes marks the boundary
of the exclusion region for this case. The monotonic solid lines
indicate that along the major axis there is no sharp edge of
the exclusion region for certain lens configurations, which can be seen
from the magnification map (Fig.\,\ref{fig:splmag}). The lens with
$h=3$ has similar properties as that of $h=2$ and is not shown here.

\begin{figure*}
  \centerline{
  \includegraphics[width=7.0cm]{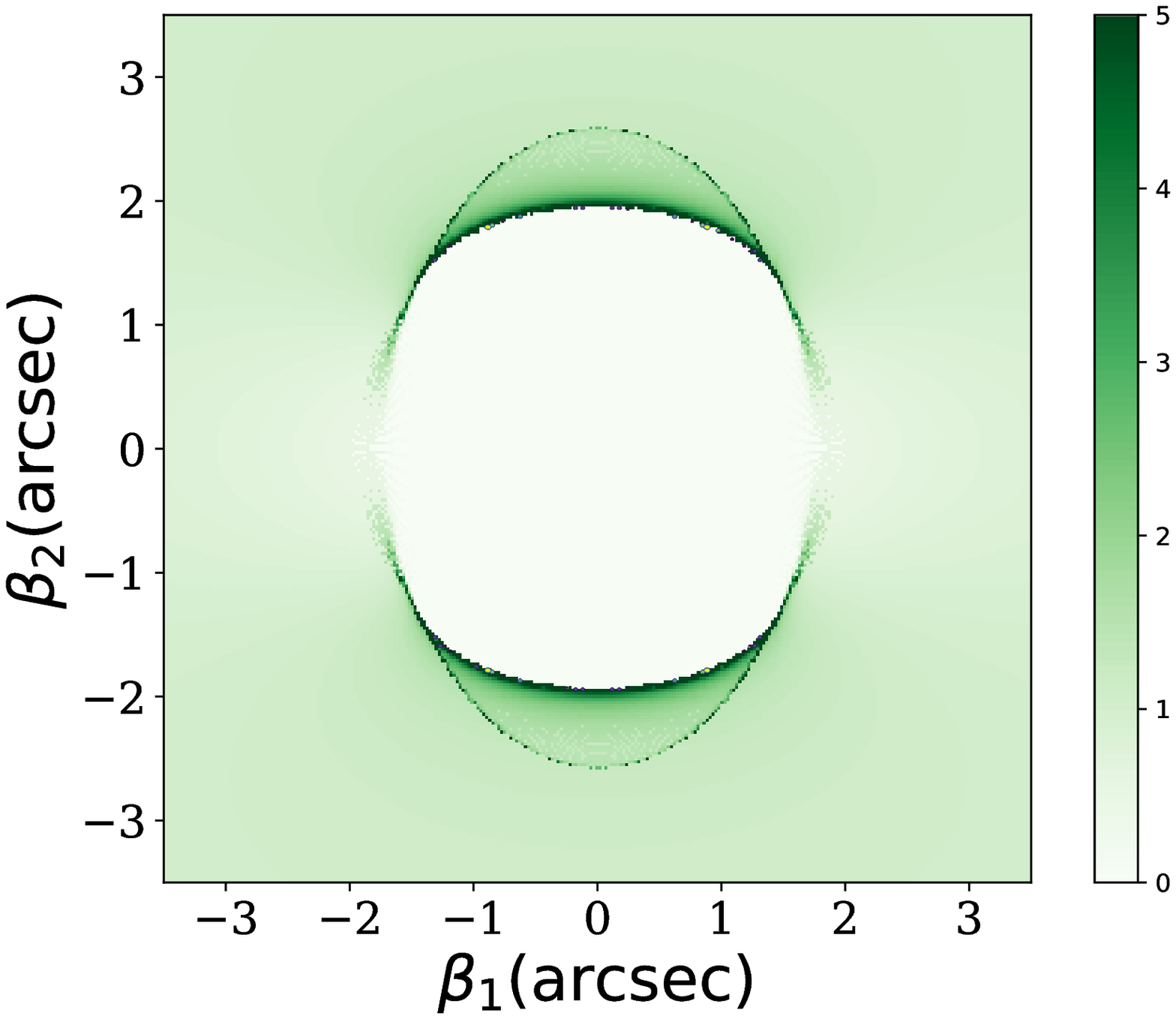}\hspace{-0.6cm}
  \includegraphics[width=7.0cm]{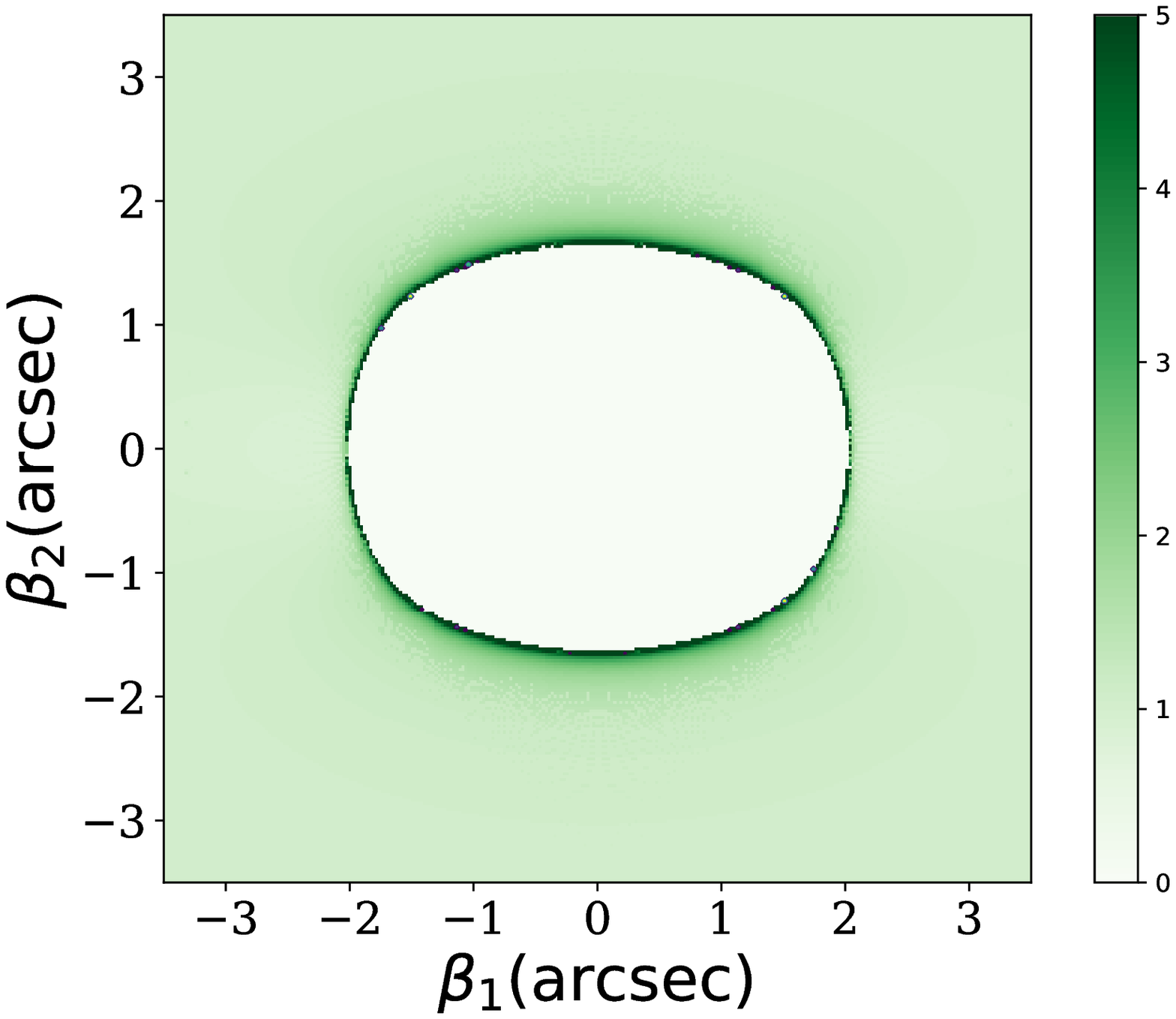}\hspace{-0.6cm}
  \includegraphics[width=7.0cm]{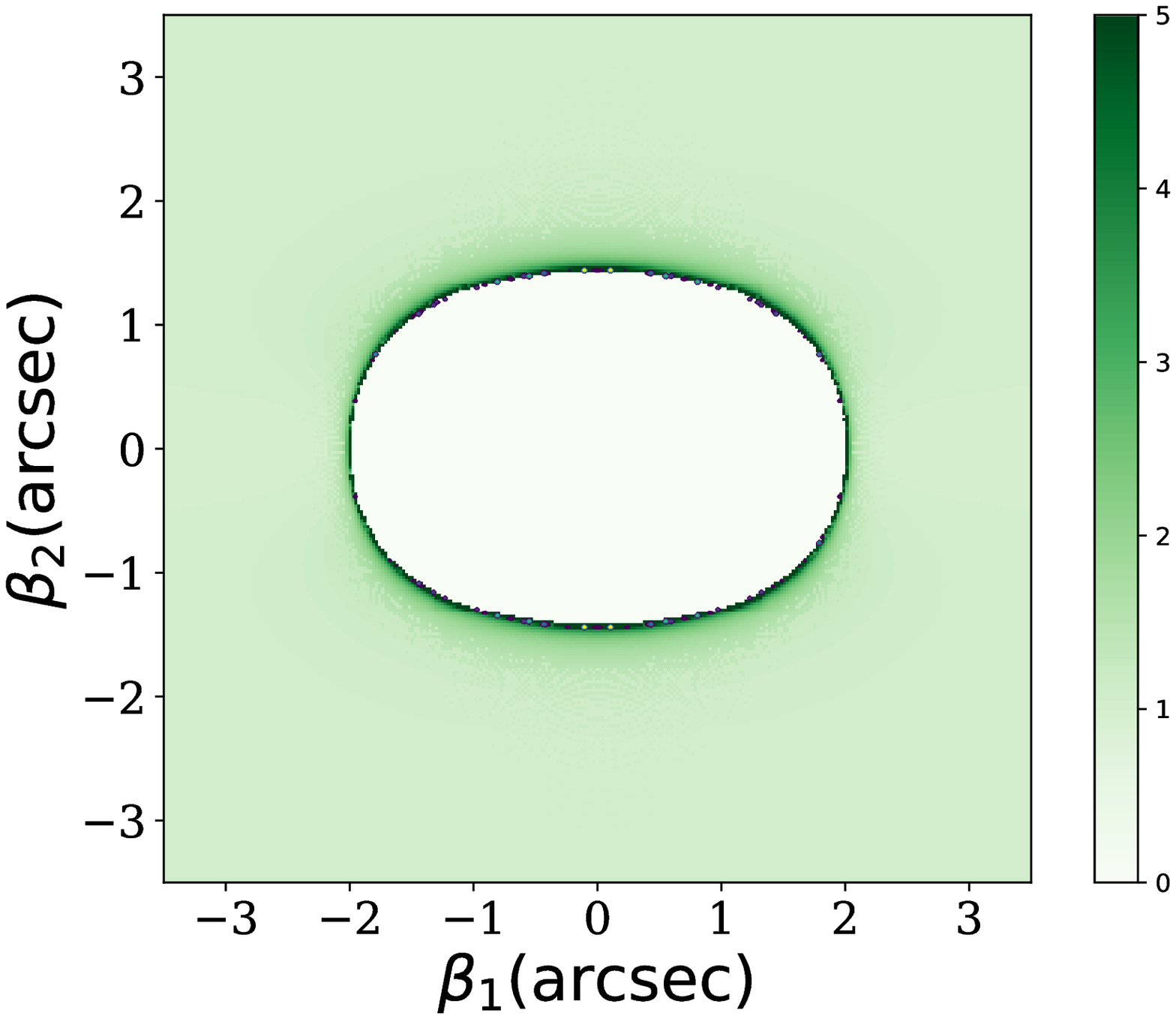}}
  \caption{The magnification map of the ESPL model on the source plane
    with the same parameters as in Fig.\,\ref{fig:splmag}. The
    magnification is truncated at $5$, and the maps are slightly zoomed
    into the center region for better visibility.}
  \label{fig:spl-musource}
\end{figure*}

The magnification maps for ESPL are shown in Fig.\,\ref{fig:splmag}.
They show distinct patterns compared with that of the exponential
lenses, although the critical curves show similar shapes between the
exponential lenses and the ESPL lenses. In all the ESPL lenses, the
demagnification region is elongated along the major axis. Especially
for the $h=1$ lens, the criticals and magnification regions locate separately
at the two sides of the major axis, while the demagnification region
continues along the major axis. Also in the lenses of $h=2,3$, there
exists weak demagnification out of the criticals along the major axis.
This can also be seen from the source plane. In
Fig.\,\ref{fig:spl-musource}, the corresponding magnification maps of
ESPL lens on source plane are present. In the left panel, the two
caustics and high magnification regions are on the two sides of the
major axis. Similar to the elliptical Gaussian lens, along the major
axis such a lens will only cause a demagnification effect, whereas along
the minor axis, the lens can cause four spikes on the light curve
which is similar to the case of a circular lens. The middle panel shows
a complete ellipse. The second caustic is beyond the scope of the
figure. The ESPL lens of $h=3$ has some similar properties as that of
$h=2$.

\begin{figure}
  \includegraphics[width=6.0cm]{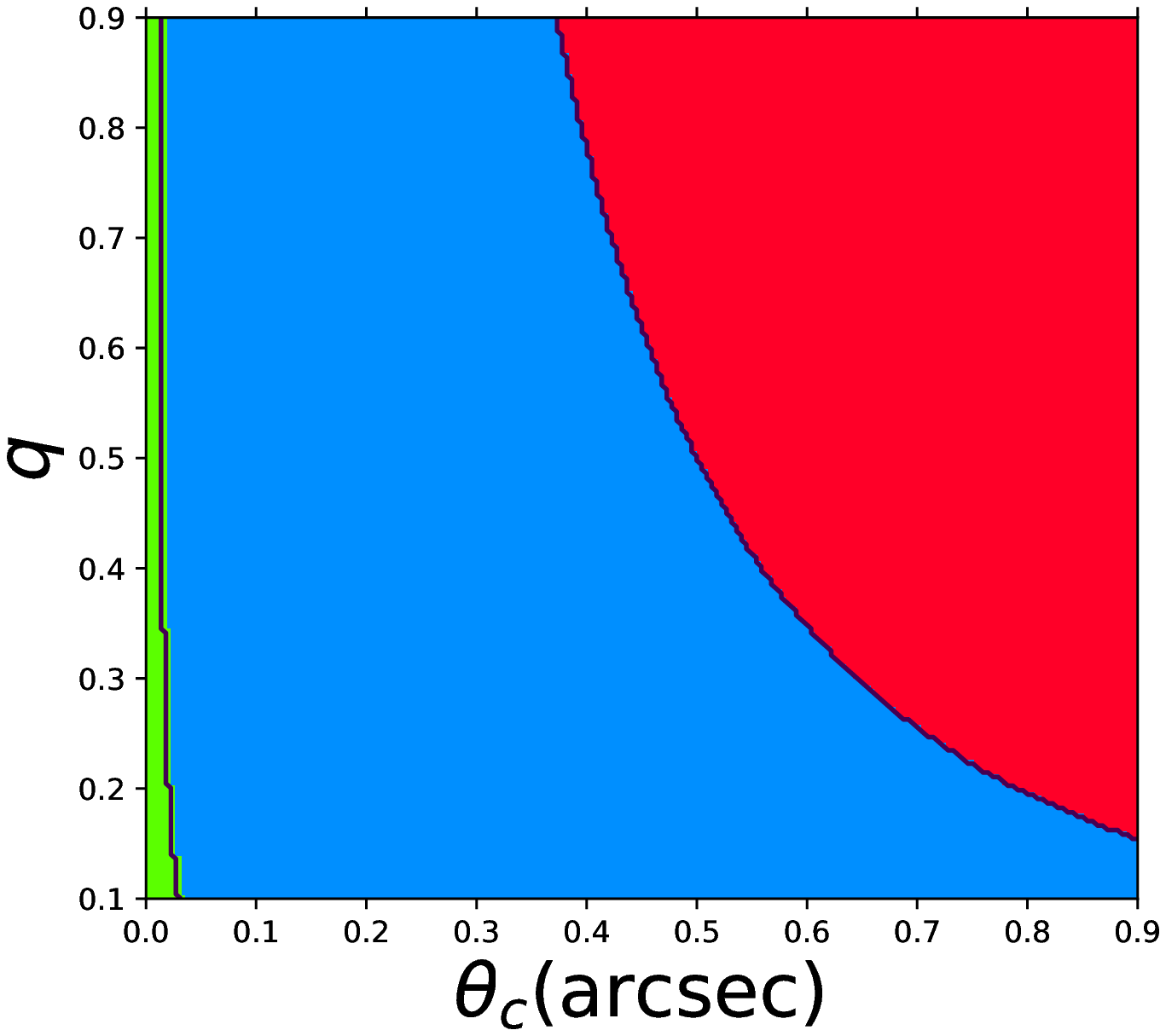}\\
  \includegraphics[width=6.0cm]{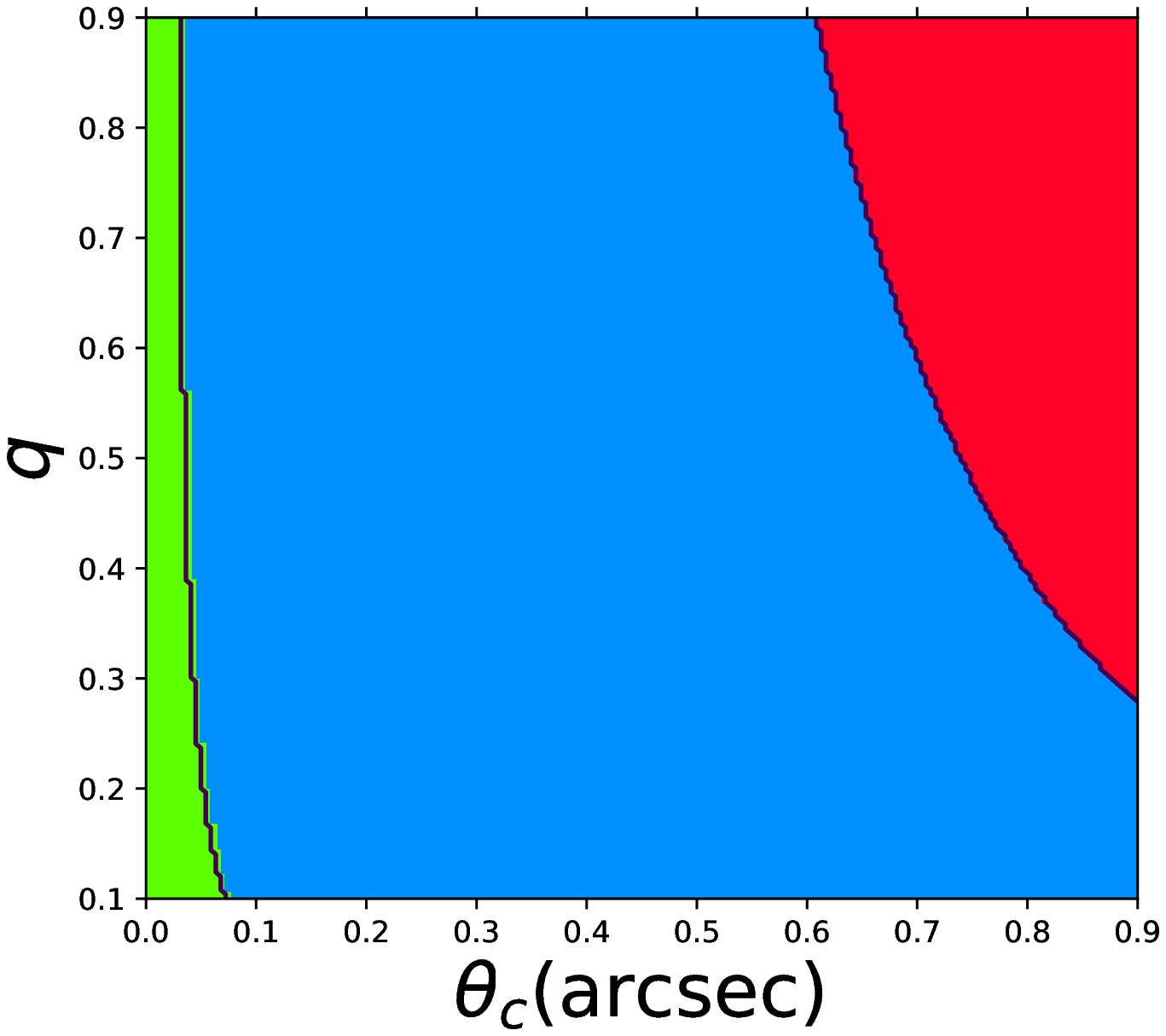}\\
  \includegraphics[width=6.0cm]{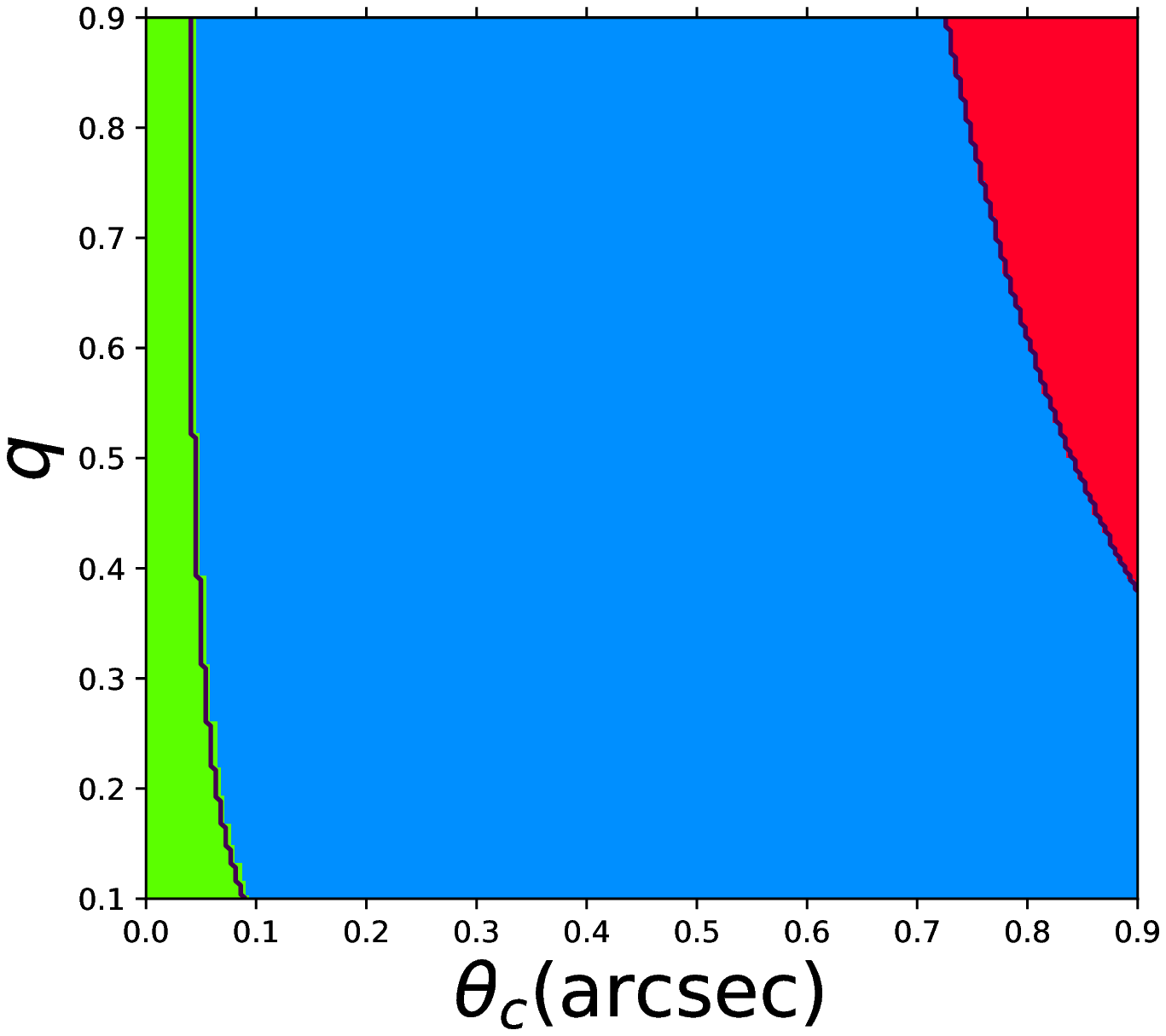}
  \caption{This figure shows the number of critical curves for the
    elliptical ESPL lenses as a function of angular core radius
    $\theta_c$ and axis ratio $q$ ($\theta_0=1$). The same color
    code as Fig.\,\ref{fig:expccs} is used in this figure. From
      top to bottom, the lens families of $h=1,2,3$ are present.}
    \label{fig:splccs}
\end{figure}

In Fig.\,\ref{fig:splccs}, we plot the number of criticals for ESPL
lenses. The parameters $q$ and $\theta_c$ are explored, and the same
colour code is applied here as for the exponential lenses. First of
all, same as that in the exponential lenses, the ellipticity increases
the lensing efficiency, i.e. the core radius of sub-critical lenses grows
with the decreasing of $q$, and the highly elongated lens even with
very large core radius ($\theta_c\sim\theta_0$) can create
criticals. It is also interesting to see that the singular lens, or
the ESPL lens with extremely small core radius, can generate
one critical curve (Fig.\,\ref{fig:splc0h123}).  In general, the ESPL
lenses with higher power index $h$ are more efficient in generating
criticals, i.e. the area of the red region becomes smaller in the bottom
panel. The shape of the critical curves generated by the lens family of
$h=3$ is similar to that of $h=2$. We also provide a gallery of the
critical curves and caustics for elliptical ESPL lenses in the appendix.

\begin{figure}
  \includegraphics[width=7.0cm]{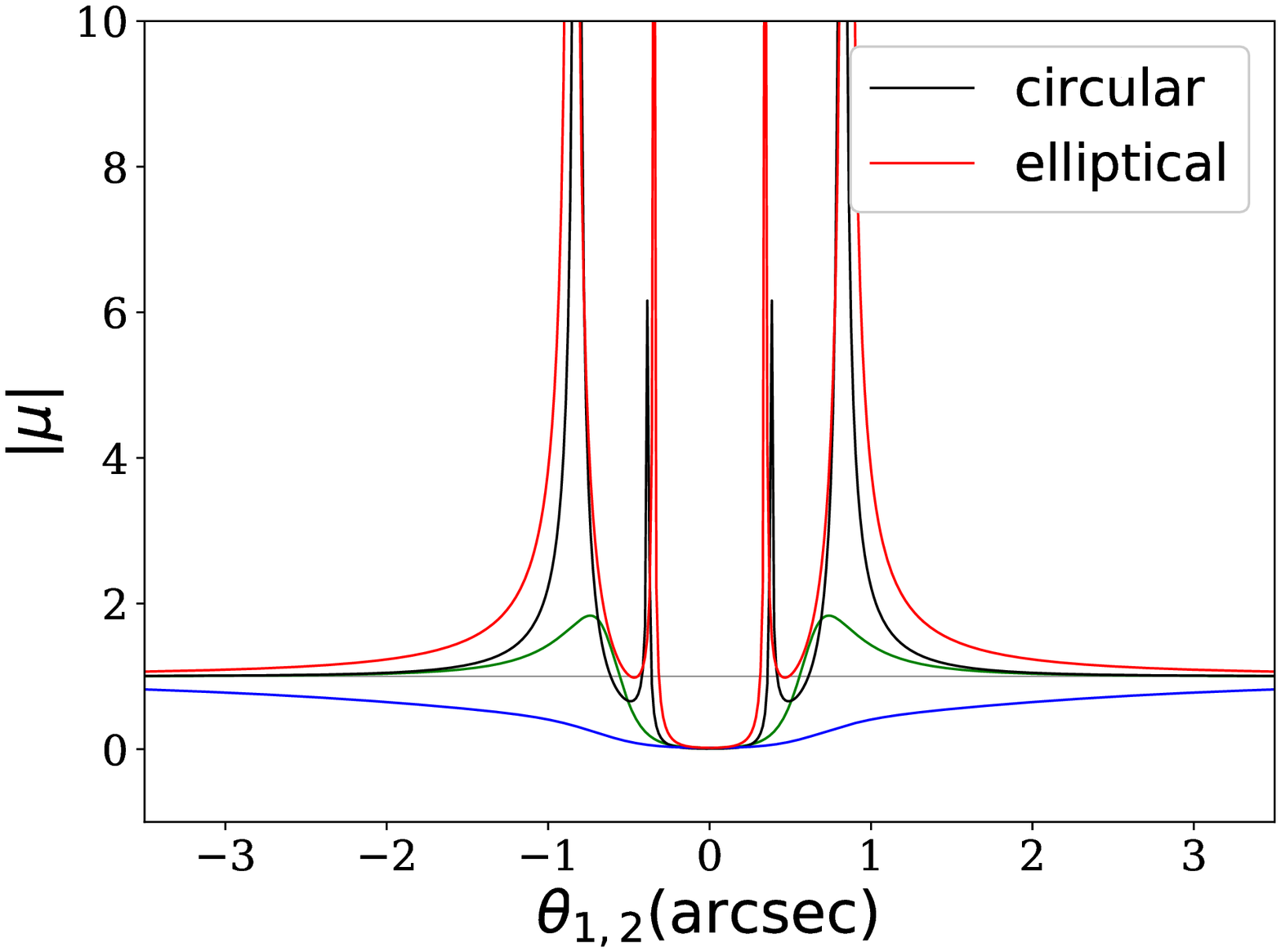}\\
  \includegraphics[width=7.0cm]{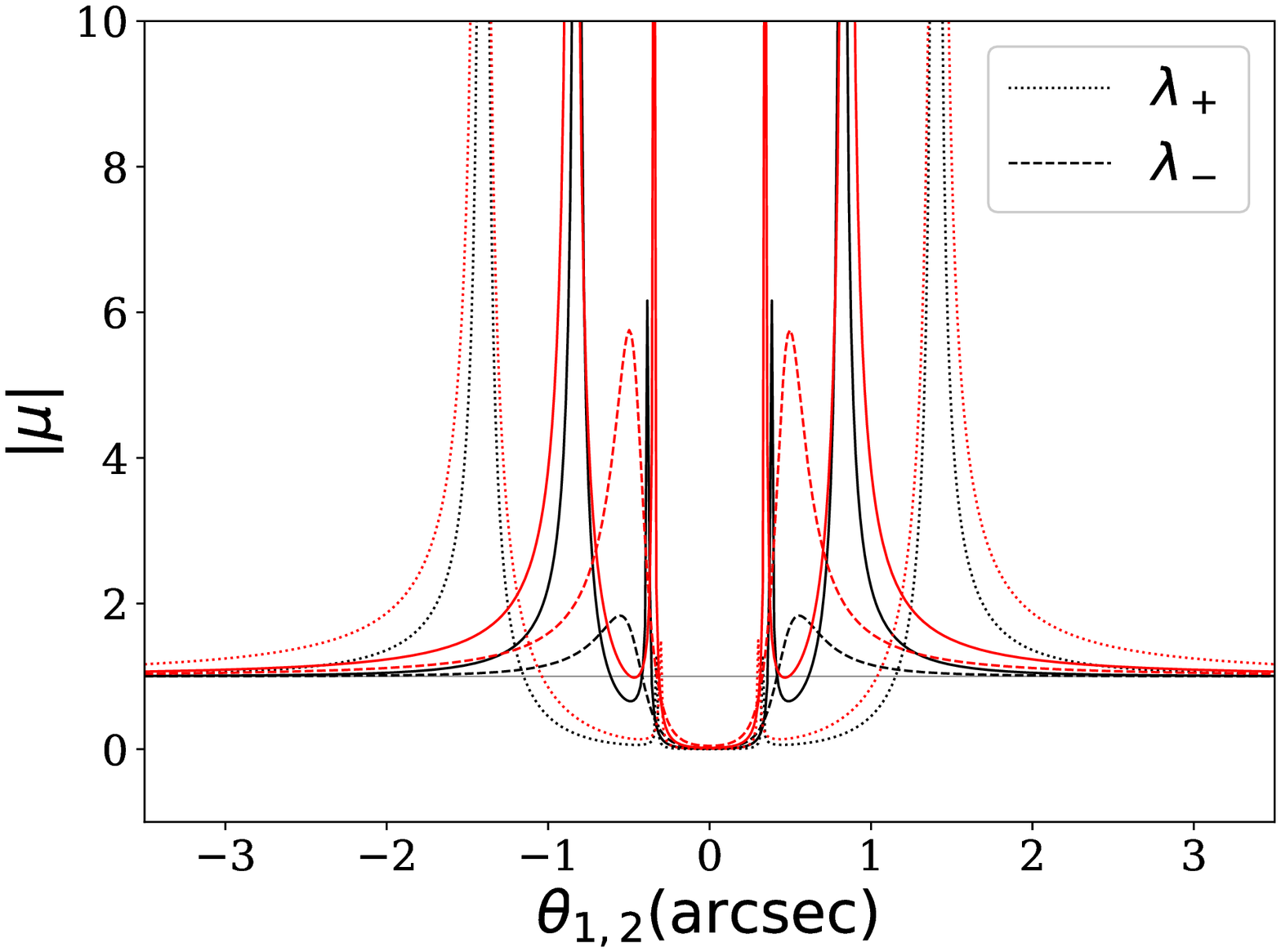}
  \caption{Comparison of magnification curves between circular ($q=1$)
    and elliptical ($q=0.5$) lens for a softened power law model with
    parameters: $h=1$. In the top panel, we use $\theta_0=1$ for all the
    lens models. The green (black) line presents the curve of SPL lens
    with $\theta_c=0.4 (0.3)$. The red (blue) line presents the curve
    of ESPL lens along the minor (major) axis. In the bottom panel,
    the solid red and black lines are identical to that in the top
    panel. The dashed (dotted) lines present the curves for same lens
    parameter but with shorter the wavelength $\lambda_-=0.75\lambda$
    (longer the wavelength $\lambda_+=1.5\lambda$) in the
    observation. }
  \label{fig:spl-comp}
\end{figure}

In addition, we compare the ESPL lenses between the circular model and elliptical model. In Fig.\,\ref{fig:spl-comp} we show the magnification curve along the axis of the lens. The same $\theta_0(=1$) is employed for all the lenses. In the top panel, the green and black line presents the magnification curve for two SPL lens with $\theta_c=0.4$ and $\theta_c=0.3$ respectively, while the red and blue line presents the curve for a ESPL lens with $\theta_c=0.4$ and $q=0.5$ along the minor and major axis. The SPL lens with $\theta_c=0.4$ cannot generate critical curve (i.e. sub-critical lens), while the magnification curve along the minor axis of ESPL lens shows similar behaviour to the super-critical lens. In the bottom panel, the solid lines are identical to those in the top panel. Besides that we present the magnifications if we perform the observations with different wavelengthes. The dotted lines present the magnifications if the wavelength in the observation is increased by 50$\%$, while the dashed lines present that if the wavelength is decreased by $25\%$. In case of the shorter wavelength, both SPL and ESPL lenses become sub-critical, but the curve of SPL lens shows taming behaviour and can be distinguished from ESPL lens. Thus, the multi-bands observations can provide important constrains to the plasma lens models.

\section{Conclusions}
\label{sec:conclusions}
In this work, we generalize two spherical plasma lens models to
produce elliptical lens models. The elliptical models are interesting
since they represent a more general distribution for the free electrons
in the ISM. Moreover, extremely elongated distributions, which are not
a realistic model for the mass distribution in gravitational lenses,
can be useful as plasma lenses since they provide a model for edge-on
plasma sheets and filaments. We demonstrate that the details of the
density profile play an important role in the lensing effect due to
the density gradient and the ellipticity. We start from an elliptical
plasma lensing potential, and show the analytical lensing expressions
for the elliptical exponential and the softened power-law families. We
performed numerical studies for each of the lens models, producing
maps that catalogue the production of critical curves as a function of
the lens parameters. We present these critical curve maps for each of
the lensing families. We found that the ellipticity can significantly
improve the lensing efficiency in generating critical curves. The
elliptical lens with the sub-critical condition given by the spherical
lens can also generate strong magnification variations, a marked
difference from the spherical lens behaviour. This may also help in
explaining the overpressure problem in the ESEs. A quantitative
analysis of how strongly the ellipticity can improve the lensing
efficiency with real observations is of interest. Moreover for the
lens families with higher power index, the ellipticity of the lens has
a larger impact in generating the critical curves.

%
In several cases, the lens can generate an extended demagnification
region along the major axis. Along the minor axis, the
magnification curve shows a behaviour similar to that of the circular lens.

The elliptical lenses offer a rich variety of magnifications for
background sources. Therefore, model degeneracy may exist in fitting
one dimensional light curves. In addition, our study only considers a
single frequency.  Radio observations contain multi-frequency power
spectra. The magnification of plasma lensing strongly depends on the
observing frequency ($\mu\sim \lambda^4$). Thus, the magnification
curve over a frequency band may show dramatic changes, especially at
the critical points. Therefore, the multi-frequency light curve of the
background source should be used in plasma lensing modelling and
can provide tight constraints on the lens parameters and even break
the model degeneracy altogether. In addition, wide frequency observations are also useful for constraining pulsar secondary spectra as discussed by \citet{kerr2018}.

Our work opens several other questions for future work, especially in
terms of the highly asymmetrical distributions. Besides the
ellipticity, the arc-shape or the small scale variations,
i.e. clumpiness will also introduce a large variety of magnification and image properties. In addition, the elliptical lenses
require further study and comparison with real ESE observations, a
topic we plan to explore in future work. Polarization can also provide
a wealth of information on both the lens itself as well as the
magnetic field along the line of sight. Detailed studies on the
polarization pattern will also be useful and interesting.

\appendix
\section{A gallery of critical curves and caustics}
In the appendix, we present the critical curves and caustics for the
elliptical models in this work. We have tried to display critical
curves and caustics from each region shown in Fig.\,\ref{fig:expccs}
and \ref{fig:splccs}. Some of the caustics extend beyond the scope of
the panel. We omit those large caustics in order to present better
resolution of the inner region.

\subsection{Elliptical exponential lenses}
\begin{figure*}
  \centerline{\includegraphics[width=5cm]{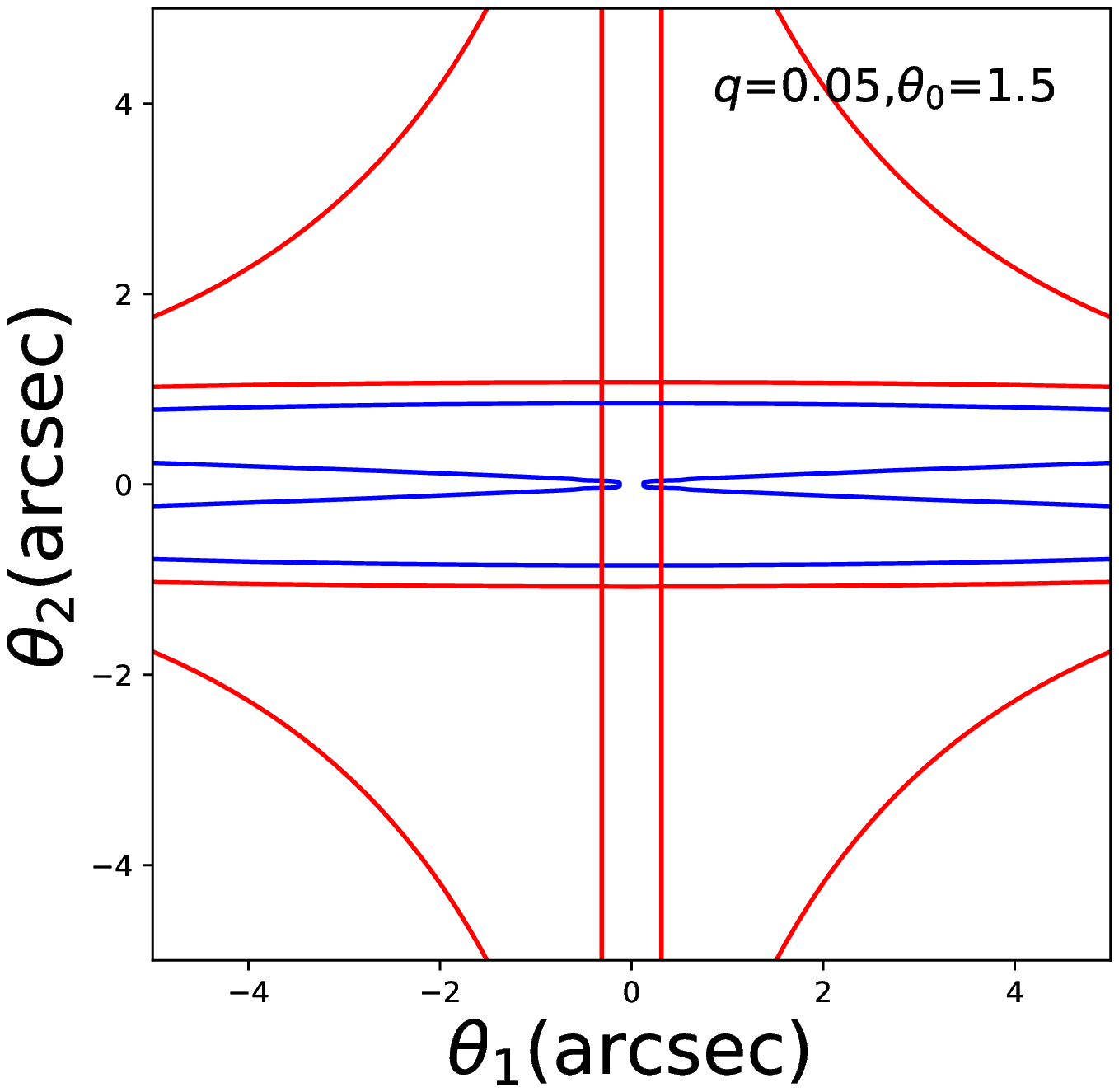}
    \includegraphics[width=5cm]{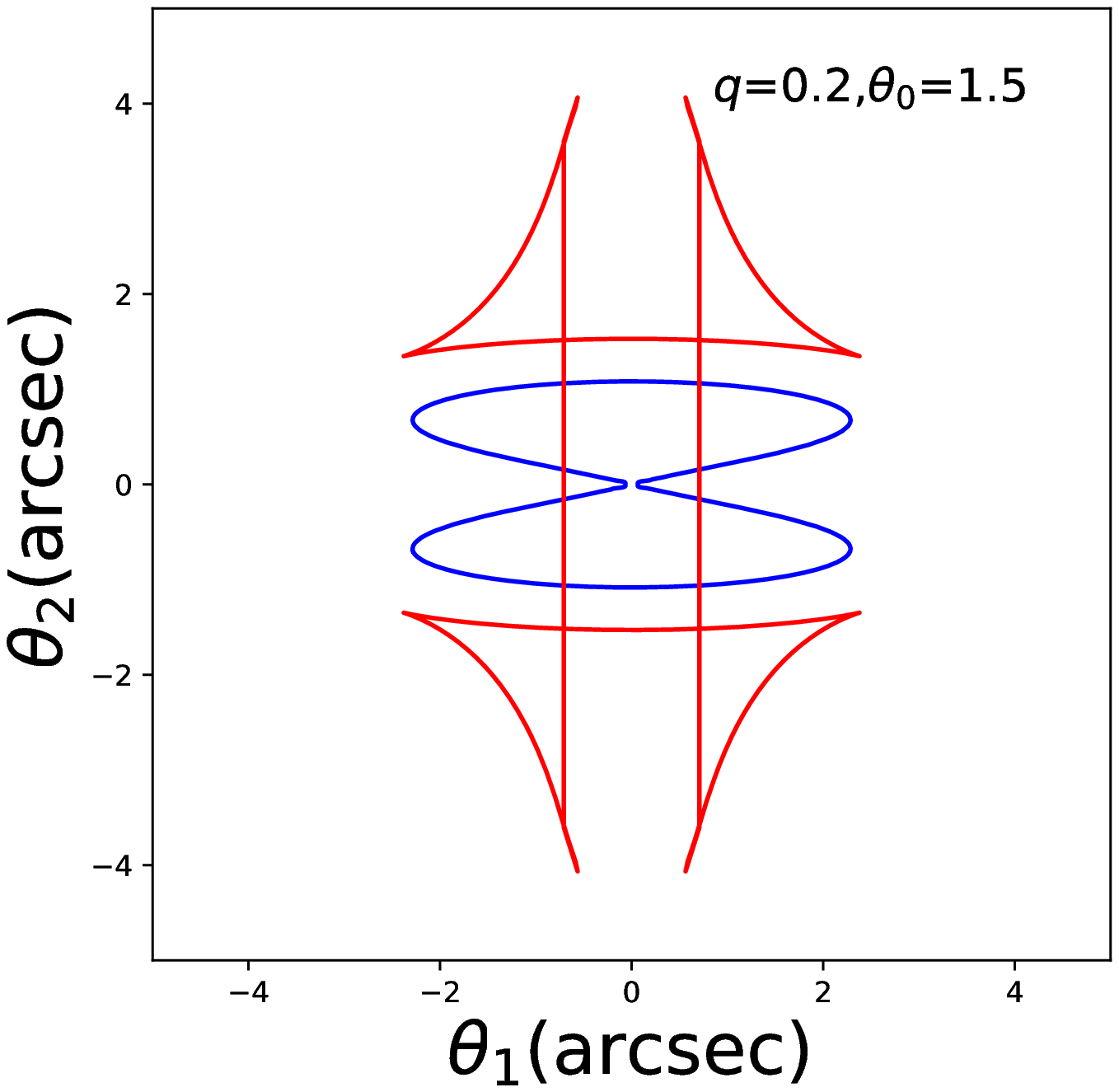}
    \includegraphics[width=5cm]{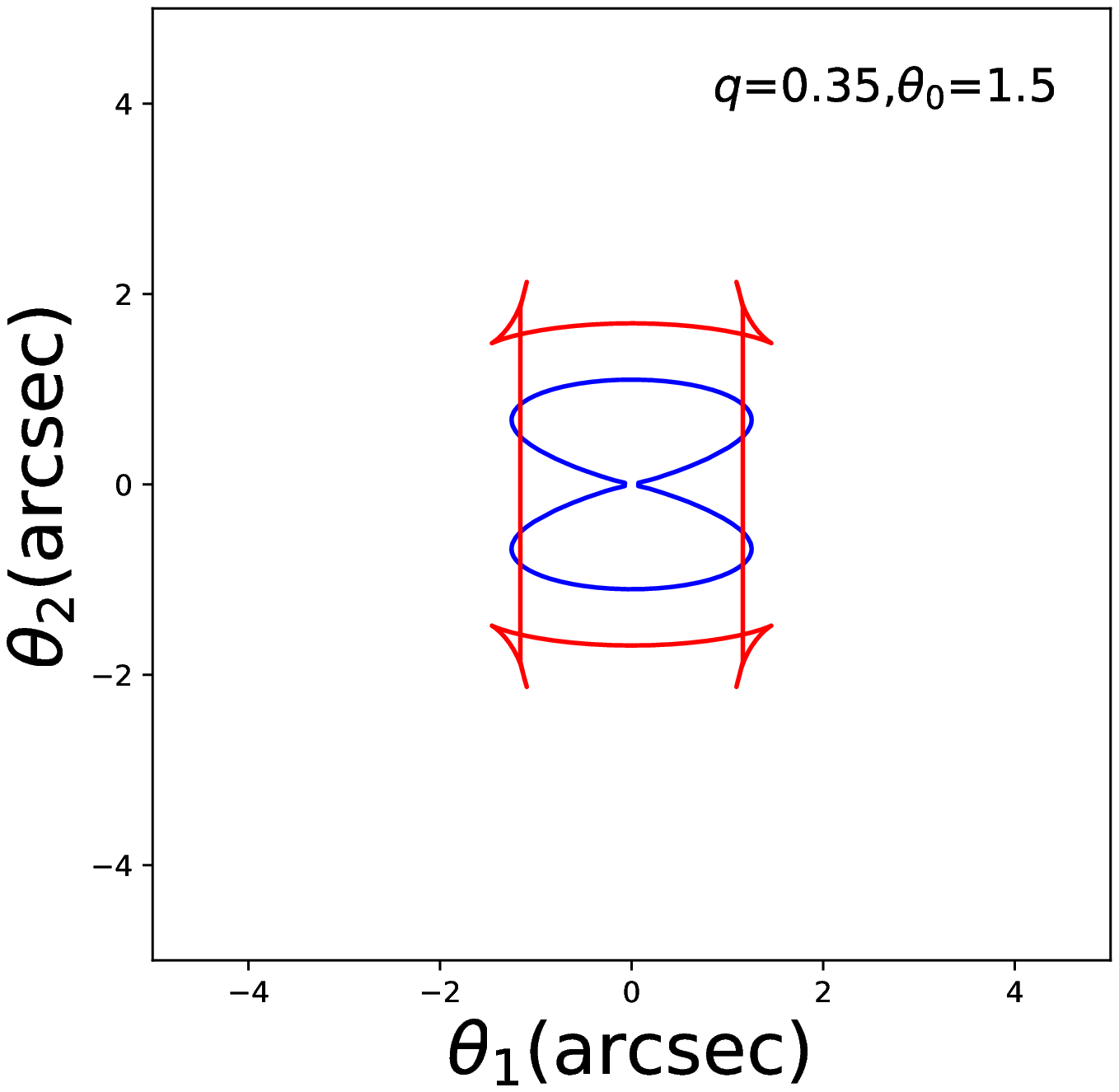}}
  \centerline{\includegraphics[width=5cm]{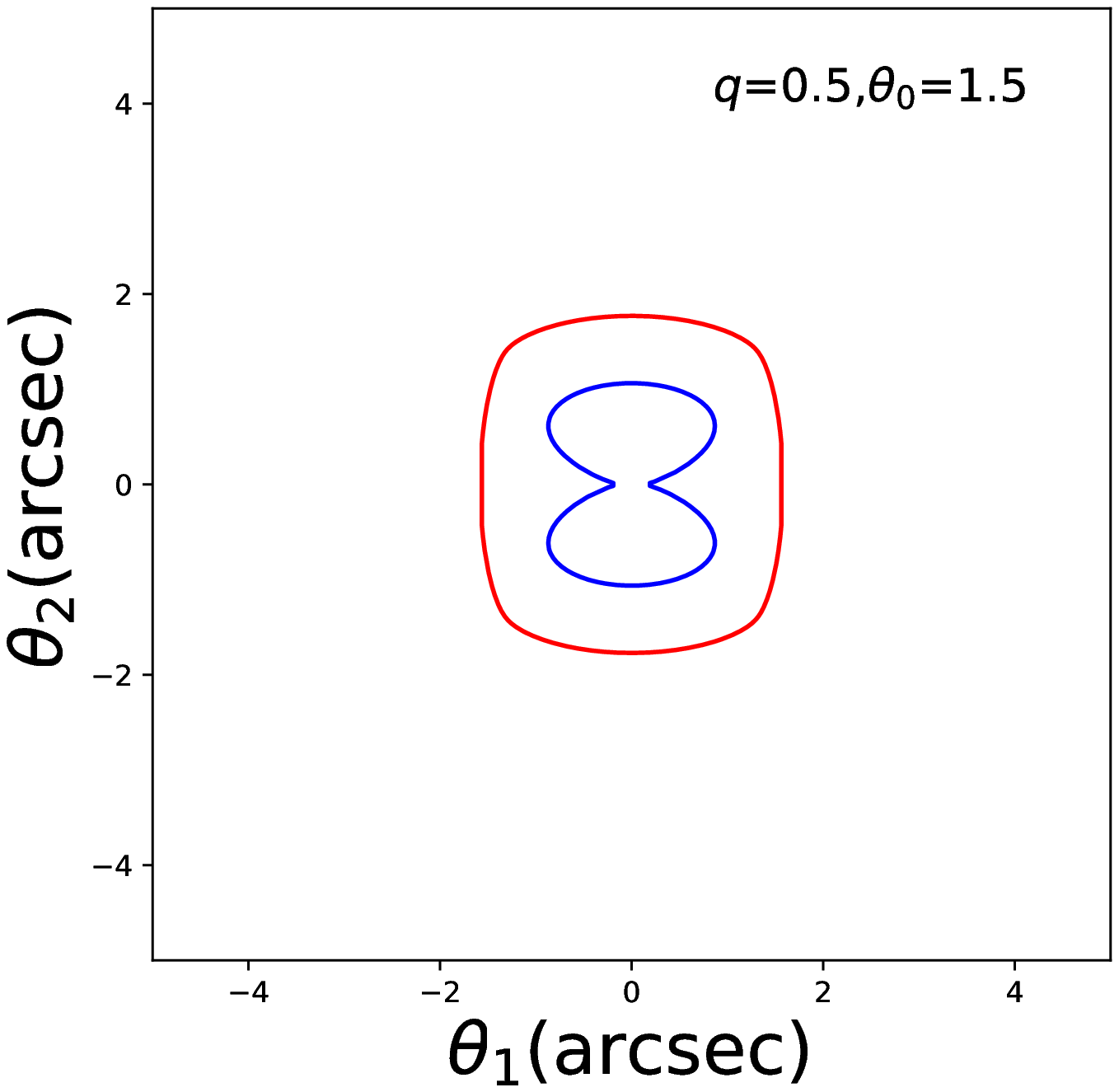}
    \includegraphics[width=5cm]{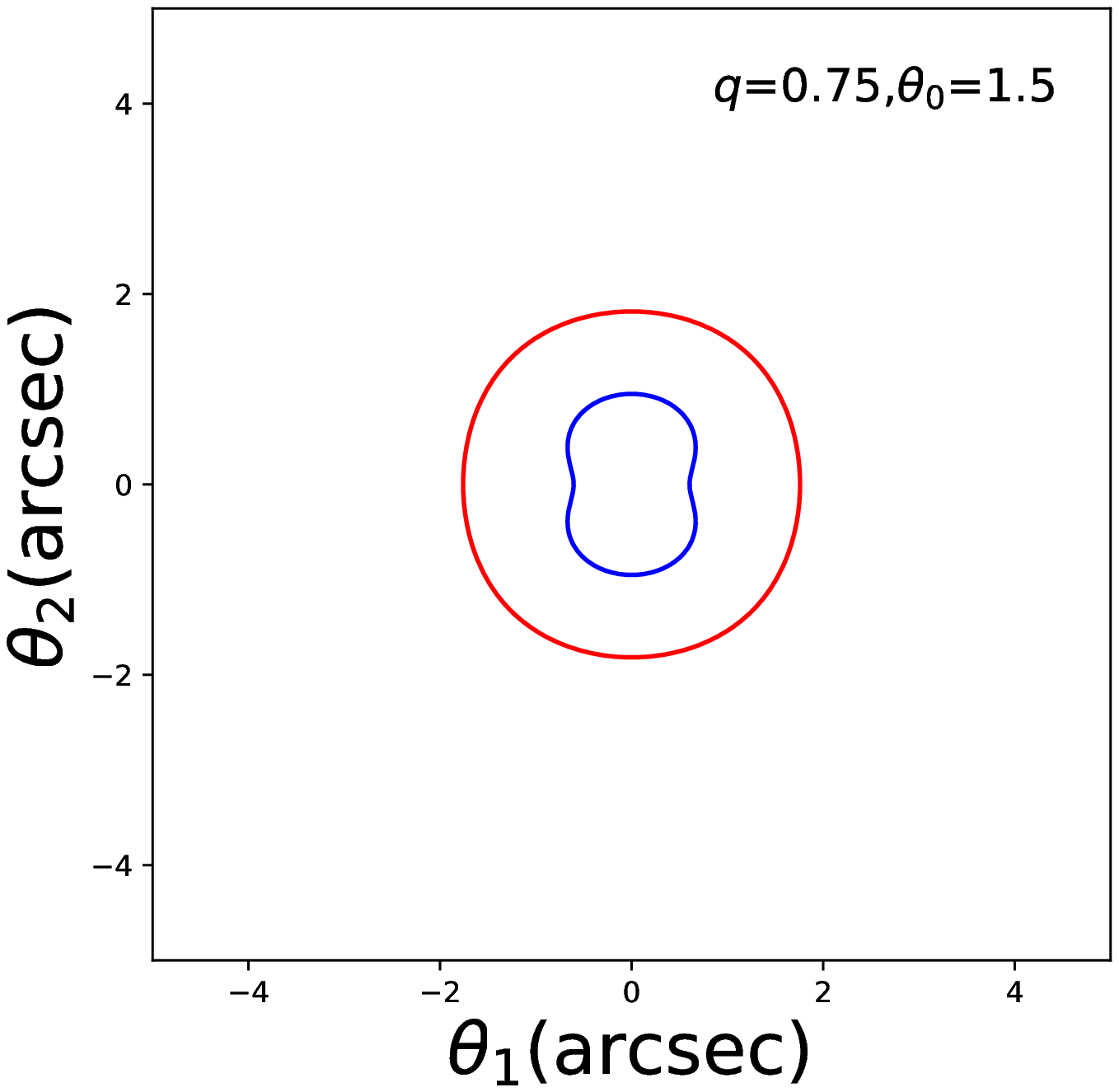}
    \includegraphics[width=5cm]{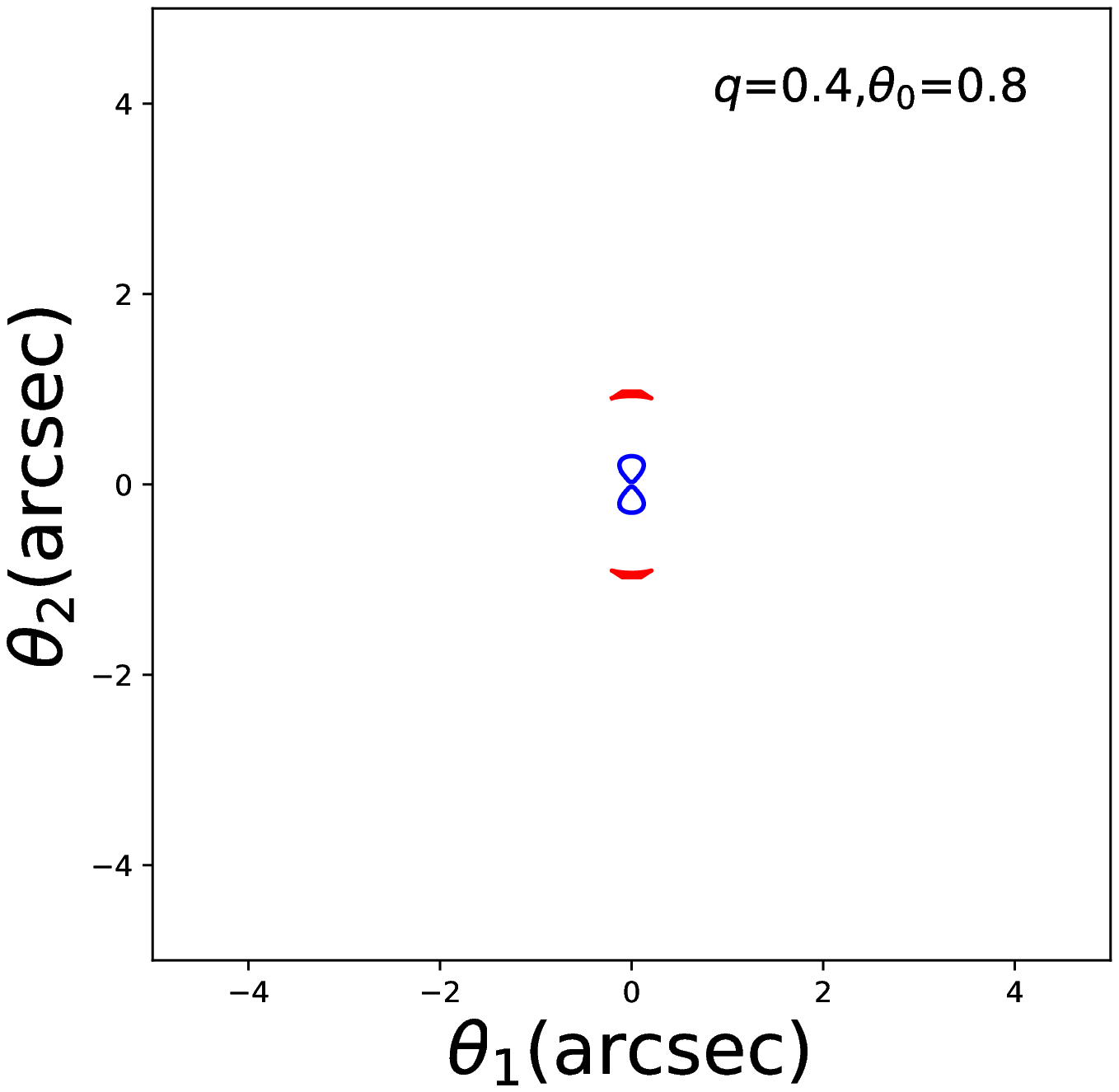}}
  \caption{The critical curves (blue) and caustics (red) of the elliptical
    exponential lens with $h=1$ and $\sigma=1$. The axis ratio and the
    characteristic radius are given at the top right corner in each
    panel.}
  \label{fig:expcch1}
\end{figure*}

\begin{figure*}
  \centerline{\includegraphics[width=5cm]{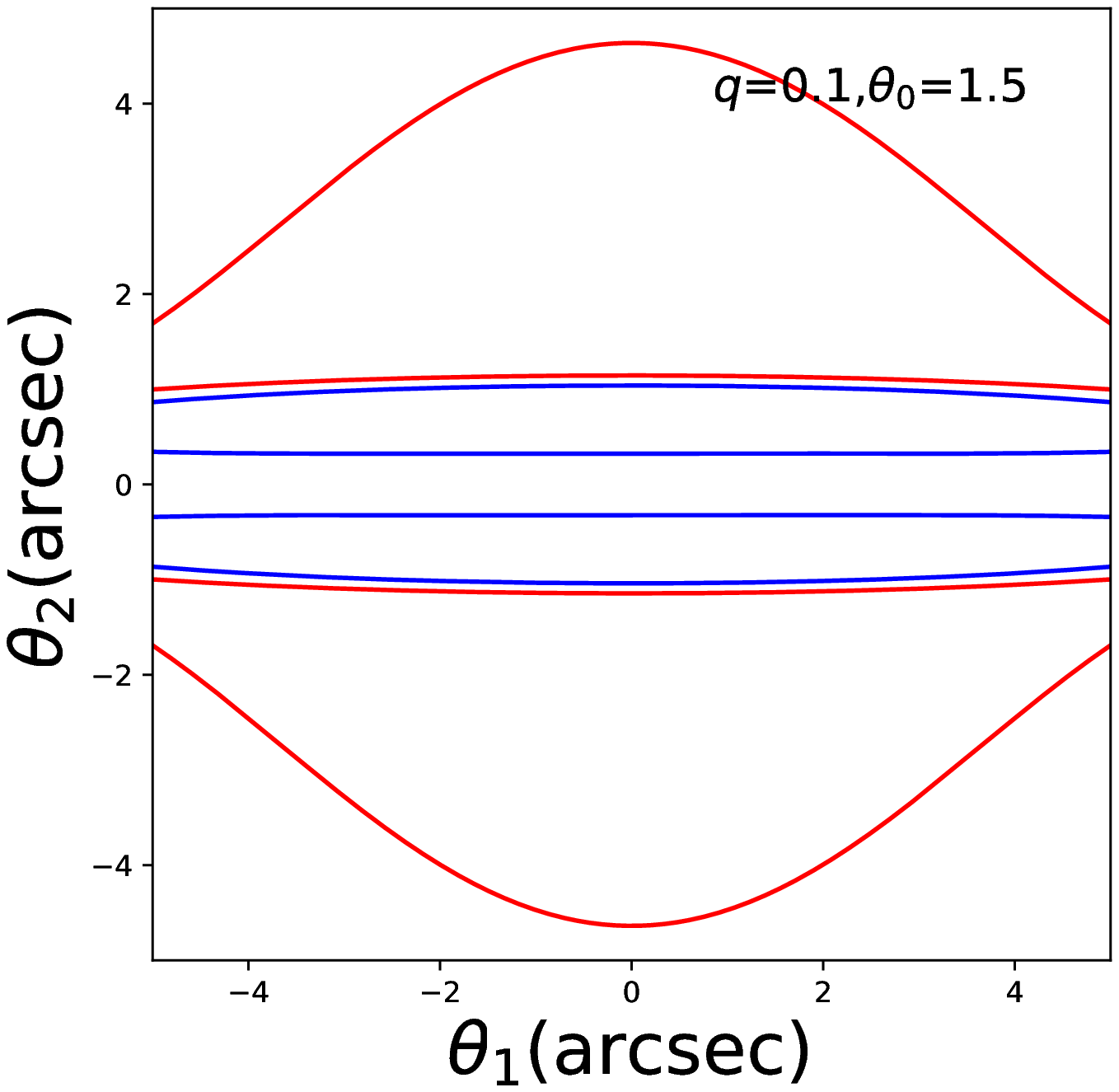}
    \includegraphics[width=5cm]{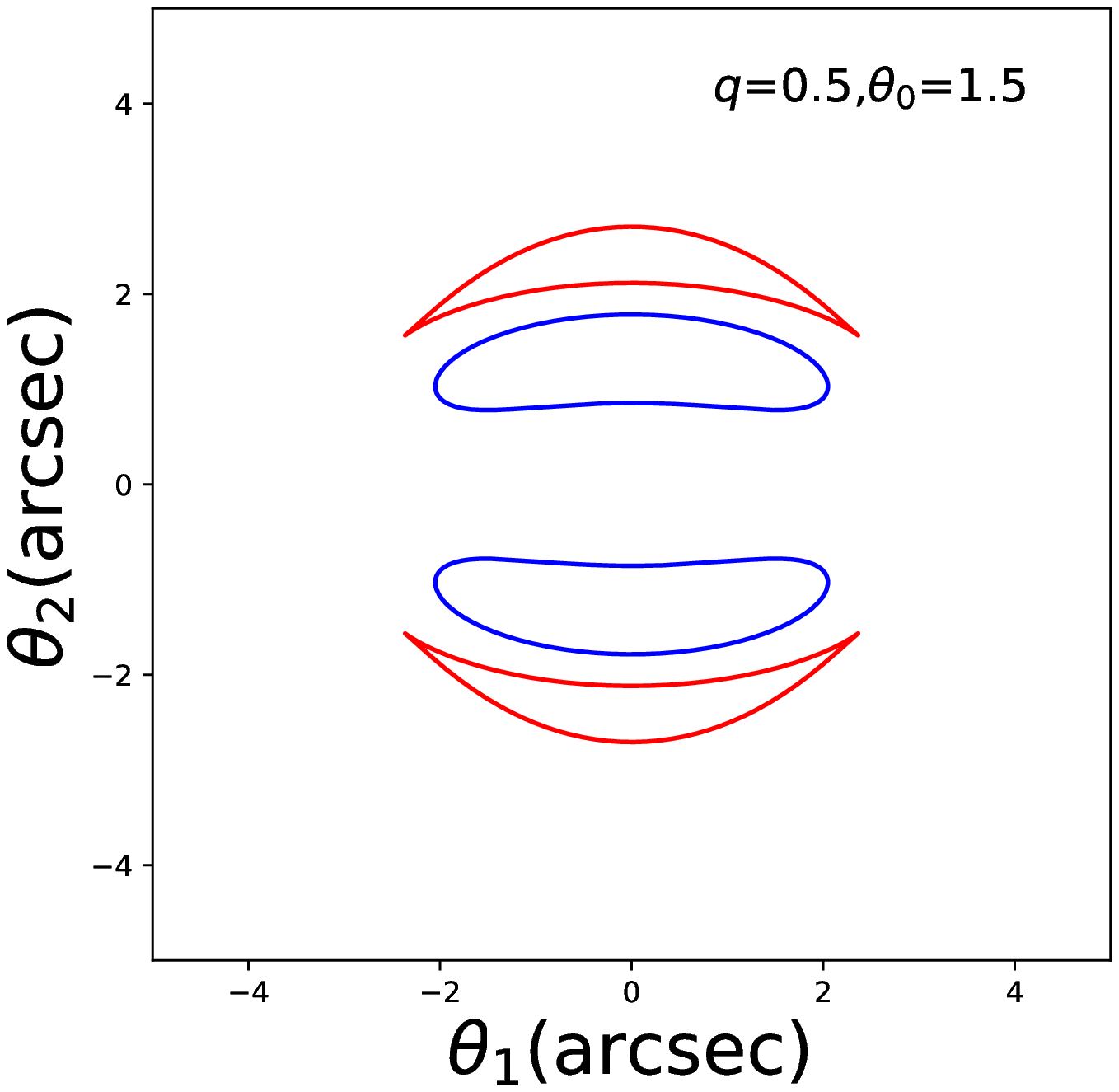}
    \includegraphics[width=5cm]{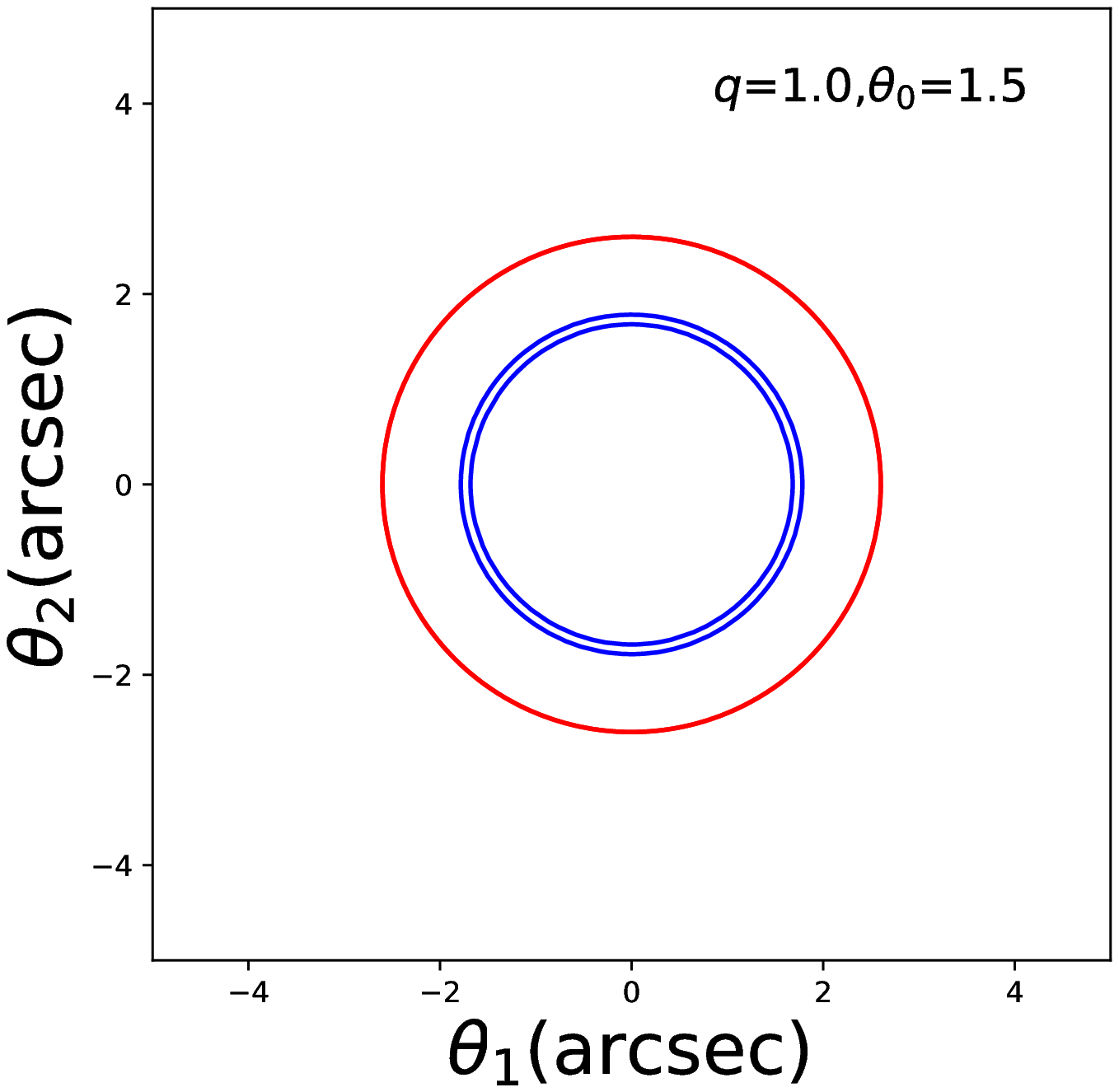}}
  \centerline{\includegraphics[width=5cm]{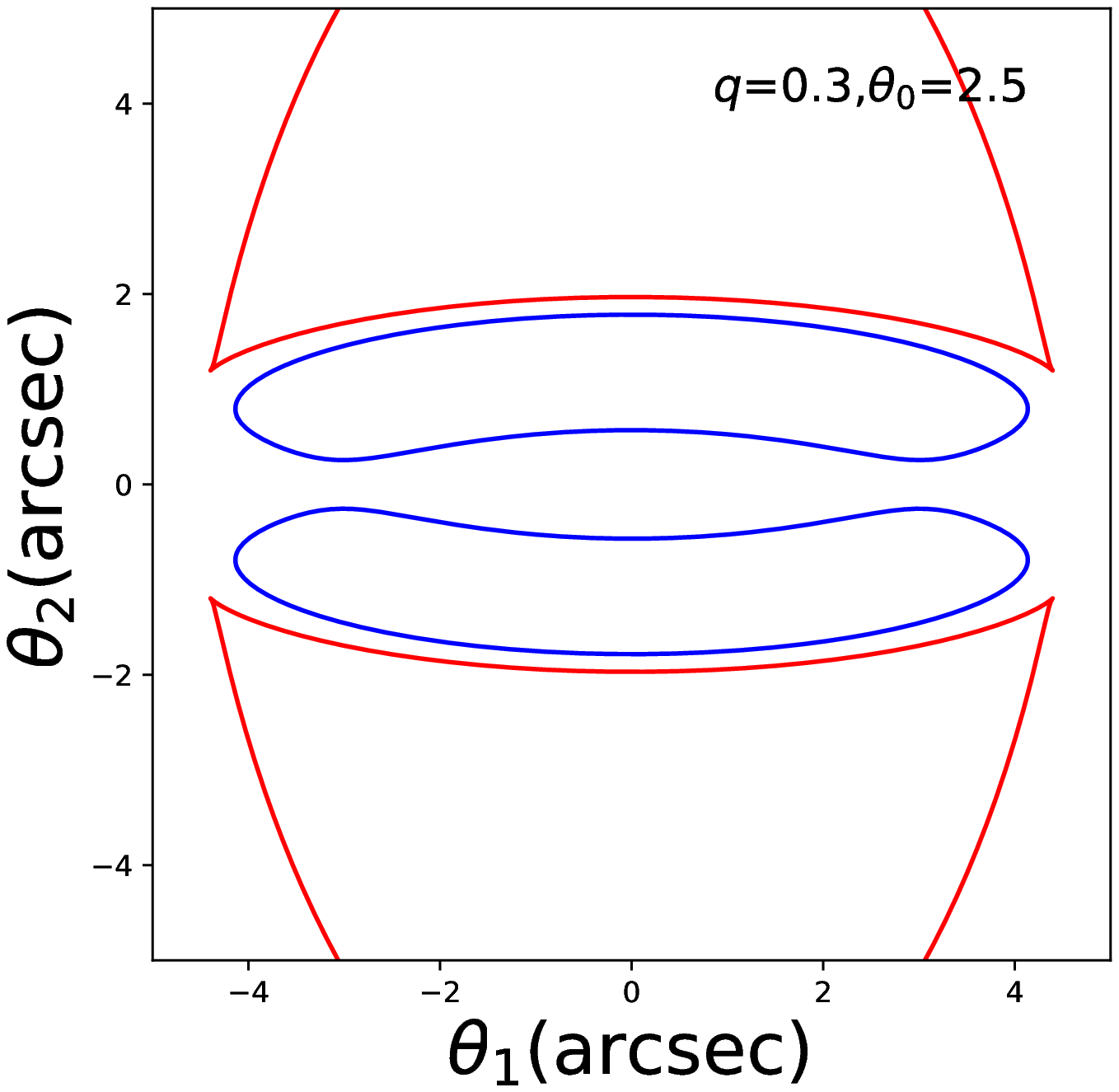}
    \includegraphics[width=5cm]{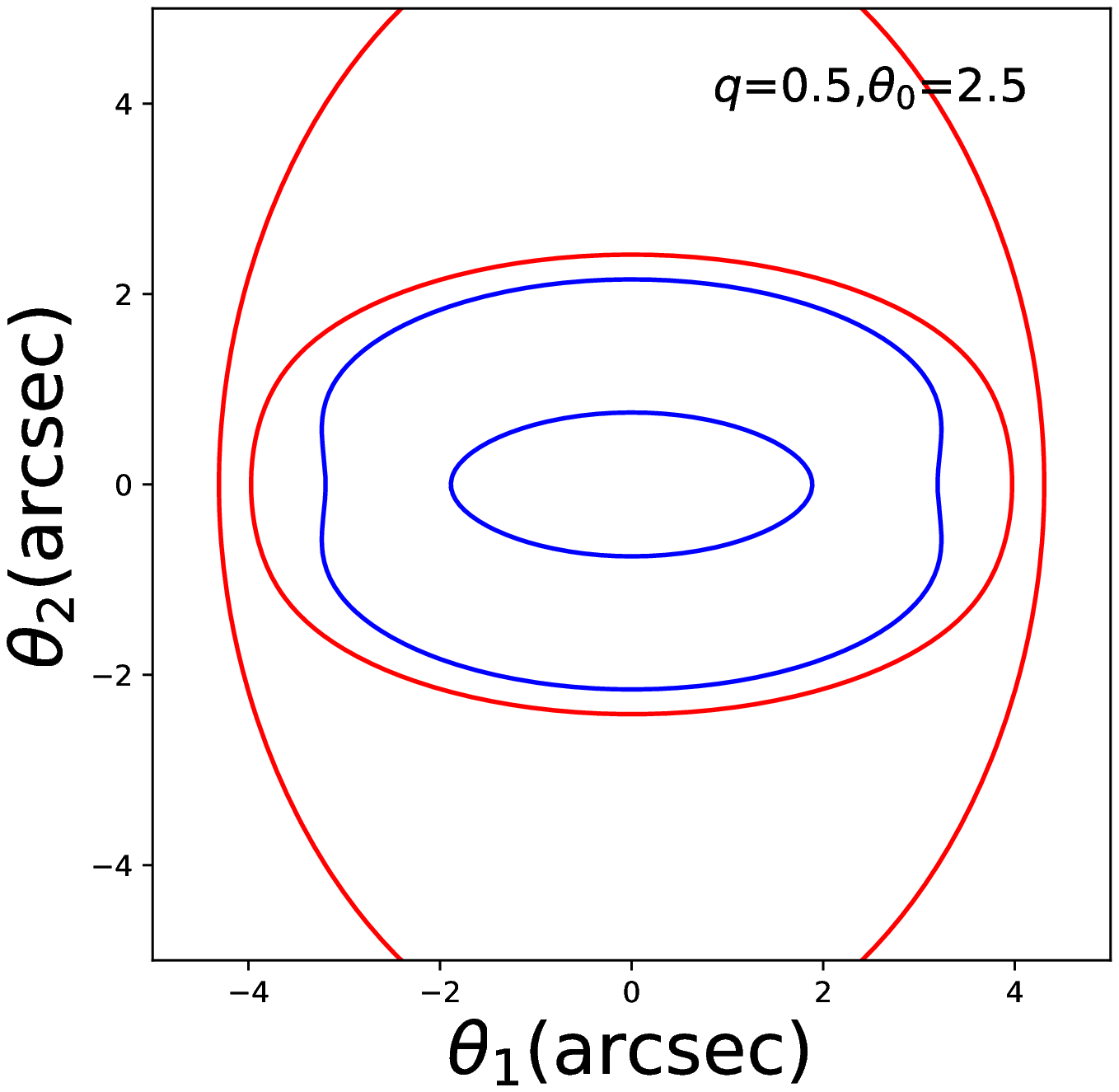}
    \includegraphics[width=5cm]{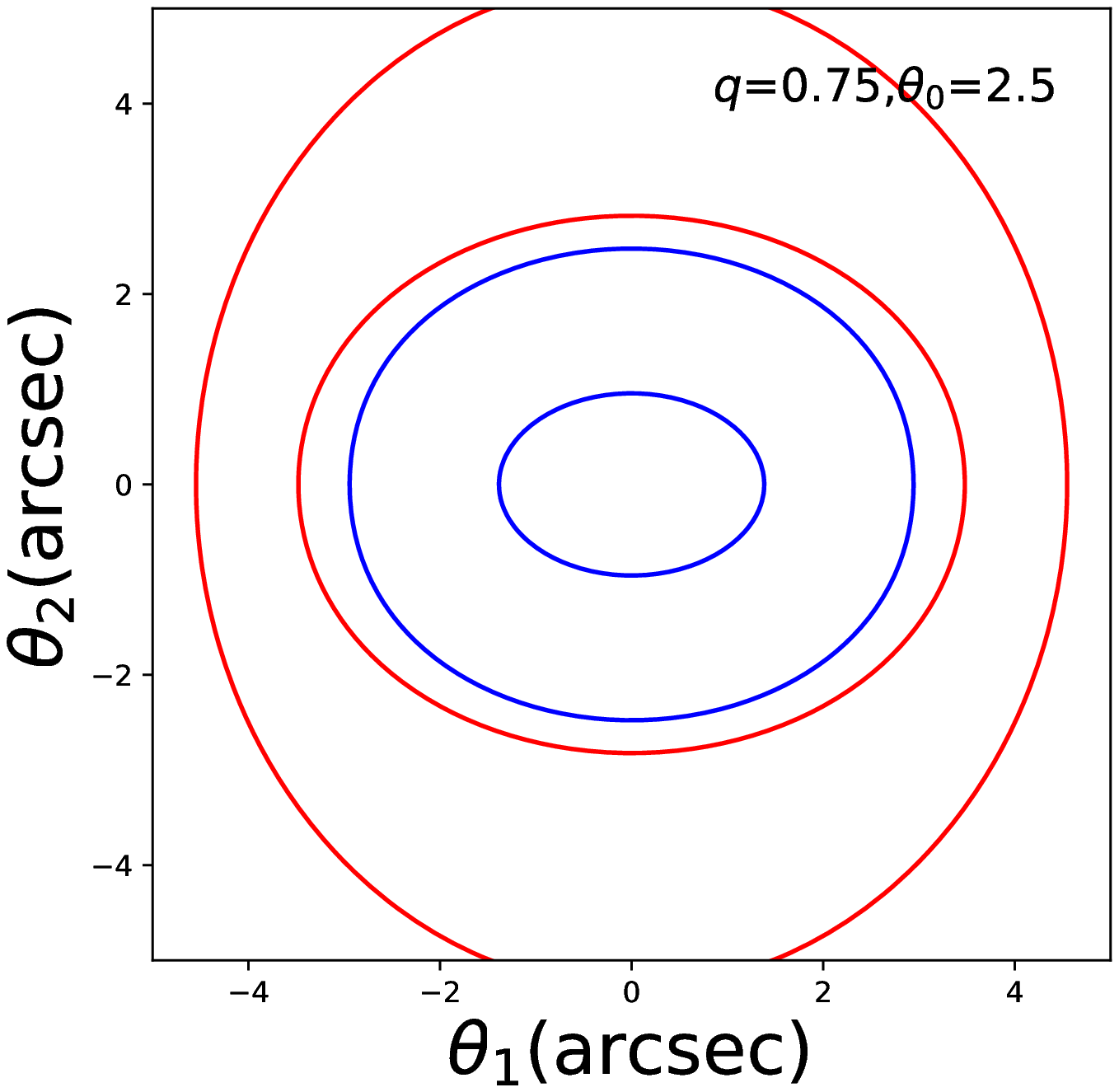}}
    \caption{The critical curves (blue) and caustics (red) of the elliptical
    exponential lens with $h=2$ and $\sigma=1$.}
  \label{fig:expcch2}
\end{figure*}

We show the critical curves (blue) and caustics (red) for the
elliptical exponential lenses with $h=1,2$ in Fig.\,\ref{fig:expcch1}
and \ref{fig:expcch2} respectively. We calculate the critical curves
by increasing the axis ratio from $0.05$ to $0.75$. The left and
middle panel in the magnification map (Fig.\,\ref{fig:expmag})
correspond to the bottom left in Fig.\,\ref{fig:expcch1} and top
middle panel in Fig.\,\ref{fig:expcch2}. For $h=1$, the dual-arrow
shape curves evolve into an elliptical shape as the axis ratio
increases. In most cases, the lenses with $h=1$ can generate one
critical curve, while the Gaussian models ($h=2$) can generate
two. The bottom right panel of Fig.\,\ref{fig:expcch1} present the
transitionary case of two criticals. In Fig.\,\ref{fig:expcch2}, for
sufficiently large $\theta_0$, the two leaf-shaped curves on the two
sides of the major axis will merge to two ellipses as the axis ratio
increases. The inner critical curve maps to the outer caustics, and
the major axis of the outer caustics aligns with the minor axis of the
lens. Moreover, the two leaves generated by the Gaussian model (top
middle panel) show similarity with that of the dual-component Gaussian
lens model, which may also cause model degeneracy.  The critical
curves of lenses with $h=3$ show similar behaviours as with $h=2$, and
for brevity we do not present them here.

\subsection{Softened power-law lenses}
\begin{figure*}
  \centerline{\includegraphics[width=5cm]{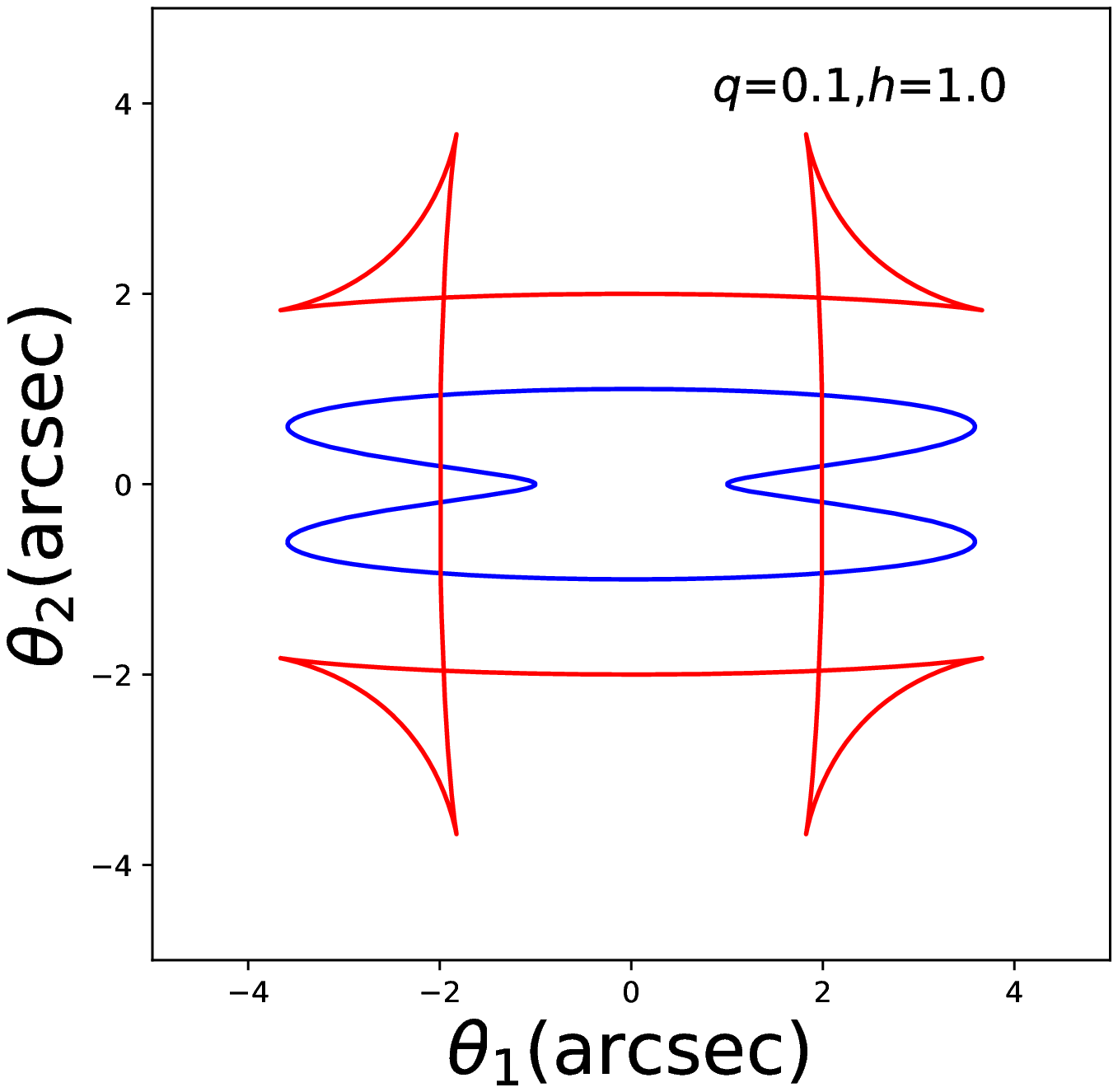}
    \includegraphics[width=5cm]{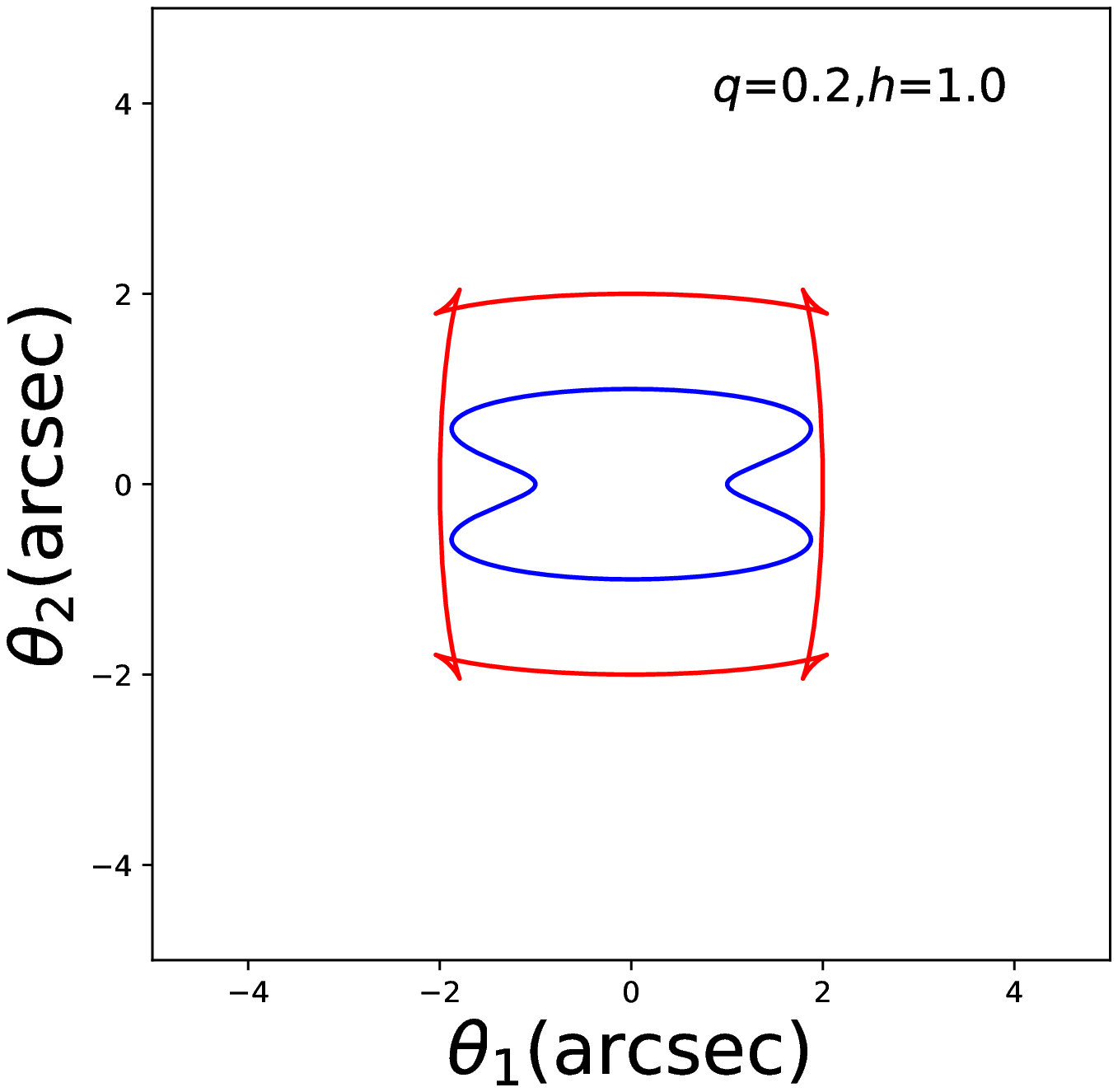}
    \includegraphics[width=5cm]{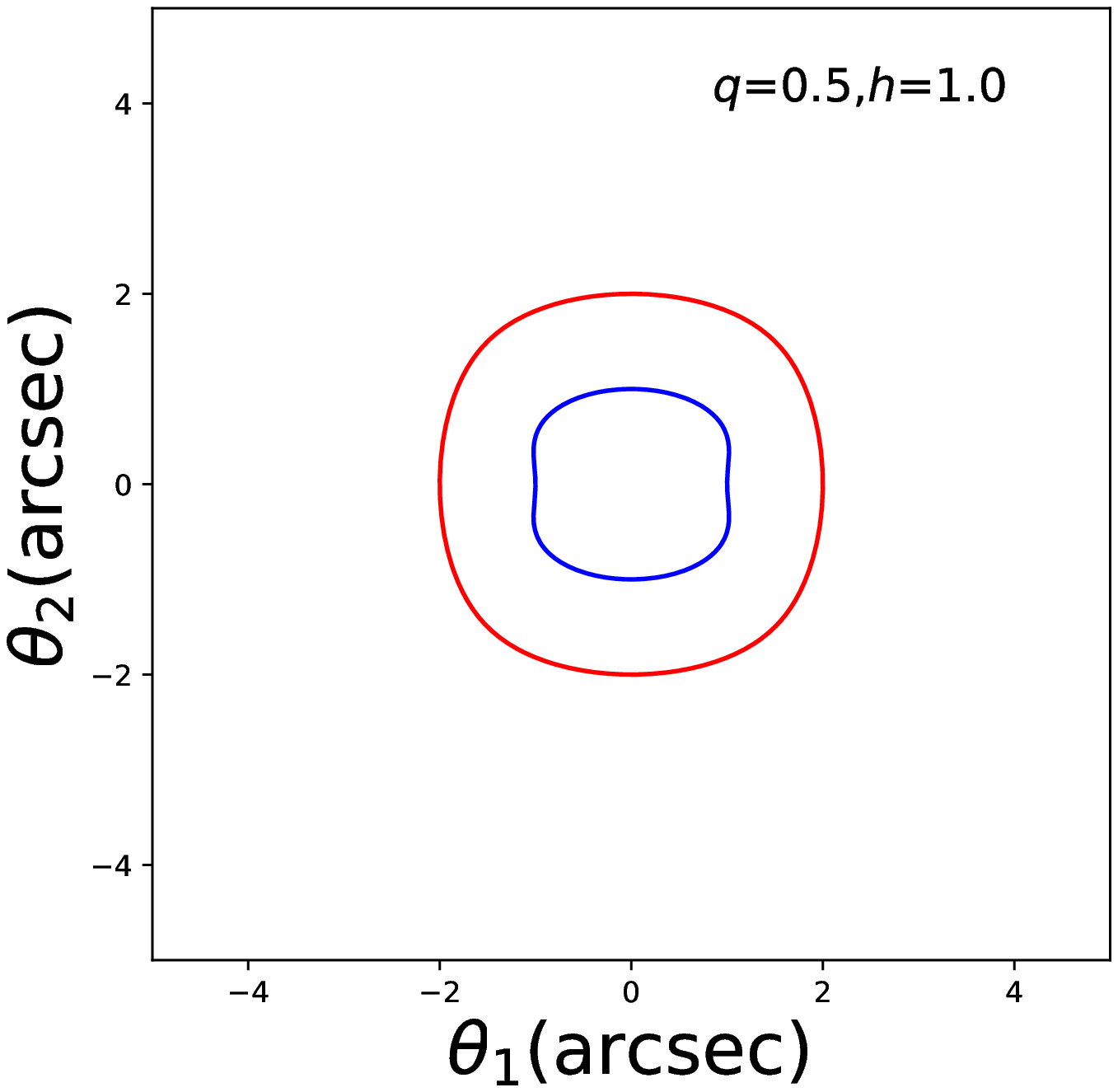}}
  \centerline{\includegraphics[width=5cm]{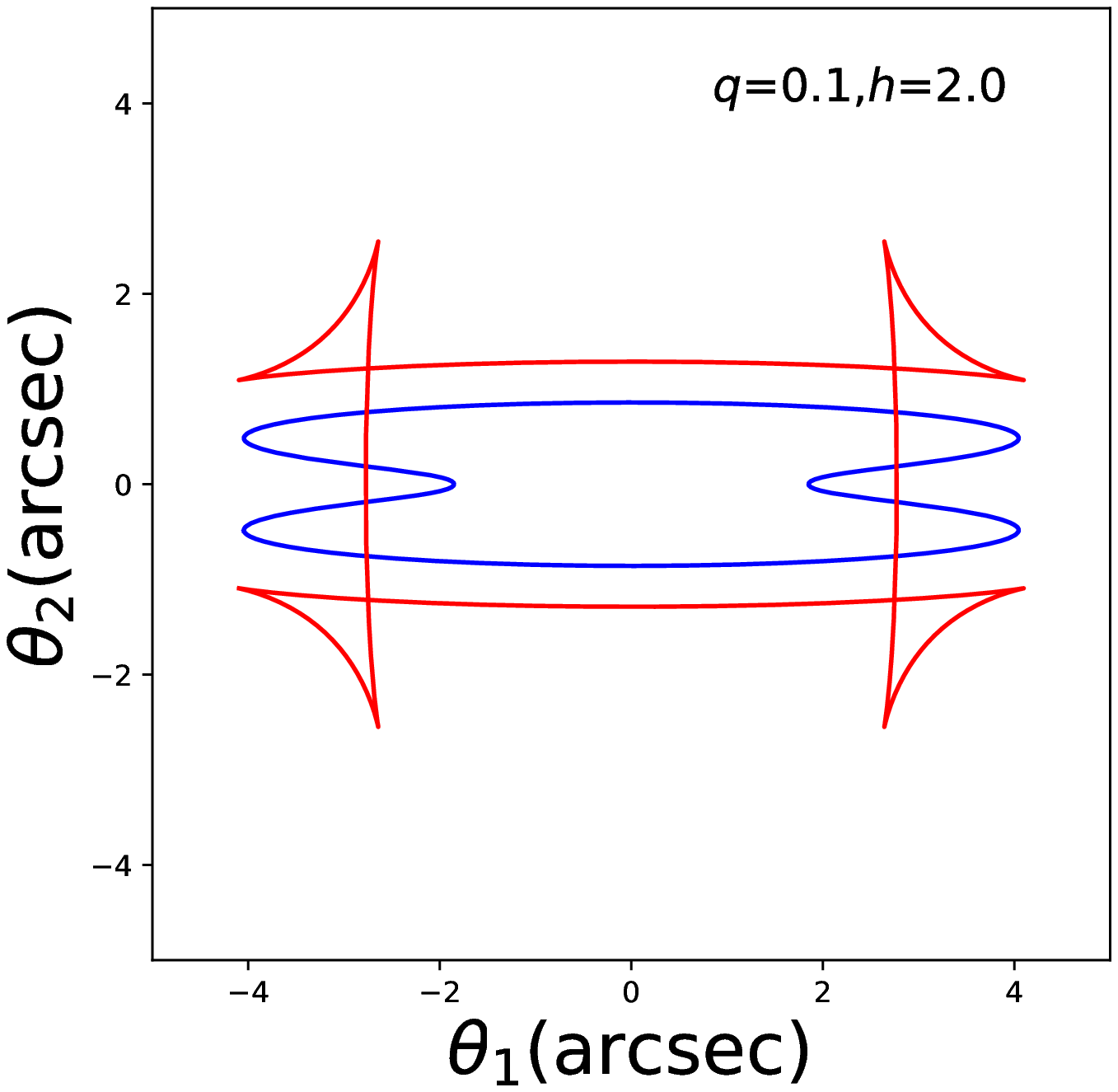}
    \includegraphics[width=5cm]{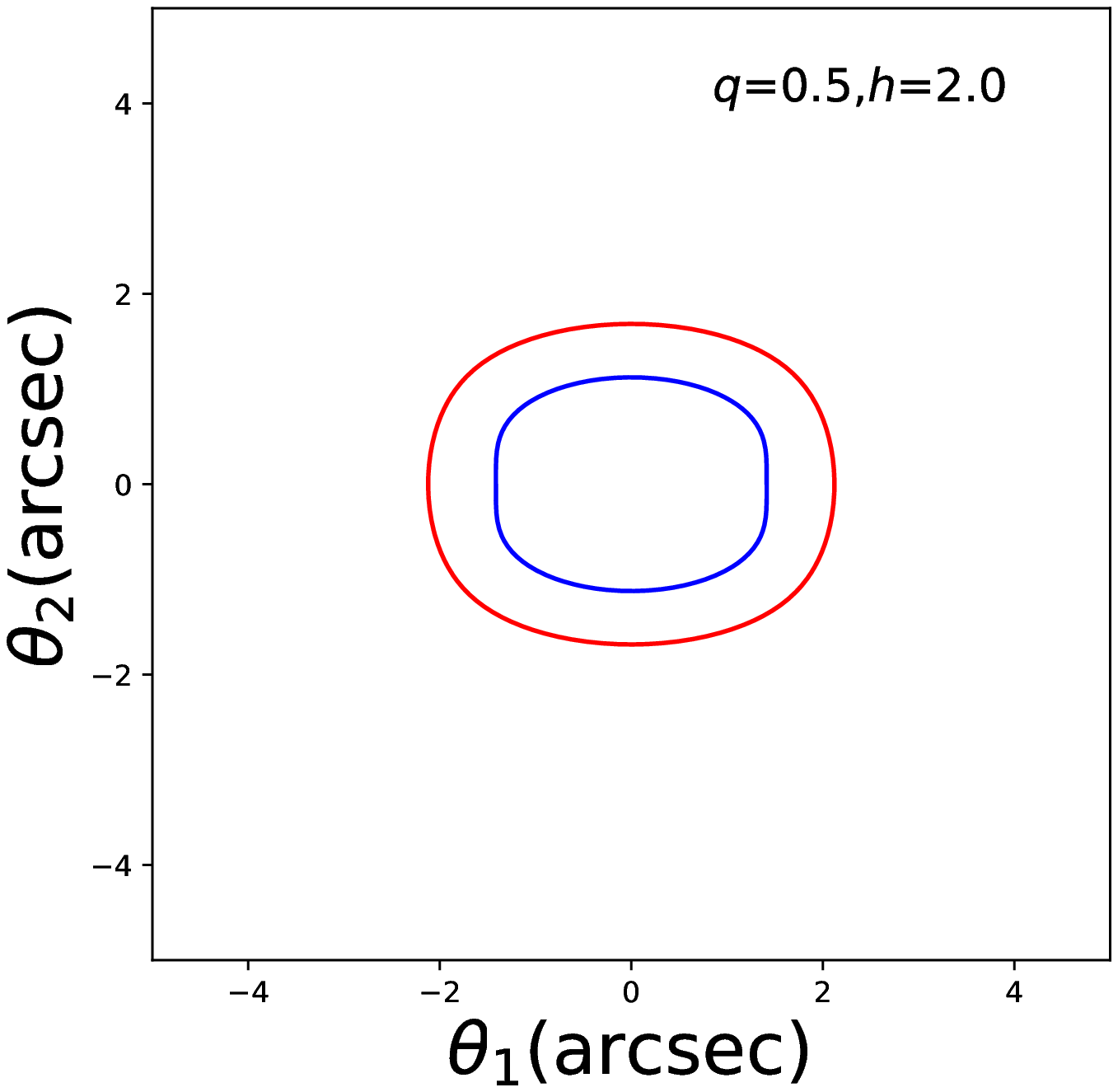}
    \includegraphics[width=5cm]{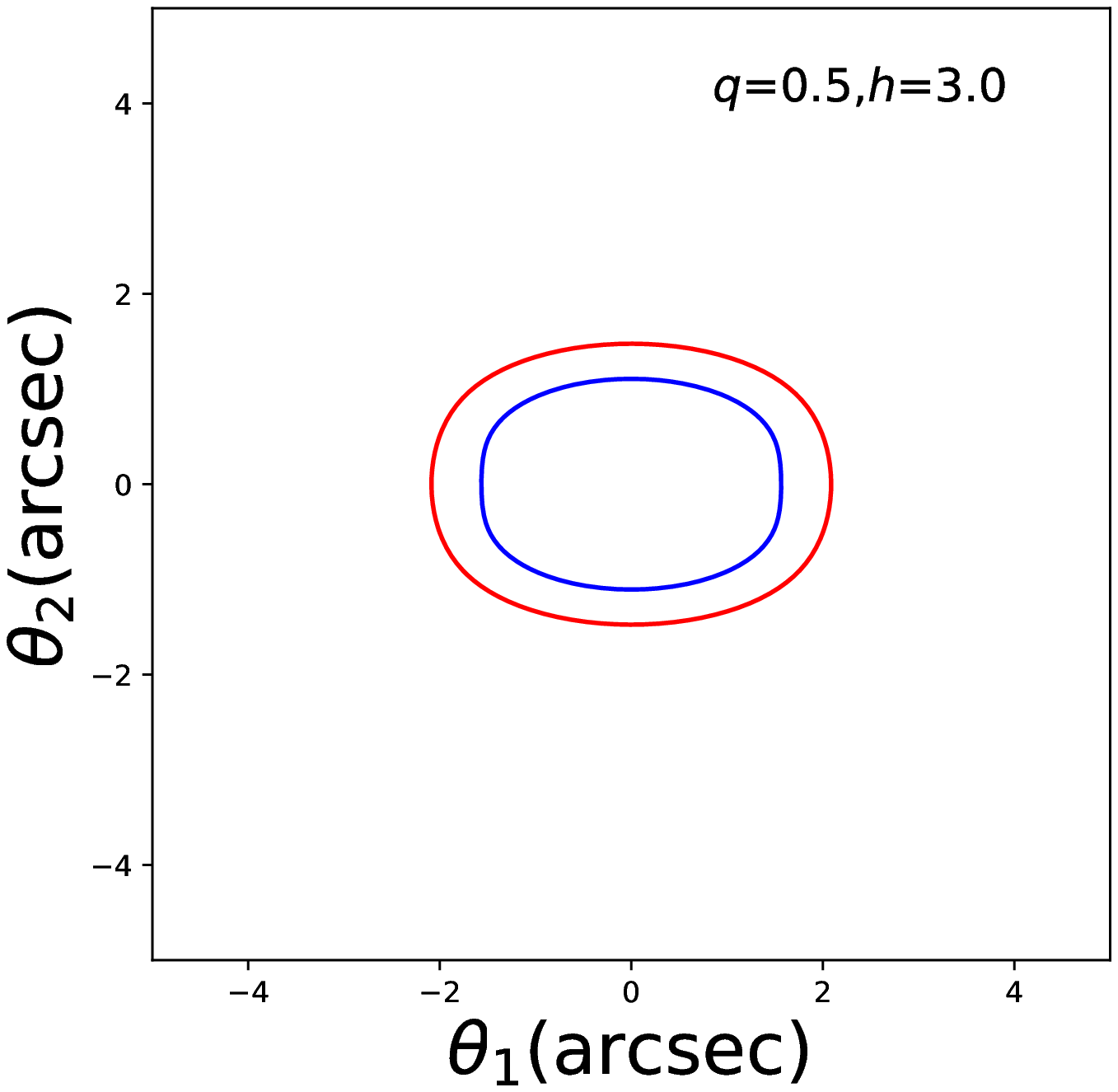}}
    \caption{The critical curves (blue) and caustics (red) of the
      elliptical singular power-law lenses ($\theta_c=0$) with $\theta_0=1$
      arcsec. The power index and axis ratio are given at top corner
      of each panel.}
  \label{fig:splc0h123}
\end{figure*}

In this section, we show examples of criticals and caustics of
the elliptical ESPL lenses. Firstly, Fig.\,\ref{fig:splc0h123} presents
the singular lens evolving from an elongated shape to a circular shape. The
singular lens can only generate one critical even for highly
elongated cases. For the lens of $h=1$, the critical curve crosses the
two axes at the same length ($\theta_1=\theta_2=\theta_0$) for all $q$, while this
is not the case of the lenses of $h=2,3$. Moreover, from the three
panels of $q=0.5$, one can see that the elliptical shape becomes more
significant with large value of $h$.

\begin{figure*}
  \centerline{\includegraphics[width=5cm]{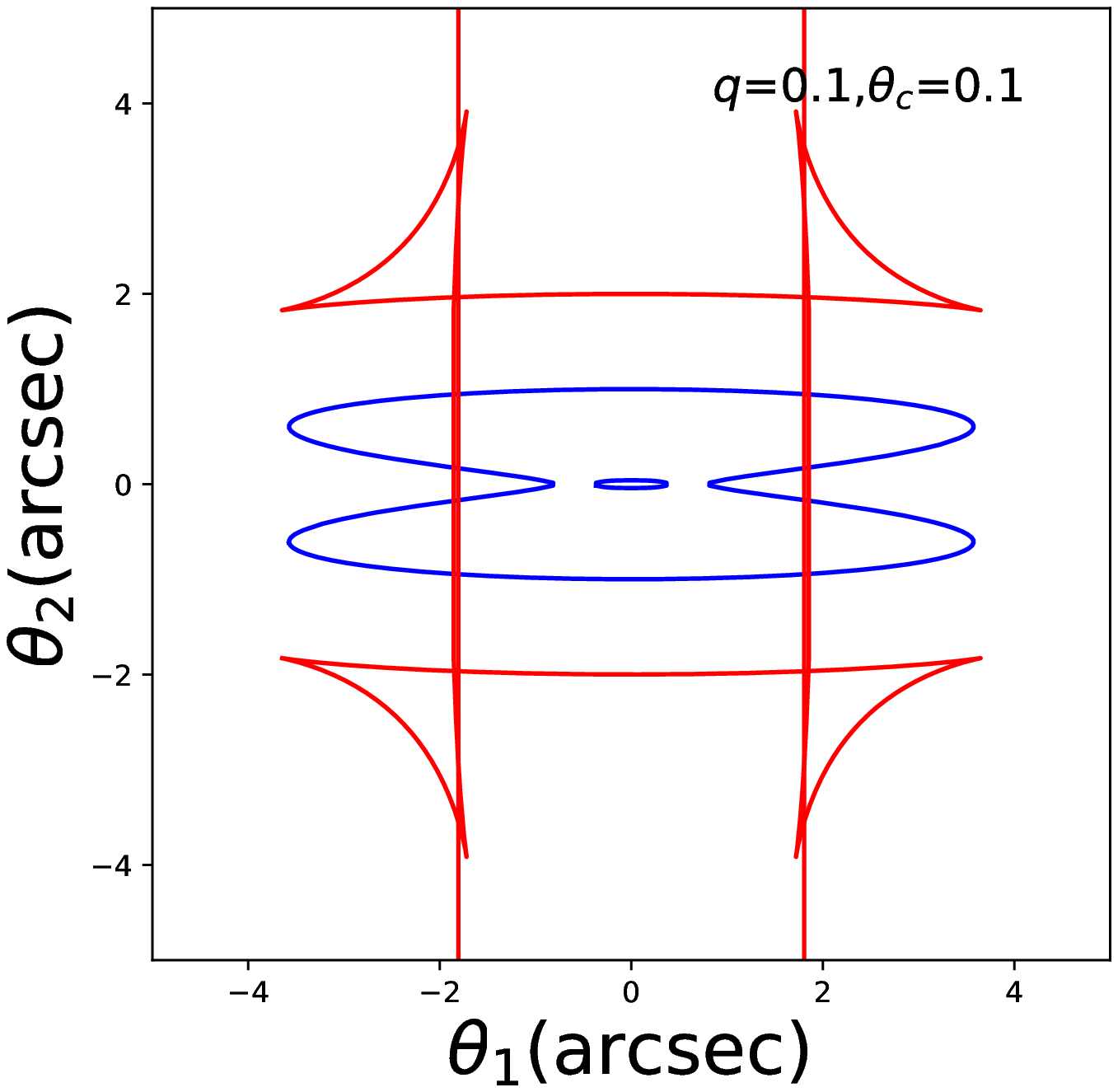}
    \includegraphics[width=5cm]{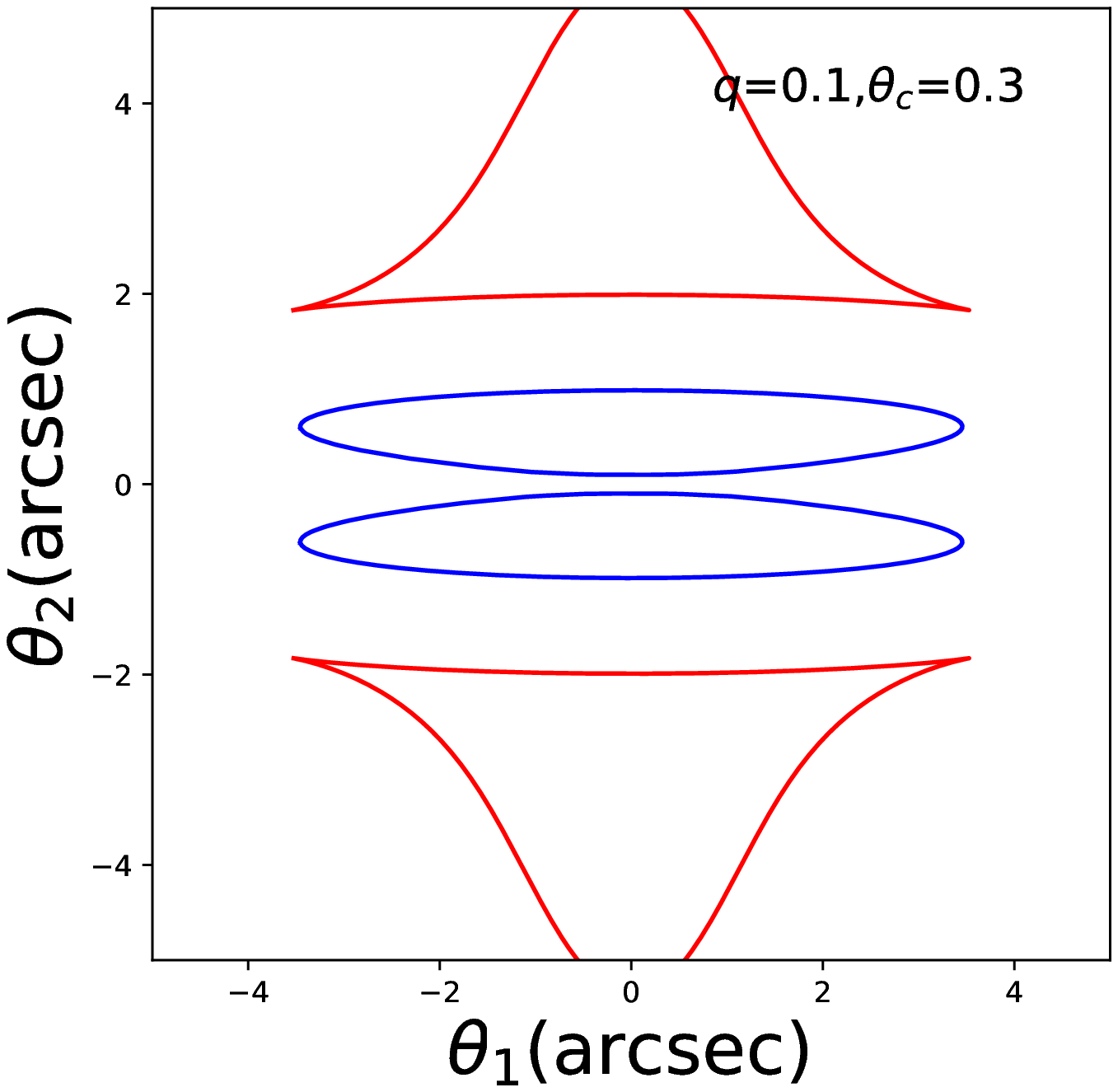}
    \includegraphics[width=5cm]{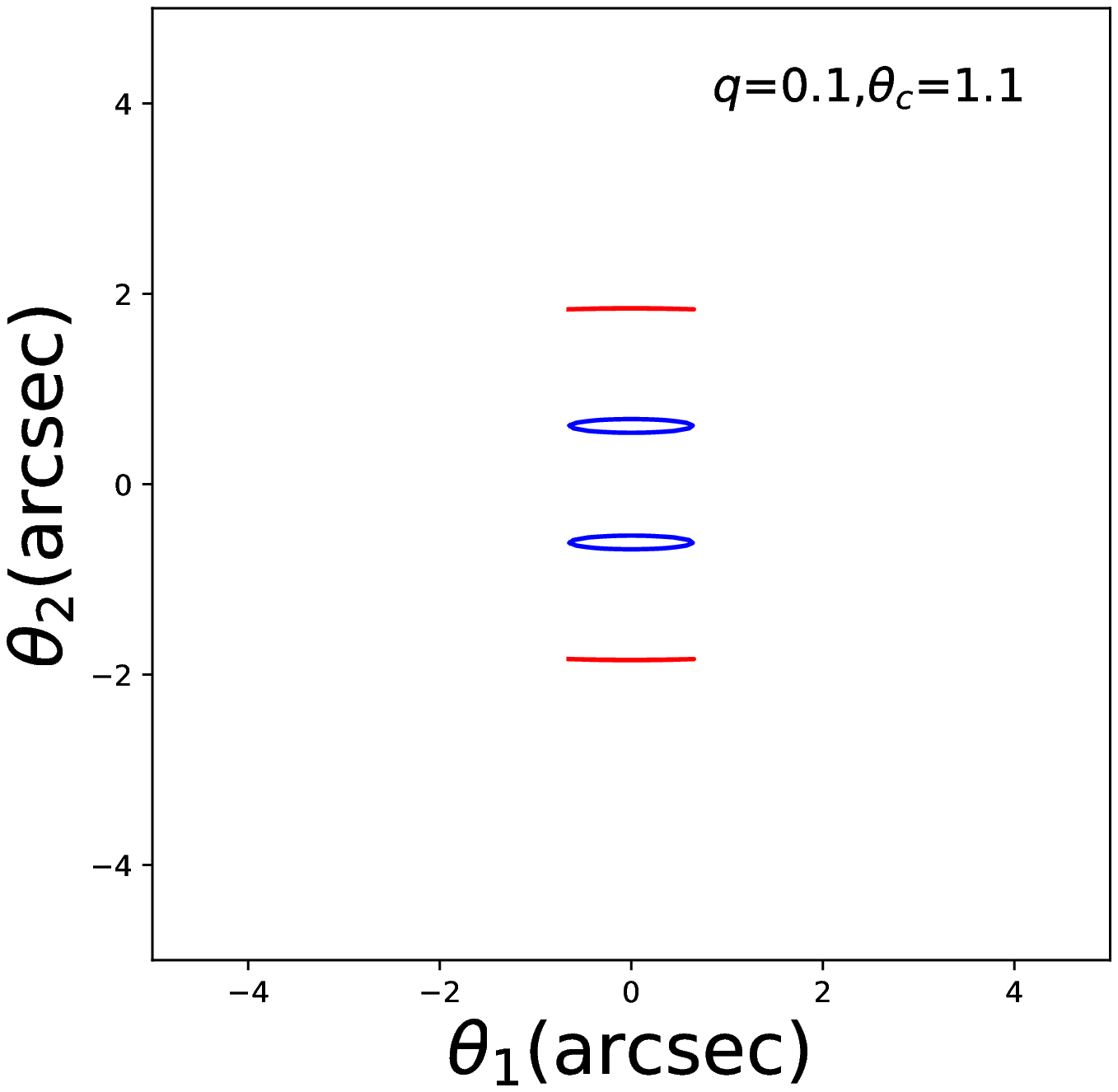}}
  \centerline{\includegraphics[width=5cm]{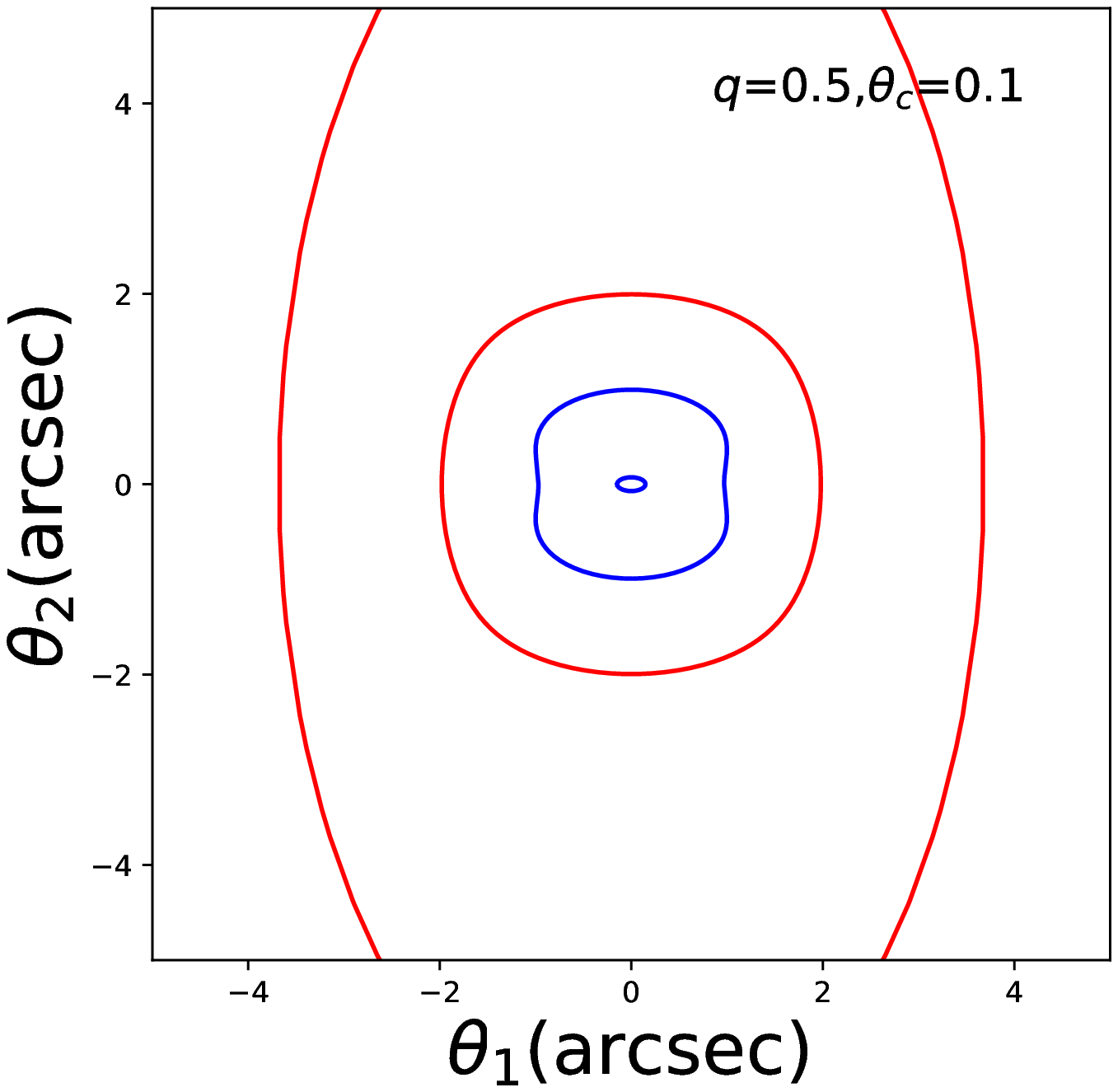}
    \includegraphics[width=5cm]{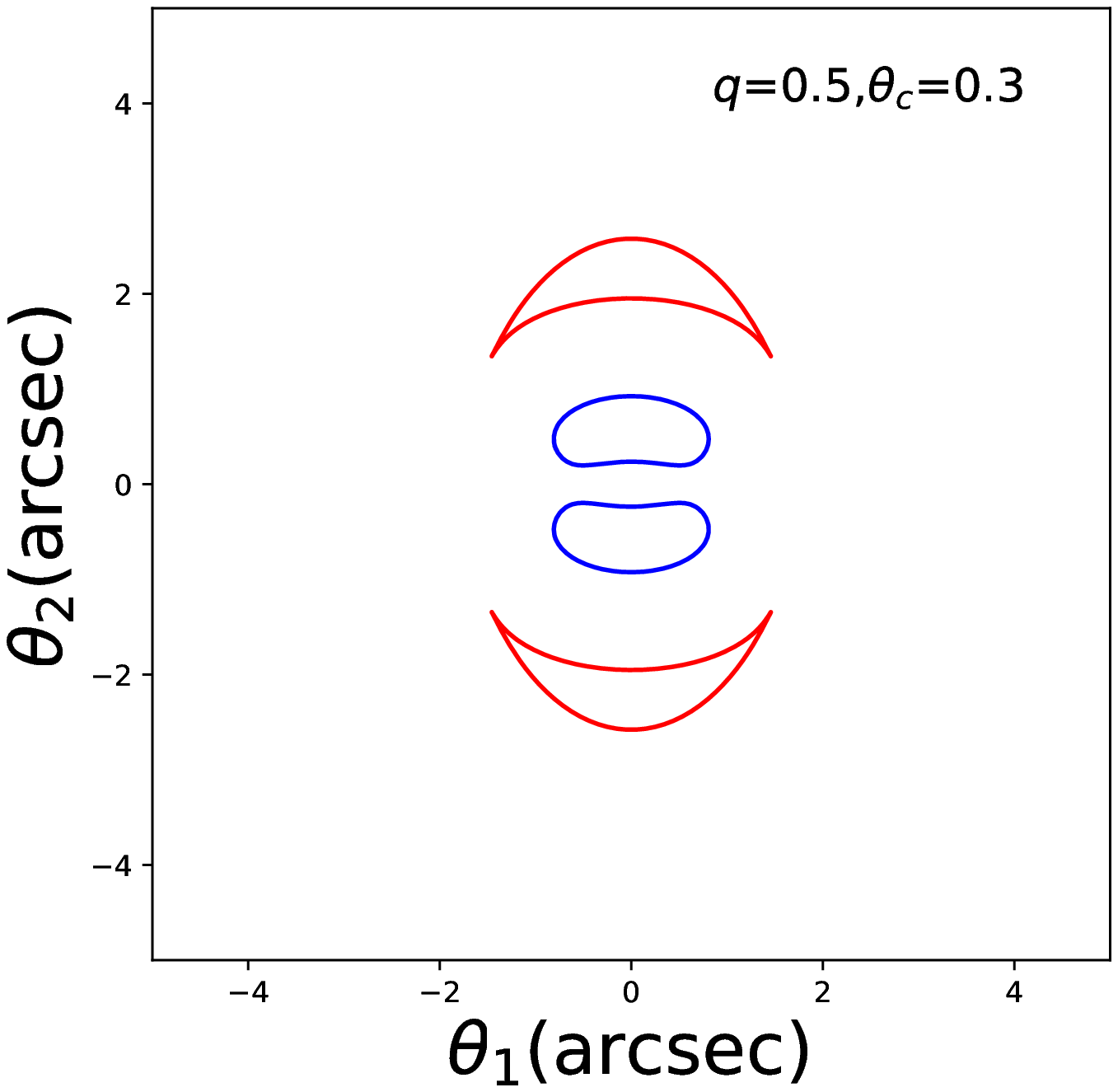}
    \includegraphics[width=5cm]{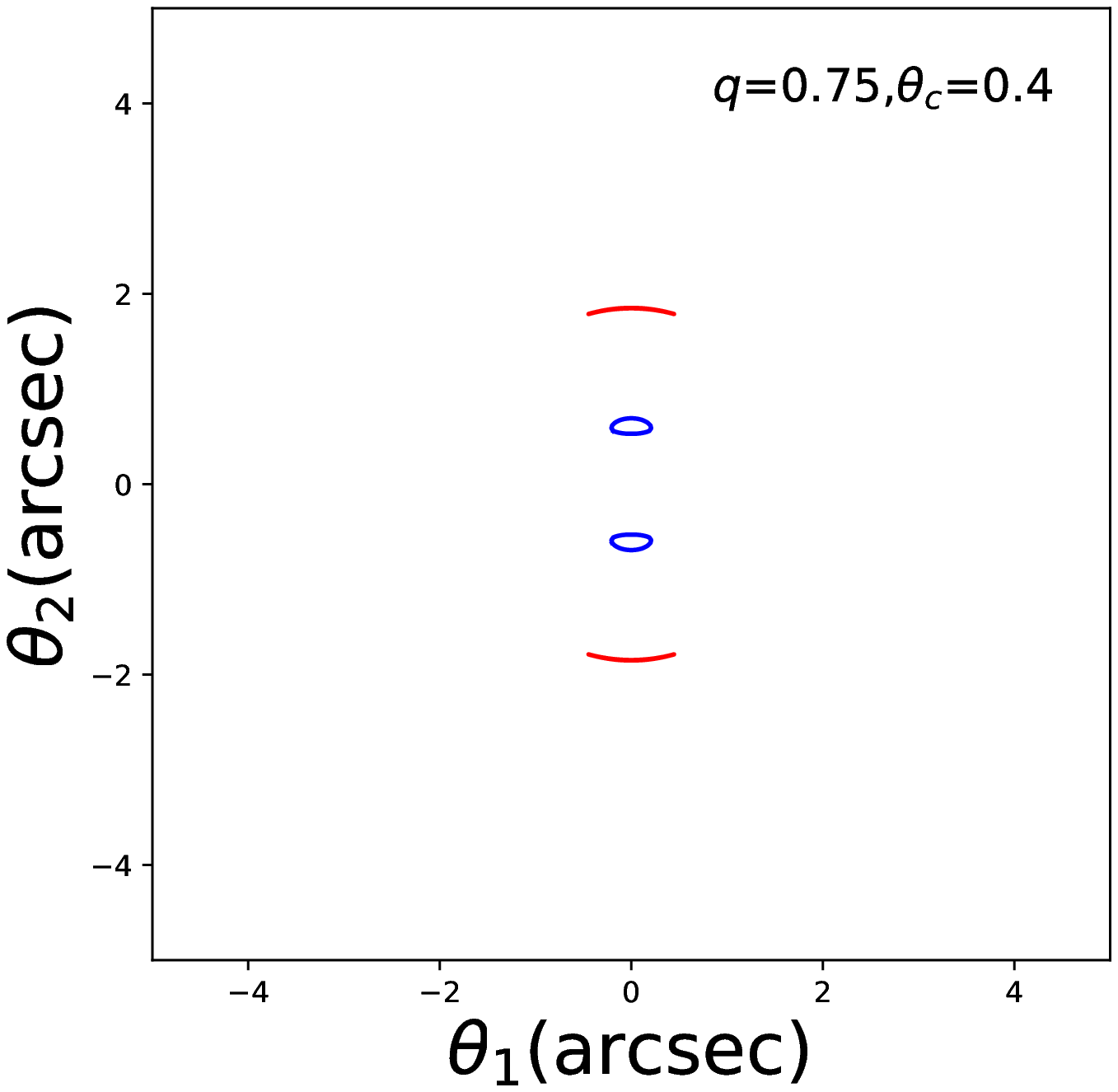}}
    \caption{The critical curves (blue) and the caustics (red) of the
      ESPL lens with parameters: $\theta_0=1$ arcsec and
      $h=1$. The core radius and the axis ratio are given at the top
      right corner of each panel.}
    \label{fig:splh1c}
\end{figure*}

\begin{figure*}
  \centerline{\includegraphics[width=5cm]{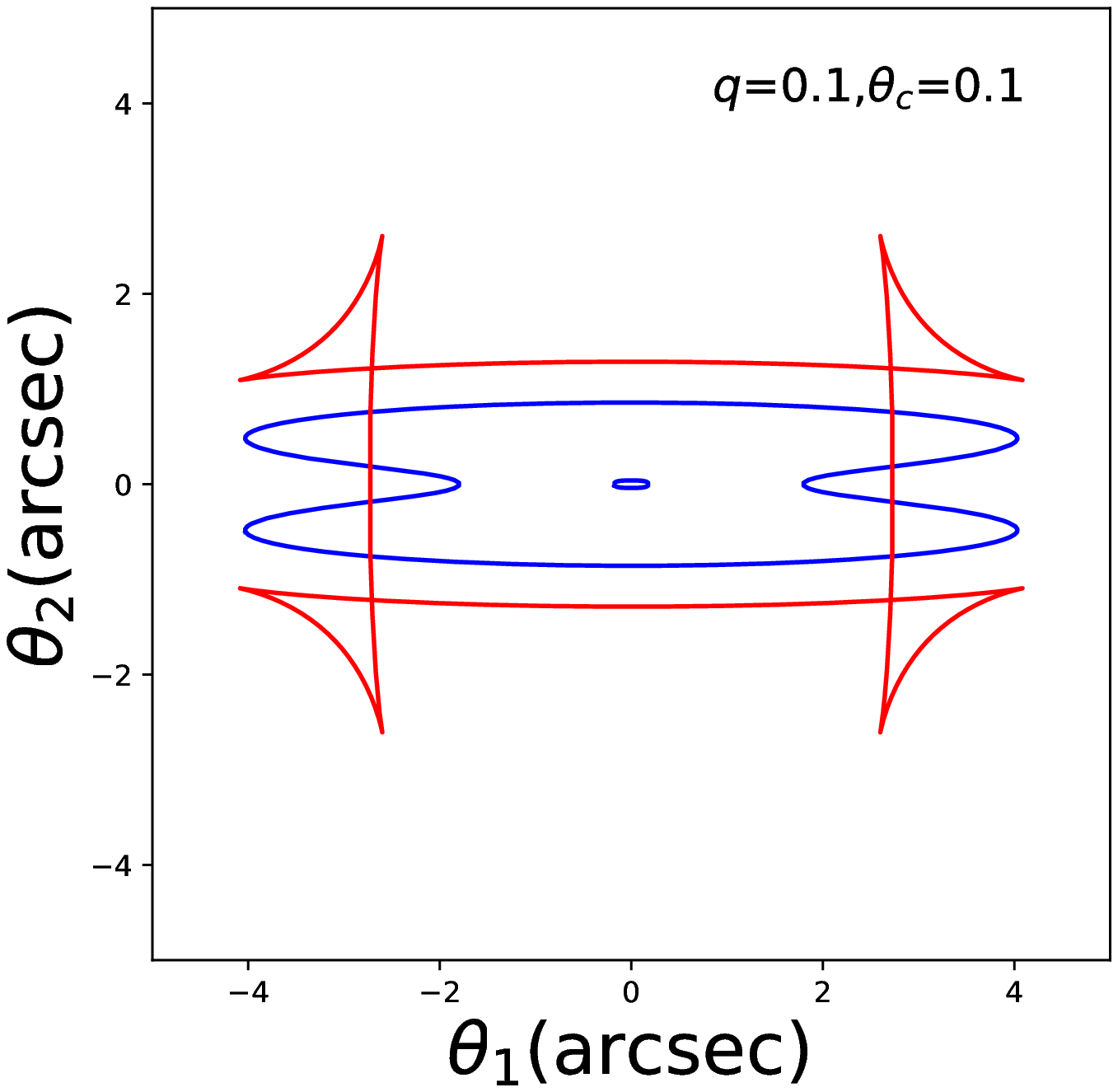}
    \includegraphics[width=5cm]{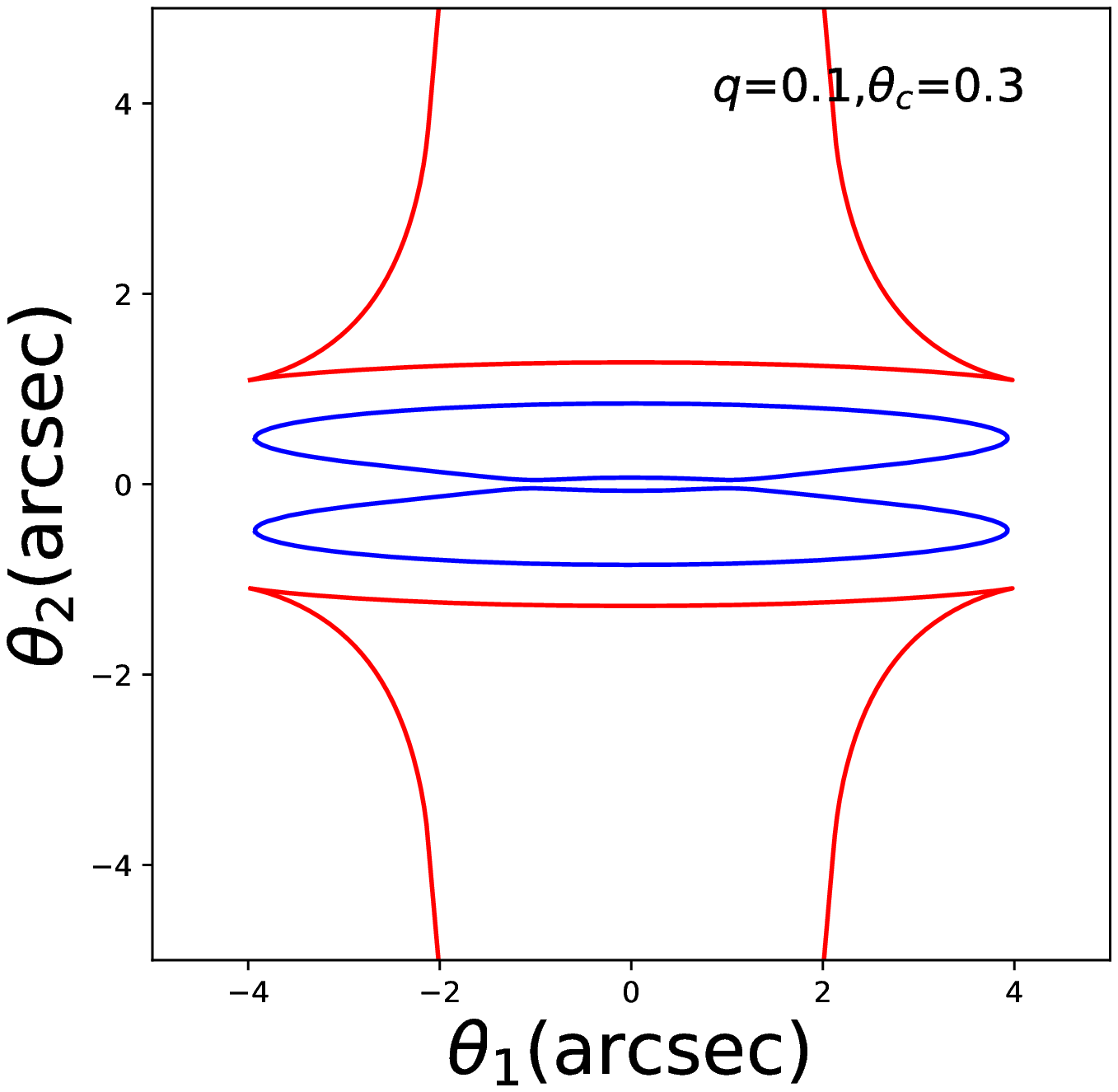}
    \includegraphics[width=5cm]{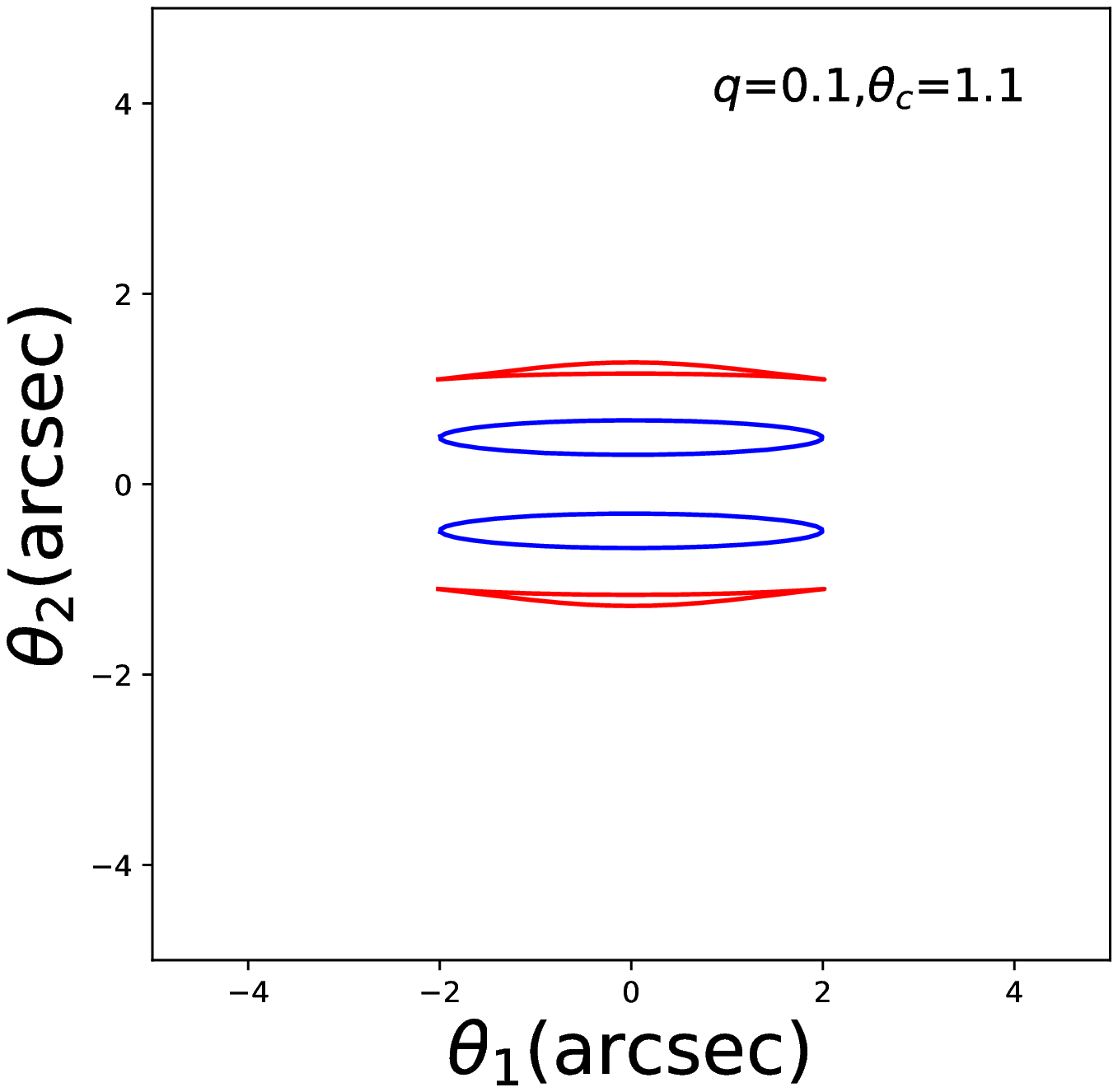}}
  \centerline{\includegraphics[width=5cm]{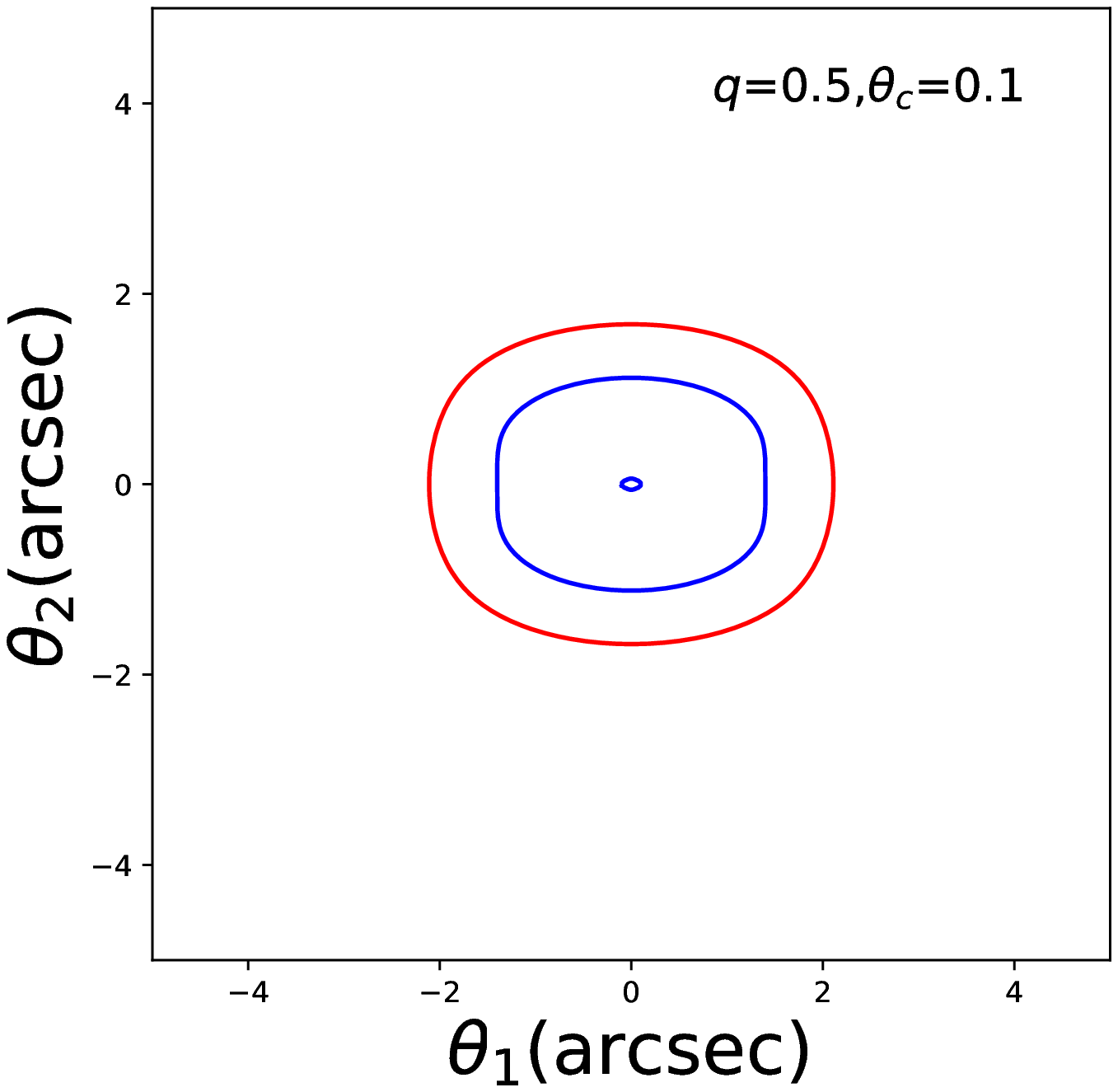}
    \includegraphics[width=5cm]{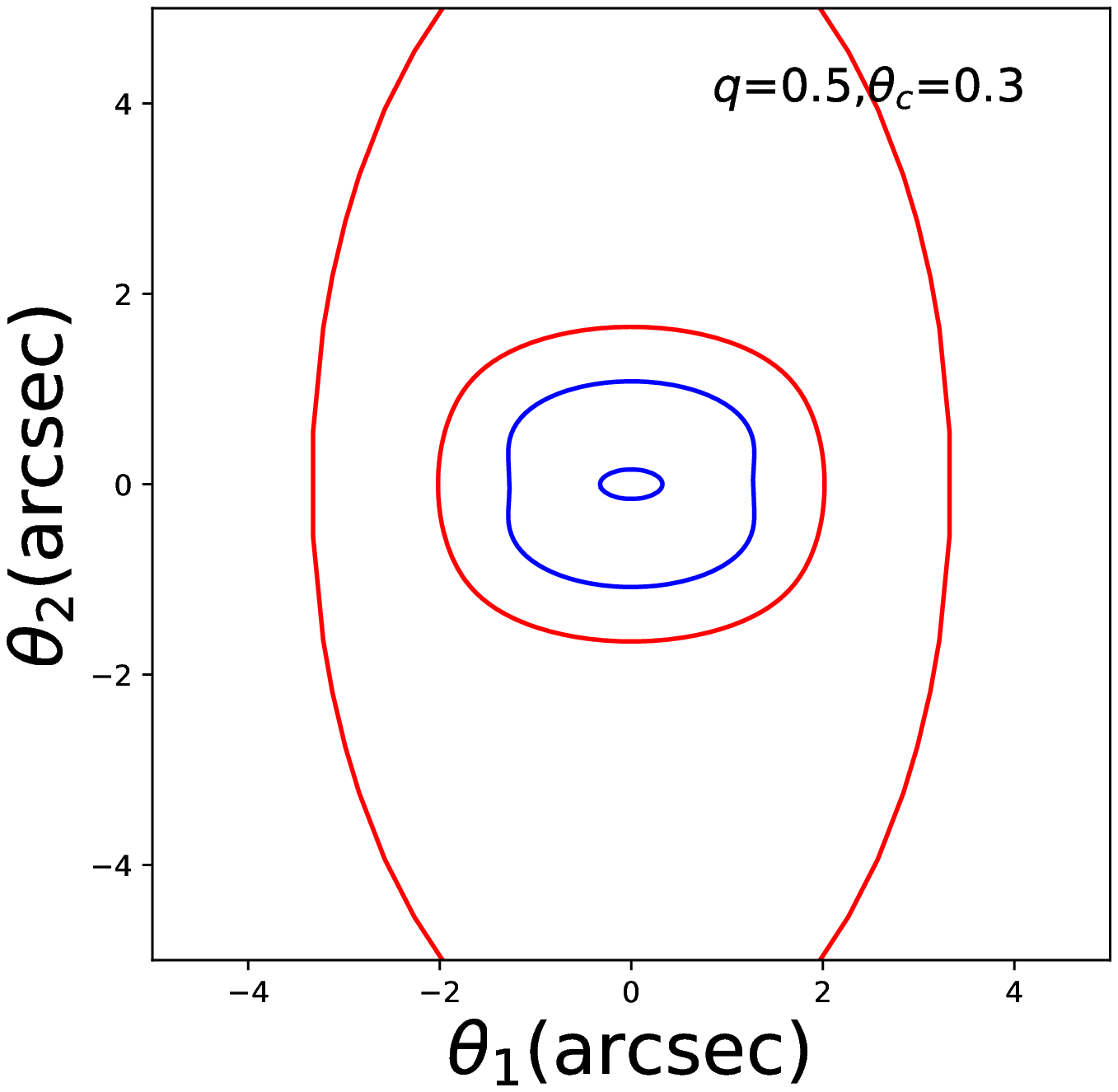}
    \includegraphics[width=5cm]{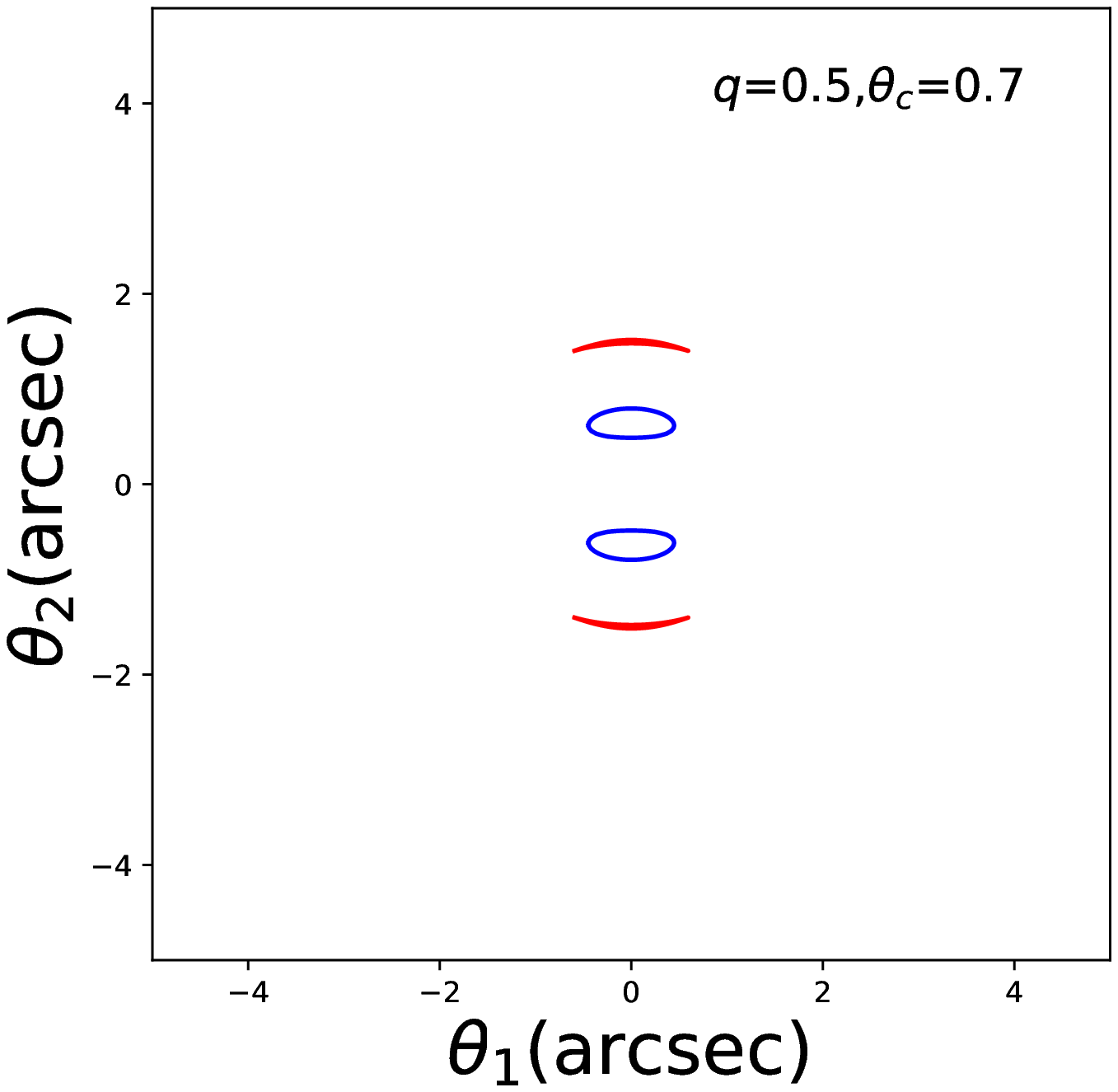}}
  \caption{Same as Fig.\,\ref{fig:splh1c} with $h=2$.}
    \label{fig:splh2c}
\end{figure*}

In Figs.\,\ref{fig:splh1c} and \ref{fig:splh2c}, we present the
critical curves and caustics for the ESPL lens of index $h=1$ and $h=2$
respectively. The top and middle panel in the magnification map
(Fig.\,\ref{fig:splmag}) correspond to the bottom middle panel in
Fig.\,\ref{fig:splh1c} and \ref{fig:splh2c}. The core radius causes
complex behaviour, i.e. the emergence of the second inner
critical curve in the core region of the lens. Increasing the core radius $\theta_c$ will cause the merger of the inner and outer critical, and forming two separated curves on both sides of the major axis.


\section*{Acknowledgments}
We thank the referee for constructive and valuable comments which increase
the scope of our work significantly and improve the overall quality even in some details of our paper.
We thank Artem Tuntsov, Oleg Yu. Tsupko and Shude Mao for interesting
discussions. XE is supported by NSFC Grant No. 11473032 and
No. 11873006. 
AR wishes to acknowledge the contribution of the Brandon University Research Committee.
\bibliographystyle{mnras}
\bibliography{eplasma}

\end{document}